\documentclass[12pt, twocolumn, numberedappendix, tighten]{aastex701}
\usepackage{natbib}
\usepackage{color} 
\usepackage{morefloats}
\usepackage{gensymb} 
\usepackage{longtable}
\usepackage{amsmath}
\usepackage{graphicx}
\usepackage{subfigure}
\usepackage{siunitx}
\usepackage{appendix}

\newcommand{\kms}{\ifmmode {\rm km\ s}^{-1} \else km s$^{-1}$\fi}
\newcommand{\ergs}{\ifmmode {\rm erg\ s}^{-1} \else erg s$^{-1}$\fi}
\newcommand{\ergscm}{\ifmmode {\rm erg\ s}^{-1} \else erg s$^{-1}$ cm$^{-2}$\fi}
\newcommand{\Msun}{\ifmmode {\rm M}_{\odot} \else $M_{\odot}$\fi }
\newcommand{\Lsun}{\ifmmode {\rm L}_{\odot} \else L$_{\odot}$\fi}
\newcommand{\qo}{\ifmmode q_{\rm o} \else $q_{\rm o}$\fi}
\newcommand{\Ho}{\ifmmode H_{\rm o} \else $H_{\rm o}$\fi}
\newcommand{\ho}{\ifmmode h_{\rm o} \else $h_{\rm o}$\fi}

\newcommand{\vFWHM}{\ifmmode v_{\mbox{\tiny FWHM}} \else
                    $v_{\mbox{\tiny FWHM}}$\fi}
\newcommand{\CCF}{\ifmmode F_{\it CCF} \else $F_{\it CCF}$\fi}
\newcommand{\ACF}{\ifmmode F_{\it ACF} \else $F_{\it ACF}$\fi}
\newcommand{\Halpha}{\ifmmode {\rm H}\alpha \else H$\alpha$\fi}
\newcommand{\Hbeta}{\ifmmode {\rm H}\beta \else H$\beta$\fi}
\newcommand{\Hgamma}{\ifmmode {\rm H}\gamma \else H$\gamma$\fi}
\newcommand{\Hdelta}{\ifmmode {\rm H}\delta \else H$\delta$\fi}
\newcommand{\Lya}{\ifmmode {\rm Ly}\alpha \else Ly$\alpha$\fi}
\newcommand{\Lyb}{\ifmmode {\rm Ly}\beta \else Ly$\beta$\fi}
\newcommand{\HeI}{\ifmmode {\rm He}\,{\sc i}\,\lambda5876 \else 
	          He\,{\sc i}\,$\lambda5876$\fi}
\newcommand{\HeII}{\ifmmode {\rm He}\,{\sc ii}\,\lambda4686 \else 
	           He\,{\sc ii}\,$\lambda4686$\fi}

\newcommand{\ciii}{\ifmmode {\rm C}\,{\sc iii} \else C\,{\sc iii}\fi}
\newcommand{\civ}{C\,{\sc iv}\ }

\newcommand{\siiv}{Si\,{\sc iv}}

\newcommand{\mbh}{$M_{\rm BH}$\ }

\newcommand{\javelin}{{\tt JAVELIN}\ }

\newcommand{\pyroa}{{\tt PyROA}\ }
\newcommand{\pyccf}{{\tt PyCCF}\ }

\shortauthors{Roessler et al.}

\begin{document}

\title{The SDSS-V Black Hole Mapper Reverberation Mapping Project: Time-delayed Variability of a Quasar's C IV Broad Absorption Troughs}

\author{Kaleb Roessler}
\email{kroessler04@gmail.com}
\affiliation{Department of Astronomy, University of Wisconsin-Madison, 475 N. Charter St, Madison, WI 53706, USA} 

\author[orcid=0000-0001-9920-6057]{Catherine J. Grier}
\email{kate.grier@astro.wisc.edu}
\affiliation{Department of Astronomy, University of Wisconsin-Madison, 475 N. Charter St, Madison, WI 53706, USA} 

\author[orcid=0000-0003-0796-362X]{Loren Robinson}
\email{ljrobinson4@wisc.edu}
\affiliation{Department of Astronomy, University of Wisconsin-Madison, 475 N. Charter St, Madison, WI 53706, USA} 

\author[orcid=0000-0002-0167-2453]{W.~N. Brandt}
\affiliation{Department of Astronomy and Astrophysics, 525 Davey Lab, The Pennsylvania State University, University Park, PA 16802, USA}
\affiliation{Institute for Gravitation and the Cosmos, The Pennsylvania State University, University Park, PA 16802, USA}
\affiliation{Department of Physics, 104 Davey Laboratory, The Pennsylvania State University, University Park, PA 16802, USA}
\email{wnbrandt@gmail.com}

\author[orcid=0000-0002-1763-5825]{Patrick B.\ Hall}
\affiliation{Department of Physics and Astronomy, York University, 4700 Keele St., Toronto, ON M3J 1P3, Canada}
\email{phall@yorku.ca}

\author[orcid=0000-0003-1659-7035]{Yue Shen}
\affiliation{Department of Astronomy, University of Illinois at Urbana-Champaign, Urbana, IL 61801, USA}
\affiliation{National Center for Supercomputing Applications, University of Illinois at Urbana-Champaign, Urbana, IL 61801, USA}
\email{shenyue@illinois.edu}

\author[orcid=0000-0002-1410-0470]{Jonathan R. Trump}
\affiliation{Department of Physics, 196A Auditorium Road, Unit 3046, University of Connecticut, Storrs, CT 06269, USA}
\email{jonathan.trump@uconn.edu}

\author[orcid=0000-0003-1728-0304]{Keith Horne}
\affiliation{SUPA Physics and Astronomy, University of St Andrews, Fife, KY16 9SS, UK}
\email{kdh1@st-andrews.ac.uk}

\author{William J. Nichols} 
\email{wjnichols2@wisc.edu} 
\affiliation{Department of Astronomy, University of Wisconsin-Madison, 475 N. Charter St, Madison, WI 53706, USA} 

\author[orcid=0000-0001-9776-9227]{Megan C. Davis}
\affiliation{Department of Physics, 196A Auditorium Road, Unit 3046, University of Connecticut, Storrs, CT 06269, USA}
\email{megan.c.davis@uconn.edu}

\author{Adrian Amaya} 
\affiliation{Department of Physics and Astronomy, York University, 4700 Keele St., Toronto, ON M3J 1P3, Canada}
\email{adriantonalliamaya@gmail.com}

\author{Halima Mohamud} 
\affiliation{Department of Physics and Astronomy, York University, 4700 Keele St., Toronto, ON M3J 1P3, Canada}
\email{halima04@my.yorku.ca}

\author{Lucas M. Seaton}
\affiliation{Department of Physics and Astronomy, York University, 4700 Keele St., Toronto, ON M3J 1P3, Canada}
\email{lucas.seaton@hotmail.com}


\author[orcid=0000-0002-6404-9562]{Scott F. Anderson}
\email{sfander@uw.edu}
\affiliation{Astronomy Department, University of Washington, Box 351580, Seattle, WA 98195, USA}

\author{Carolina Andonie}
\email{cpandonie@mpe.mpg.de}
\affiliation{Max-Planck-Institut für Extraterrestrische Physik, Giessenbachstraße, 85748 Garching, Germany}


\author[orcid=0000-0002-3601-133X]{Dmitry Bizyaev}
\affiliation{Apache Point Observatory and New Mexico State
University, P.O. Box 59, Sunspot, NM, 88349-0059, USA} 
\affiliation{Sternberg Astronomical Institute, Moscow State
University, Moscow, 119234, Russia}
\email{dmbiz@apo.nmsu.edu}


\author{Mauricio Elías-Chávez} 
\affiliation{Universidad Nacional Autónoma de México, Instituto de Astronomía, A.P. 106, E-22800 Ensenada, B.C., México}
\email{dgonzalez@astro.unam.mx} 

\author[orcid=0000-0001-9203-2808]{Xinyu Dai}
\affil{Homer L.\ Dodge Department of Physics and Astronomy, University of Oklahoma, Norman, OK 73019, USA}
\email{xdai@ou.edu}


\author[0000-0002-6770-2627]{Sean Morrison}
\affiliation{Department of Astronomy, University of Illinois at Urbana-Champaign, Urbana, IL 61801, USA}
\email{smorris0@illinois.edu}

\author[orcid=0000-0002-2835-2556]{Kaike Pan}
\affiliation{Apache Point Observatory and New Mexico State
University, P.O. Box 59, Sunspot, NM, 88349-0059, USA} 
\email{kpan@apo.nmsu.edu}


\author[orcid=0000-0003-0677-785X]{Paola Rodr\'iguez Hidalgo}
\affiliation{Physical Sciences Division, School of STEM, University of Washington Bothell, WA, 98011, USA}
\email{paola@uw.edu}

\author[orcid=0000-0001-7240-7449]{Donald P. Schneider}
\affiliation{Department of Astronomy and Astrophysics, 525 Davey Lab, The Pennsylvania State University, University Park, PA 16802, USA}
\affiliation{Institute for Gravitation and the Cosmos, The Pennsylvania State University, University Park, PA 16802, USA}
\email{dps7@psu.edu}

\author[orcid=0000-0002-8501-3518]{Zachary Stone}
\affiliation{Department of Astronomy, University of Illinois at Urbana-Champaign, Urbana, IL 61801, USA}
\email{stone28@illinois.edu}


\begin{abstract}
Broad absorption line (BAL) quasars offer a powerful view into the structure, kinematics, and feedback mechanics of active galactic nuclei. In this study, we track the dramatic behavior of intrinsic \civ outflows of the quasar SDSS J142129.40+522752.0 at $z = 3.228$. We analyze 199 high-signal-to-noise optical spectra collected over a 12-year baseline via the Sloan Digital Sky Survey III/IV Reverberation Mapping and the SDSS-V Black Hole Mapper Reverberation Mapping programs. We quantify multi-year variability across two prominent absorption features: a primary high-velocity \civ BAL trough (Trough A) and a lower-velocity mini-BAL (Trough B). Both troughs exhibit rapid, significant variations in equivalent width and average depth, culminating in temporary disappearances during the 2022–2023 observing seasons before reappearing at their original velocities. We find that the behavior of the two troughs is highly correlated with both the $g$-band flux and each other, strongly indicating that the observed variability is driven by ionization state changes of the gas. 
The equivalent width of the mini-BAL Trough B is found to lag that of Trough A by $\sim$ 4 days in the quasar rest frame, suggesting that the equilibrium timescales of the two absorbers are different and possibly indicative of different outflow densities or distances. 

\end{abstract}


\section{INTRODUCTION}
\label{introduction}

Most massive galaxies host a supermassive black hole (SMBH) in their centers, and observations of strong correlations between the mass of the SMBH ($M_{\rm BH}$) and various properties of their host galaxies suggest that the growth of these SMBHs is likely connected with the growth of the galaxy in which it resides (e.g., \citealt{Gebhardt00a, Kormendy13}). For example, strong correlations have been observed between \mbh and the stellar velocity dispersion within the host bulge ($\sigma_*$; known as the $M_{\rm BH}-\sigma$ relation; e.g., \citealt{Gebhardt00a, Ferrarese00}), and the mass of the host-galaxy bulge ($M_{\rm bulge}$, known as the $M_{\rm BH}-M_{\rm bulge}$ relation; e.g., \citealt{Magorrian98, Gultekin09, Kormendy13}). Some SMBHs are actively accreting matter, releasing enormous amounts of energy in the accretion process; these are called active galactic nuclei (AGN), or quasars. 
AGNs provide a means for potential interaction between SMBHs and their galaxies through feedback (e.g., \citealt{Dimatteo05, Moll07, KingPounds15}). 
Some quasars have been observed to produce fast outflows that blow gas and dust into our line of sight and into the galaxy; if these outflows contain enough energy, they may produce feedback in the host galaxy and disrupt or induce the formation of stars. Quasar outflows can thus potentially have a significant impact on the host galaxy. Quasar outflows are thought to be launched from the accretion disk (e.g., \citealt{Murray95, Proga00, Higginbottom14}) and/or from a dusty torus (\citealt{Chan16, Choi22}) and appear along our line of sight as absorption lines in the observed spectra of quasars. Broad absorption lines (BALs) are defined as absorption features reaching a depth of $\leq$~0.9 times the continuum-normalized flux, with a velocity width greater than 2000 km~s$^{-1}$ (\citealt{Weymann91}). Understanding the impact of these outflows on their host galaxies is crucial to understanding the effect a SMBH has on its galaxy.
 
Throughout a BAL trough's life, it is possible to see large variability (e.g., \citealt{Barlow93, Lundgren07, Gibson08b, Capellupo11, Filizak13}; in some cases, BALs can even fully disappear and reappear again (\citealt{Filizak12}). Previous studies have reported BALs at a variety of different ionization states, the most common being the \civ ion ($\lambda\lambda 1548,1550)$. Most previous studies of BAL variability in samples of quasars have been limited in sample size, numbers of spectra, or in time baseline covered. However, many of these studies have demonstrated that the spectral variability of BAL quasars can give insight into the physical parameters of the outflow: By examining the spectral variability of BAL quasars, one can learn about the energy, structure, and physical environment of the outflows.  

The Sloan Digital Sky Survey (SDSS) provides a unique opportunity to investigate both short and long-timescale variability in BAL quasars. In 2014, the SDSS Reverberation Mapping project (SDSS-RM; e.g., \citealt{Shen15a}) began monitoring a field of 849 quasars, accumulating a total of 90 epochs of spectroscopy over seven years as a part of the SDSS-III and SDSS-IV surveys (\citealt{Blanton17}). In 2021, the SDSS-V Black Hole Mapper Reverberation Mapping program (BHM-RM; \citealt{Kollmeier26}) continued monitoring a subset of quasars in this field and has thus far acquired more than 100 additional epochs of spectroscopy over five more years (and continues to observe through 2026). There are roughly 90 quasars with BALs (or mini-BALs, which have velocity widths between 1000--2000 km~s$^{-1}$) in the original SDSS-RM quasar sample, so this dataset thus provides an opportunity to explore both short and long-timescale variability in BAL quasars. \cite{Hemler19} carried out an investigation of short-timescale variability in SDSS-RM BAL quasars, finding that rapid variability (on timescales of $<$ 10 days in the quasar rest-frame) was common in BAL quasars. The work of \cite{Hemler19}, hereafter referred to as ``H19", included data from the first four years of SDSS-RM spectroscopic monitoring. 

While the primary goal of H19 was to characterize the short-timescale variability in a sample of quasars and make statistical inferences about BAL variability in quasars, they identified several individual sources along the way that showed particularly interesting behavior. In particular, H19 reported that a \civ BAL in the quasar SDSS\,J142129.40+522752.0 (also referred to in their study as RM\,508) was observed to show dramatic variability in strength over a very short timescale: Its equivalent width (EW) was reported to decrease at a significance of more than 20$\sigma$ on a timescale of only 4.03 days in the quasar rest frame. SDSS has continued to monitor this quasar and has now accrued more than 200 epochs of spectroscopy that span more than 12 years. With these additional years of monitoring in hand, we decided to revisit this quasar to determine whether this dramatic variability continued and to characterize its behavior in more detail.

In this work, we present the results of a systematic analysis of \civ BAL variability in the quasar SDSS\,J142129.40+522752.0 with 12 years of monitoring. In Section~\ref{sec:data} we discuss the SDSS observations that comprise our dataset and the preparation of the data for analysis. Section~\ref{sec:measurements} describes our measurements of the absorption features in this quasar and the results of our analysis, and Section~\ref{sec:discussion} discusses the interesting behavior of this source and considers the implications of our results, placing them in the context of models of BALs. Where necessary, we adopt a $\Lambda$CDM cosmology with $\Omega_{\Lambda}=0.7$, $\Omega_{M}=0.3$, and $h_{\rm 0}=0.7$.

\section{Data}
\label{sec:data}
In this investigation, we focus on the SDSS-RM quasar SDSS\,J142129.40+522752.0, hereafter referred to as J1421 (RA = 215.3725, DEC = 52.4644), which lies at a redshift of $z = 3.228$ (\citealt{Shen19}) and has an observed $i$-magnitude $m_i = 18.12$ (\citealt{Ahn14}). Our spectra were acquired by SDSS as a part of three different programs: Between 2014--2020, we obtained 90 epochs of spectra as a part of the SDSS Reverberation Mapping project (SDSS-RM; e.g., \citealt{Shen15a}), which was carried out during the SDSS-III and SDSS-IV surveys (\citealt{Blanton17}). We obtained an additional 118 epochs from 2021--2025 as a part of the SDSS-V Black Hole Mapper reverberation mapping program (BHM-RM; \citealt{Kollmeier26}). In total, we have 208 epochs of spectroscopy of J1421 , spanning from late 2013 to 2025.
All of our spectra were obtained with the BOSS spectrograph on the 2.5m SDSS telescope at Apache Point Observatory (\citealt{Gunn06, Dawson13, Smee13}). The BOSS spectrograph covers a wavelength range of roughly 3650--10400\,\AA\ in the observed frame and has a spectral resolution of $R\sim2000$ across the observed wavelength range.  

In addition to the SDSS-RM and BHM-RM spectra, there are three additional SDSS spectra of this source that were taken by earlier SDSS surveys. We have one spectrum from 2003 (MJD 52764, Plate 1326, Fiber 243) and two spectra from 2013 (MJD 56397, Plate 6717, Fiber 264; and MJD 56455, Plate 7029, Fiber 144).  

\subsection{Data Preparation}
The spectra from 2013 and 2014 were processed with the standard SDSS-III pipeline (version 5\_7\_1; \citealt{Bolton12}), and the remaining SDSS-IV spectra with the SDSS-IV eBOSS pipeline version 5\_10\_1 (\citealt{Dawson13}). The SDSS-V data (2021--2025) were processed through the standard BHM pipeline using version 6\_1\_2 (S. Morrison et al, in preparation). 
We first cropped the spectra to include only wavelengths between 3660\,\AA\ and 10000\,\AA\ (in the observed frame) to remove regions at the edges that were overly noisy and/or suffered from significant sky contamination. 
We then used an interpolation function to correct for remaining bad pixels in our data that were identified using the bitmask flags produced by the SDSS reduction pipeline. When the `ANDMASK' flag was set, we interpolated between adjacent pixels to estimate the flux in the affected pixels and replaced the flux value at that pixel with the interpolated value.  This reduced the sky contamination throughout the spectrum, though residuals still remain, particularly at the red end of the spectrum. The number of pixels affected in each spectrum ranges from zero (where no pixels had the `ANDMASK' flag set after we cropped off the edges) to about 300 pixels, with an average of 67 pixels. With 4366 total pixels in each spectrum, this corresponds to 1.5\% of the pixels on average. These pixels also tend to be more concentrated toward the blue and red edges of our spectra, with fewer pixels identified as poor in the middle of the spectrum, which is our region of interest. 

\begin{figure*}
    \centering
    \includegraphics[width=0.95\linewidth, trim = 0 0 10 0, clip]{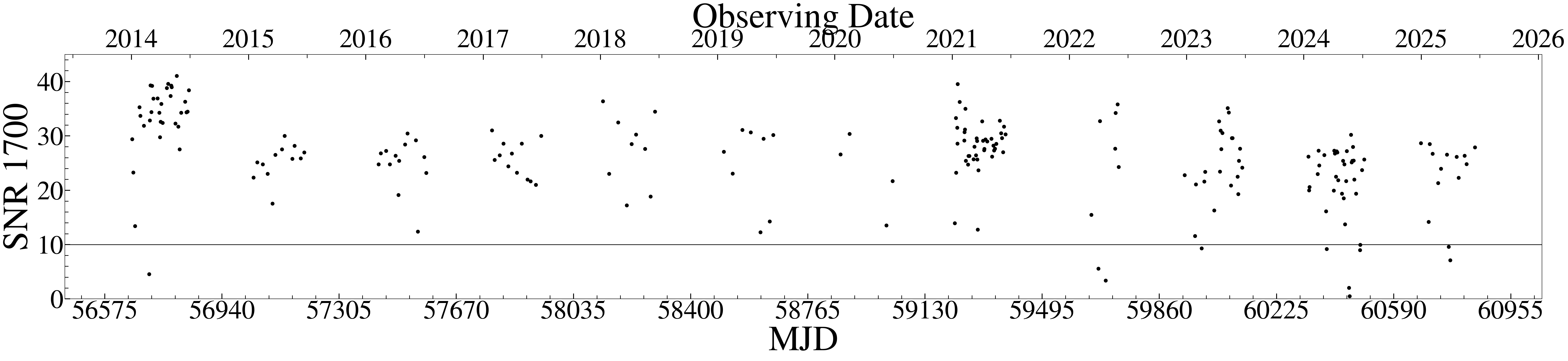}
    \caption{Measured signal-to-noise ratio at 1700 \AA\ (SNR$_{\rm 1700}$) as a function of observing date for our observations. The bottom x-axis gives the MJD of each observation, and the top x-axis shows the corresponding calendar date of the observation. The horizontal line indicates our cutoff SNR$_{\rm 1700} = 10$; epochs below this line were removed from our study.}
    \label{fig:cadence}
\end{figure*}

We then corrected our cropped and interpolated data for Galactic extinction using a Milky Way extinction model with  $R_V$~=~3.1 (\citealt{Cardelli89}), with $A_V$~=~0.0287 and $E_{BV}$~=~0.0092 (measured by \citealt{Schlafly11}\footnote{Accessed via \url{https://irsa.ipac.caltech.edu/applications/DUST/}}). Next, we shifted the observed wavelengths of our spectrum to the quasar rest frame using $z=3.228$ (\citealt{Shen19}).

\begin{figure*}
    \centering
    \includegraphics[width=.8\linewidth, trim = 0 0 0 0, clip]{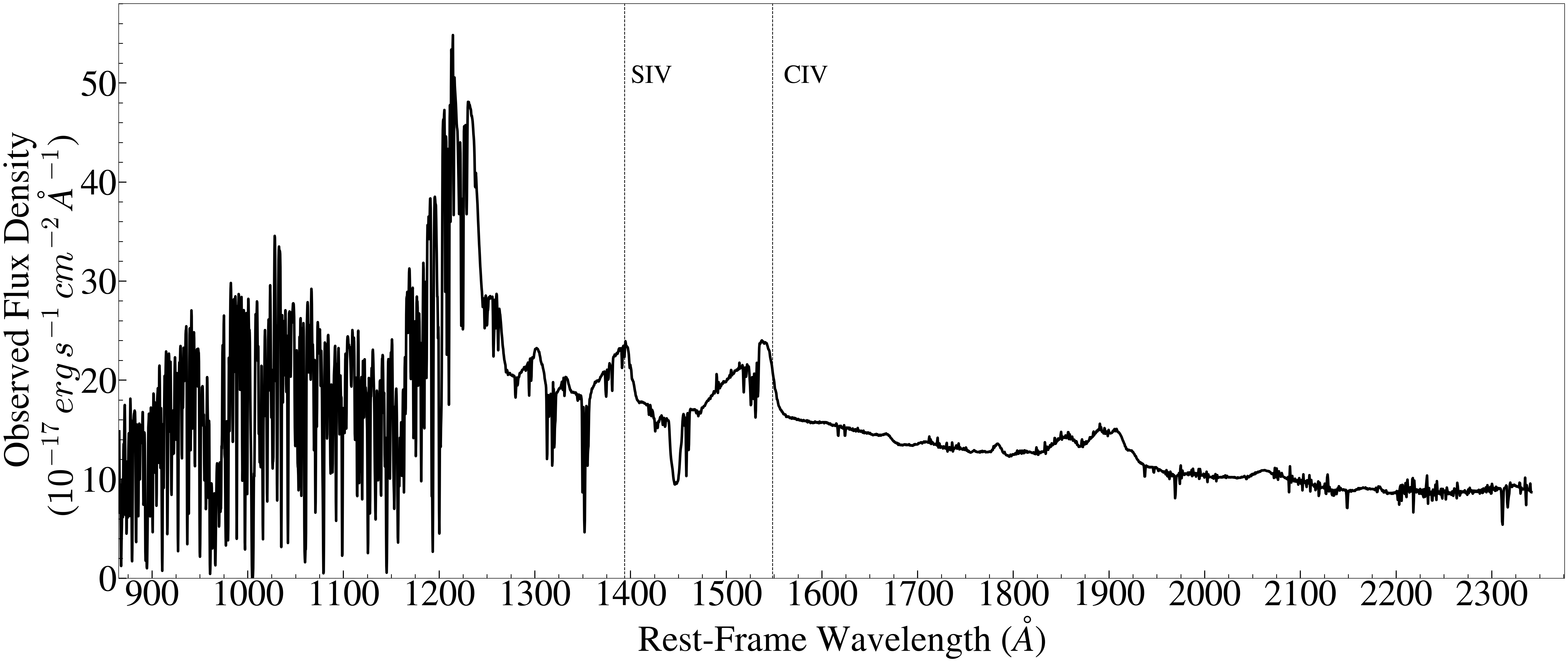}
    \caption{The mean spectrum of J1421, calculated from all 199 epochs of corrected spectra in our final dataset. Vertical gray lines indicate the rest-frame wavelengths of the \siiv\  and \civ  emission lines ($\sim$1398\,\AA\ and $\sim$1548\,\AA, respectively). }
    \label{fig:meanspec}
\end{figure*}

With our corrected spectra, we calculate the median signal to noise ratio between the rest-frame wavelengths of 1650\,\AA\ and 1750\,\AA\ (SNR$_{1700}$) in each of our corrected spectra to determine the final set of usable spectral epochs for our study (see Figure~\ref{fig:cadence}). As this is one of the brighter quasars in the SDSS-RM sample, most of the SNR were high; however, there were a few epochs where SNR$_{1700}$ was low due to insufficient exposure time and/or poor weather conditions. We removed epochs with SNR$_{1700} < 10$ from our sample, as in this work we require sensitivity to small-scale variability. We also removed an additional epoch (MJD 59971) due to bad pixels that lie directly on top of the BAL of interest. In total, we removed 9 epochs (MJDs 56713, 59670, 59693, 59992, 60381, 60451, 60486, 60766, and 59971), leaving us with 199 spectral epochs for our analysis. Figure~\ref{fig:meanspec} shows the mean spectrum produced from our final set of 199 corrected spectra.

\subsection{Continuum and Emission-Line Fits} 
\label{sec:confits} 

\begin{figure} 
    \centering
    \includegraphics[width=0.98\linewidth, trim = 5 0 5 0,clip]{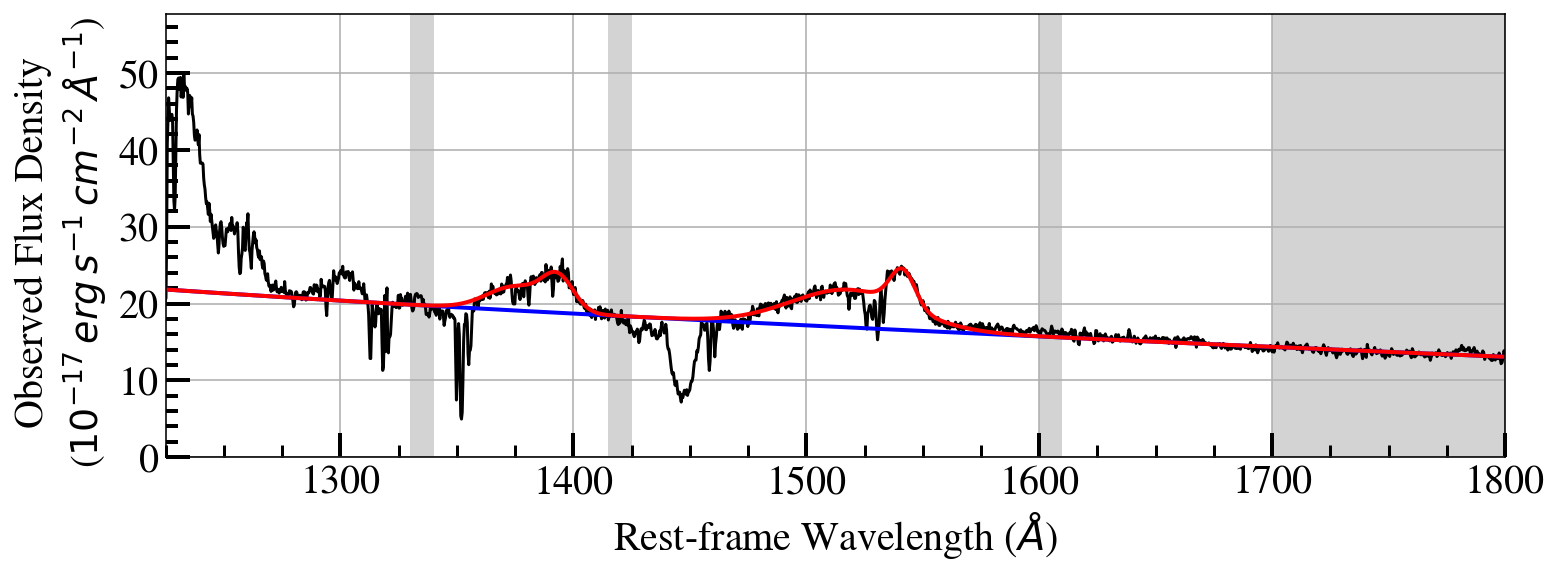}
    \caption{An example of a continuum and emission-line fit for one of our spectral epochs (MJD = 56686). The black line indicates the corrected spectrum, the solid blue line shows the continuum fit, and the red line shows the continuum+emission-line fits. Shaded gray rectangles indicate the line-free regions used to fit the continuum (see Section~\ref{sec:confits}). }
    \label{fig:Both fits}
\end{figure}

We follow previous work (e.g., \citealt{Filizak12, Grier15, Hemler19}) and fit a reddened power-law model to the continuum in each of our spectral epochs. Using a least-squares fitting algorithm, we adopt the SMC-like reddening model of \cite{Pei92} for our continuum. To fit the continuum, we chose sections that were absent of any absorption or emission features. These regions include 1330--1350\,\AA,  1415--1425\,\AA, 1600--1610\,\AA, 1700--1800\,\AA, 1950--2035\,\AA, 3950--4050\,\AA, 4140--4270\,\AA, 4400--4770\,\AA, and 5100--6400\,\AA. These regions were originally drawn from the line-free regions described by \cite{Gibson09b}, with some adaptations for this particular quasar to increase accuracy toward the blue end of the continuum while still avoiding the absorption features in the area (see Figure~\ref{fig:Both fits}). We used the same set of fitting regions for all epochs. 

\begin{figure*} 
    \centering
    \includegraphics[width=0.9\linewidth]{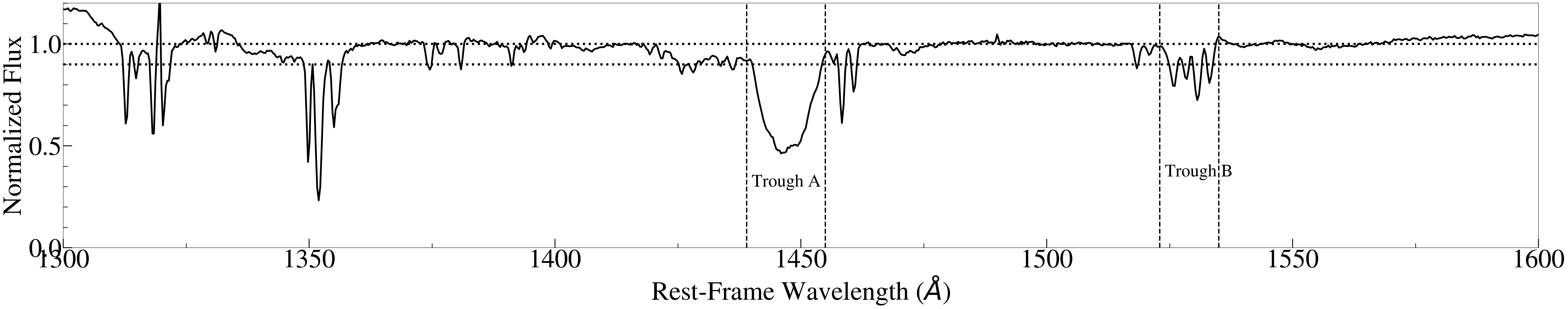}
    \caption{The mean normalized spectrum of J1421. Troughs A and B are labeled, and the wavelengths of each are marked by dotted vertical lines. We note that due to its weakness throughout much of the observations, Trough B appears in the mean spectrum more as a cluster of features than a single one;  however, inspection of the epochs in which it is much stronger suggest that it is a single feature, and it varies coherently.}
    \label{fig:Normalized meanspec}
\end{figure*}

While the primary BAL of interest in this source is detached from the \civ emission line, this quasar has additional visible absorption features that are superimposed upon the \siiv\ and \civ emission lines. To disentangle these absorption features from the emission lines, we follow \cite{Grier16} and H19 and fit an emission-line profile in addition to the continuum. We tried fitting three different line profiles to the \siiv\ and \civ emission lines: Voigt, Gauss-Hermite, and double-Gaussian (\citealt{vanderMarel93, Park13, Gibson09b}). To determine the uncertainties on the fits, we ran 100 Monte Carlo iterations where we randomly adjusted the flux at each spectral pixel by a gaussian deviate scaled by its uncertainty and re-fit the emission line. We then calculated the mean of each parameter and adopted those as our fits. To maintain consistent fits for the double-Gaussian model, we fixed the centers of each Gaussian to the mean center value determined when we calculate the mean of all parameters. After inspecting all fits, we determined that the double-Gaussian model was the best fit for both the \siiv\ and \civ emission lines in this source. We then added the double-Gaussian emission-line fit to our previously determined continuum fit for each epoch. We divided each spectrum by its combined emission-line+continuum fit to obtain a normalized spectrum. We then use these continuum and emission-line normalized spectra for all further measurements. We propagate both the spectral uncertainties and the uncertainties in the emission-line and continuum fits when calculating uncertainties in the normalized spectra. 

\begin{figure*}
    \centering
    \includegraphics[width=0.8\textwidth]{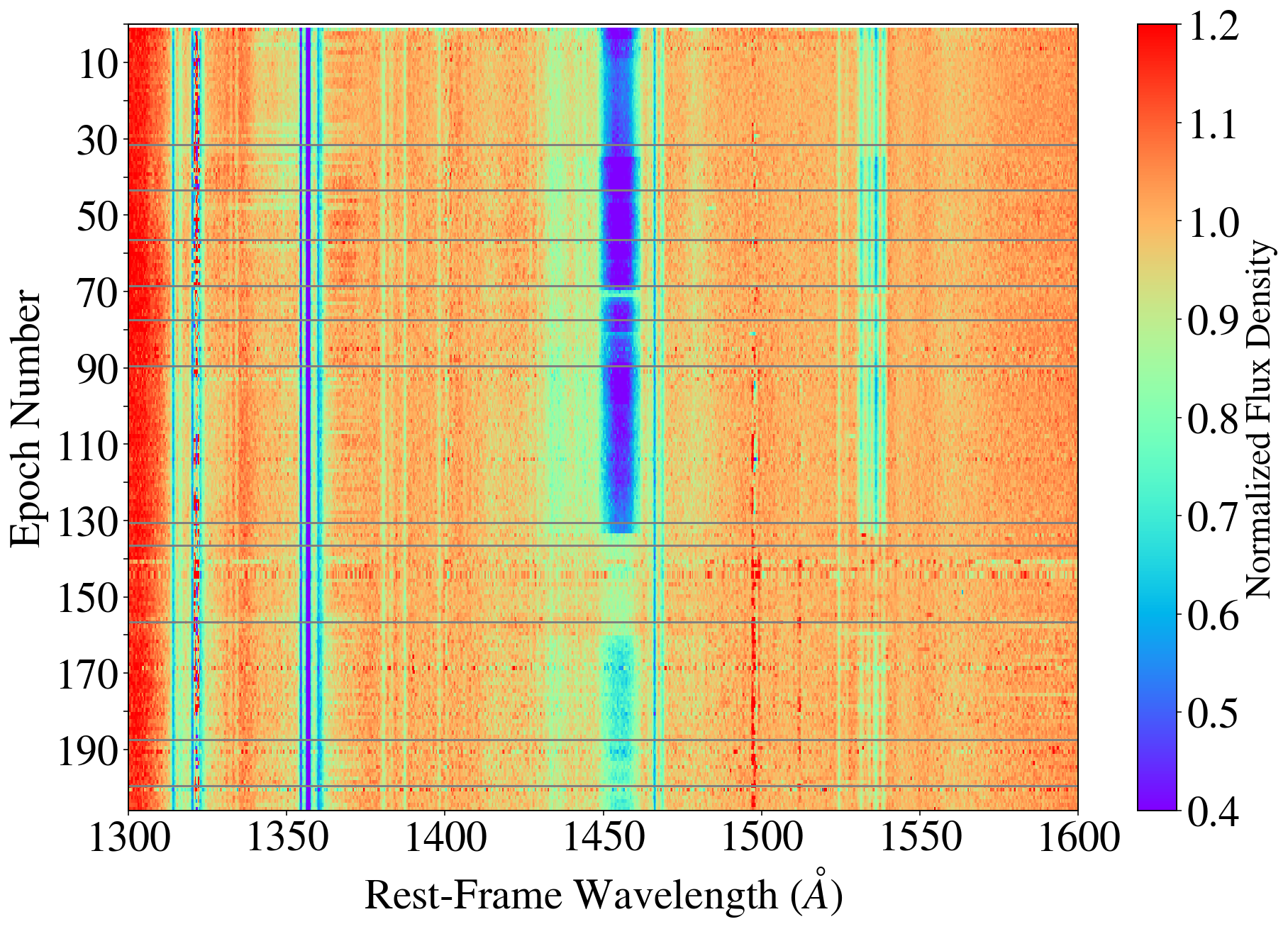}
    \caption{A trailed spectrogram showing all of our normalized spectra for J1421. Each horizontal line represents a spectrum. The first epoch is shown in the top row, and as you move down on the plot, time moves forward. The color represents the normalized flux density at each pixel, with higher fluxes shown as warm colors, and lower fluxes shown as cooler/bluer colors. In the plot above, absorption features thus appear as vertical stripes. Horizontal gray lines indicate the break between observing seasons. }
    \label{fig:trailedspec}
\end{figure*}

Figure~\ref{fig:Normalized meanspec} shows the mean spectrum from 1300\,\AA\ to 1600\,\AA\ created from all of the normalized spectra used in this study, and Figure~\ref{fig:trailedspec} shows a trailed spectrogram that includes all of these normalized spectra for J1421. In the trailed spectrogram, the primary BAL of interest appears as the blue vertical stripe around 1450\,\AA. In both the mean spectrum and the trailed spectrogram, there is also a weaker absorption feature at a lower velocity that is superimposed on the \civ emission line. We hereafter refer to the primary (high-velocity) BAL as Trough A, and the lower-velocity feature as Trough B (as labeled in Figure~\ref{fig:Normalized meanspec}). Throughout much of the monitoring period, Trough B is quite weak, and thus appears as a weak cluster of narrower features in the mean spectrum; however, in the beginning years when it is strong, it presents as a single, unified feature that varies coherently (see Figure~\ref{fig:app_mean_spec_all} in the Appendix, for example).  

There is also some significant absorption present blueward of Trough A at roughly 1315\,\AA\ and 1350\,\AA. While we originally thought this could be \siiv\ absorption, further inspection indicates that these features do not line up in velocity with the \civ absorption. We thus conclude that these are clusters of narrow intervening systems rather than BAL or mini-BAL features. We do not include these features in our analysis. We do see absorption in O\,{\sc vi} and N\,{\sc v} associated with system A, and from O\,{\sc vi}, Ly$\alpha$, and N\,{\sc v} associated with system B. These features are also beyond the scope of our analysis.

\section{Measurements and Results} 
\label{sec:measurements}
\subsection{Basic Measurements}
\label{sec:basic_measurements}
\begin{figure*}
     \centering
     \includegraphics[width=0.7\linewidth]{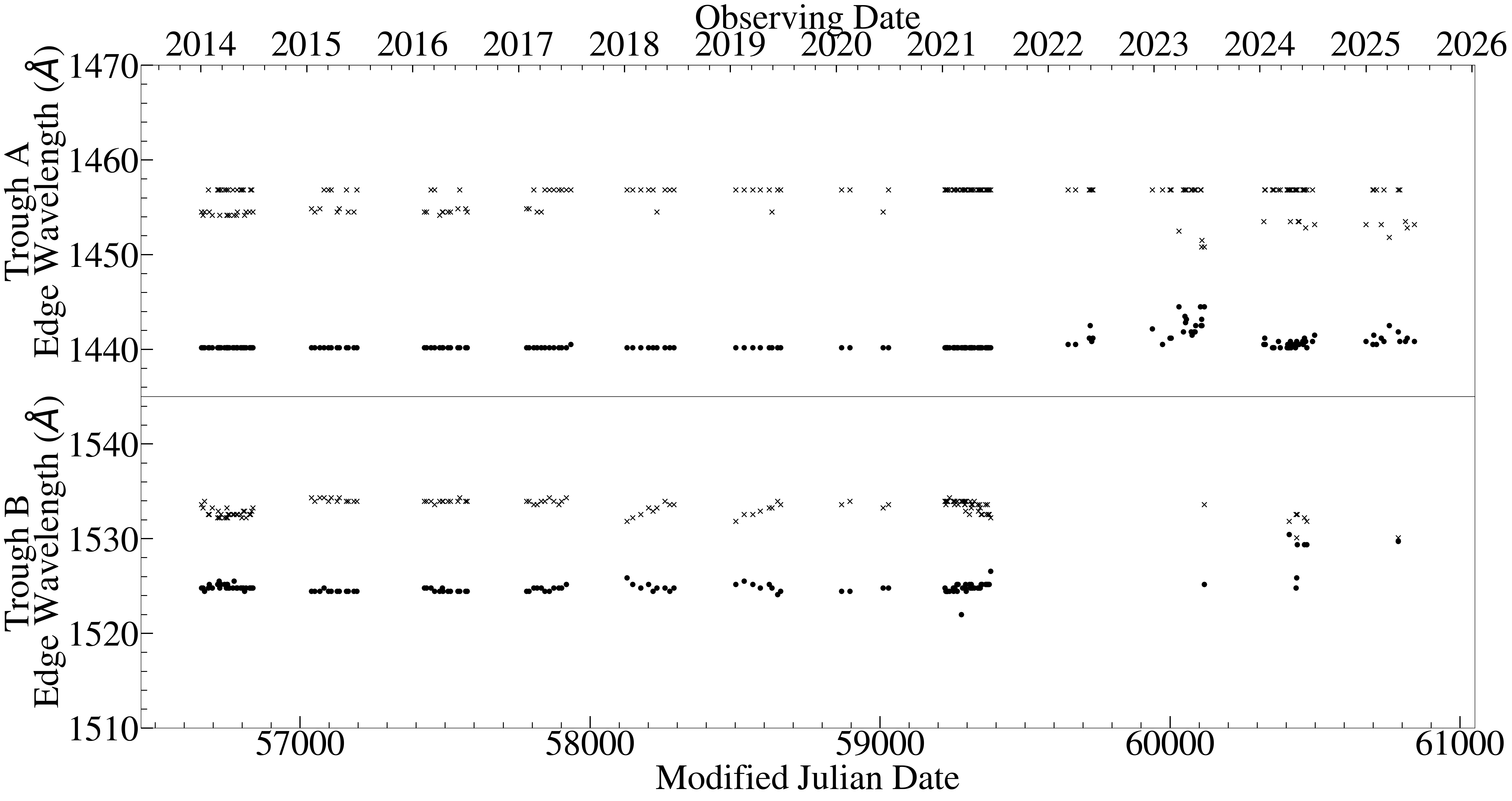}
     \caption{The upper and lower wavelength edges for Trough A (top) and Trough B (bottom). The upper and lower limits of both troughs are quite stable  most of the monitoring (excluding the years where the troughs are extremely weak and/or have disappeared). The $\sim$3\,\AA\ oscillation of the red limit of Trough A is due to a weak feature on the red side of the Trough that sometimes dips below our threshold of 0.9 and sometimes stays above (see Section~\ref{sec:basic_measurements}). }
     \label{fig:BAL Upper/lower}
 \end{figure*}

We used our normalized spectra at each epoch to identify and measure various properties of the absorption features and characterize Troughs A and B. We first identified the location of each absorption feature by determining the upper and lower limits of the absorption at each epoch. To do this, we first smoothed the spectra using a boxcar filter with a width of 8 pixels, in which each pixel's flux was replaced by the average flux over a window of 8 pixels. We then used the smoothed spectra to identify the wavelengths at which the continuum dropped below a normalized flux of 0.9 on the edges of each absorption feature. We also force the red limit of Trough A to exclude the two narrow absorption features just redward of the Trough. Figure~\ref{fig:BAL Upper/lower} shows the measured upper and lower limits of both absorption troughs as a function of time. 

The edges of both Troughs A and B are remarkably stable in velocity throughout the entire 12-year duration of monitoring; however, in the later years (2022 and beyond), both troughs have weakened considerably and/or disappeared, causing the upper and lower boundaries to be more uncertain or impossible to measure in cases of complete disappearance. The red limit of Trough A does oscillate back and forth by about 3\,\AA\ (Figure~\ref{fig:BAL Upper/lower}); this oscillation is due to a very weak, narrow absorption component at about $\sim$1456\,\AA\ whose normalized flux is very close to the limit of 0.9; this tiny component sometimes lies above 0.9 and sometimes drops below to ``join" the primary BAL, causing the measured upper limit for Trough A to oscillate. For consistency in our analysis, we thus adopt the upper and lower limits as measured from the mean spectrum to make our measurements of the troughs (see Figure~\ref{fig:Normalized meanspec}). These limits correspond to 1439--1455\,\AA\ for Trough A and 1523--1535\,\AA\ for Trough B. Trough A is formally considered a BAL, with a velocity width of 3436.19~km~s$^{-1}$ in the mean spectrum, while Trough B has a velocity width of 1449.39~km~s$^{-1}$ in the mean spectrum, thus categorizing it as a mini-BAL. 
 
After determining the upper and lower limits of Troughs A and B, we used our normalized spectra to make a variety of measurements to characterize the absorption at each epoch. We measure the Equivalent Width (EW), defined as 
\begin{equation} 
 EW = 
 \int_{\lambda_{\text{min}}}^{\lambda_{\text{max}}} (1-f_n(\lambda)) \,d\lambda. 
\end{equation}
We also measure the centroid velocity ($v_{\text{cent}}$), defined as 
\begin{equation} 
v_{\text{cent}} = 
\frac
{\int_{v_{\text{min}}}^{v_{\text{max}}} v(1-f_n(v)) \,dv} {\int_{v_{\text{min}}}^{v_{\text{max}}} (1-f_n(v)) \,dv}
\end{equation}

and the mean trough depth ($d$):
\begin{equation} 
d = \frac{\int_{\lambda_{\text{min}}}^{\lambda_{\text{max}}} (1-f_n(\lambda)) \,d\lambda }{N}
\end{equation}

\begin{figure*}
    \centering
    \includegraphics[width=0.75\linewidth]{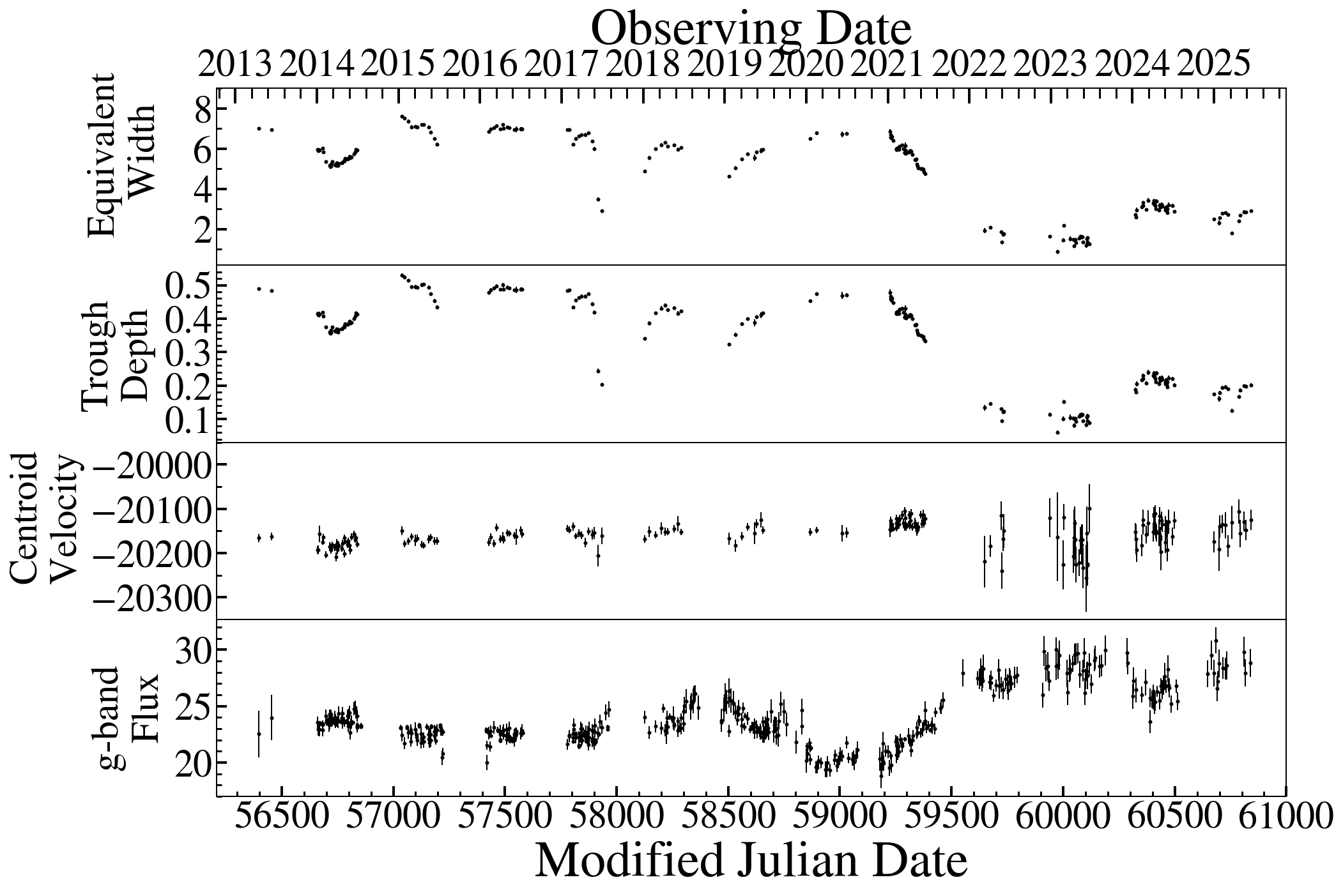}
    \caption{Measurements of Trough A as a function of time. The top panel shows the EW, the middle shows the average trough depth, and the bottom shows the centroid velocity.
        The bottom x-axis shows the MJD corresponding to each measurement, and the top x-axis labels indicate the corresponding calendar date. }
    \label{fig:Measurements}
\end{figure*}

\begin{figure*}
    \centering
    \includegraphics[width=0.75\linewidth]{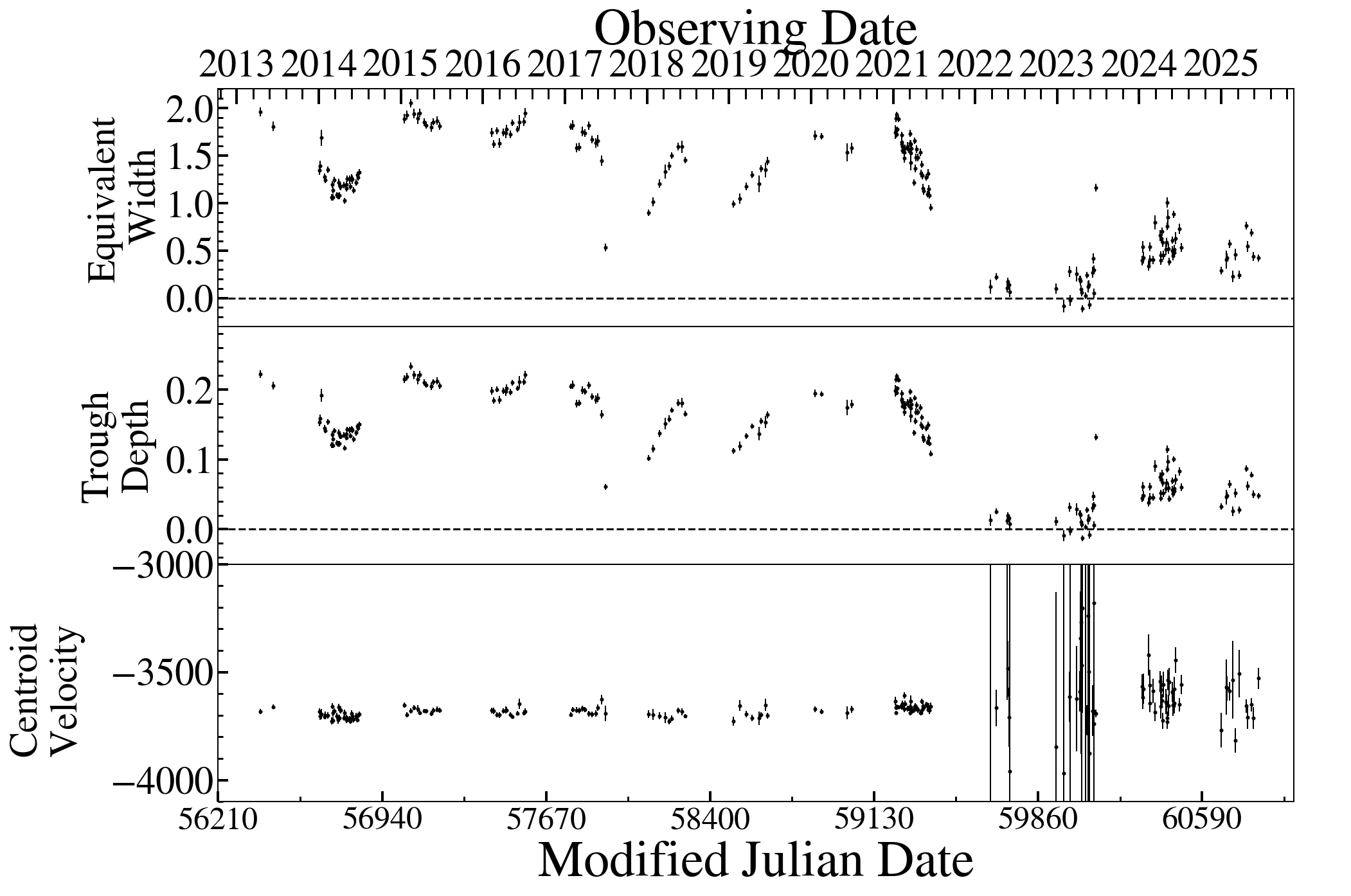}
    \caption{Measurements of mini-BAL Trough B as a function of time. The top panel shows the EW, the second-from top shows the average trough depth, the third-from top panel shows the centroid velocity, and the bottom panel shows the measured photometric $g$-band flux (\citealt{Pena26}). The bottom x-axis shows the MJD corresponding to each measurement, and the top x-axis labels indicate the corresponding calendar date. The points around MJD 60000 with large uncertainties in the centroid velocity represents a period when this absorption feature has disappeared, so the centroid velocity is essentially just measuring noise.}
    \label{fig:B Measurments}
\end{figure*}

 In all cases, $f_n$ represents the normalized flux values and $N$ is the number of pixels within the BAL regions. We again used Monte Carlo iterations to determine the uncertainties for each measurement --- this produces an uncertainty that takes into account the uncertainties in the flux density as well as the continuum+emission-line fits. Figures~\ref{fig:Measurements} and \ref{fig:B Measurments} show these measurements as a function of time for both Troughs A and B. We also report these measurements for all epochs in the Appendix in Table~\ref{tab:spectral_data}. 

 Figures~\ref{fig:Measurements} and \ref{fig:B Measurments} show a number of interesting features: In the fourth year of monitoring, we see a rapid drop in the EW and depth of both Trough A and Trough B (though at slightly different times); this dramatic short-timescale variability in Trough A was also identified by H19. We also find that both Troughs A and B disappear almost entirely during 2022 and 2023, and reappear later on, remaining fairly weak through the remainder of the campaign.   

Figures~\ref{fig:Measurements} and \ref{fig:B Measurments} suggest that the velocity centroid of both Troughs is stable throughout the monitoring; however, from 2022--2023, both troughs are weak enough that the centroid velocities are uncertain, skewing the y-axis limits of the Figure and making this hard to determine by eye. Trough A recovers enough strength in 2024 and 2025 to put useful constraints on $c_{\rm cent}$, though Trough B remains too weak. A closer look at $v_{\rm cent}$ in years when we can meaningfully constrain it suggests that $v_{\rm cent}$ in both troughs is actually decreasing (moving to smaller negative values) very slowly over time. Fitting a linear relation to all of the $v_{\rm cent}$ measurements, we find a slope of 0.0117$\pm$0.0012 km~s$^{-1}$~day$^{-1}$ (0.0031 $\pm$0.0003 cm~s$^{-2}$) for Trough A and 0.0113$\pm$0.0012 km~s$^{-1}$~day$^{-1}$ (0.0030 $\pm$0.0003 cm~s$^{-2}$) for Trough B. If we interpret these slopes as a measure of the deceleration of the outflow, we find they are roughly three orders of magnitude smaller than the acceleration reported by \cite{Wheatley24}, who report an average acceleration of $-1.37\pm0.05$ cm~s$^{-2}$ in a \civ BAL over the same time period of 8 years.  

\begin{figure}
\centering
\includegraphics[width=0.98\linewidth, trim = 0 0 8 0,clip]{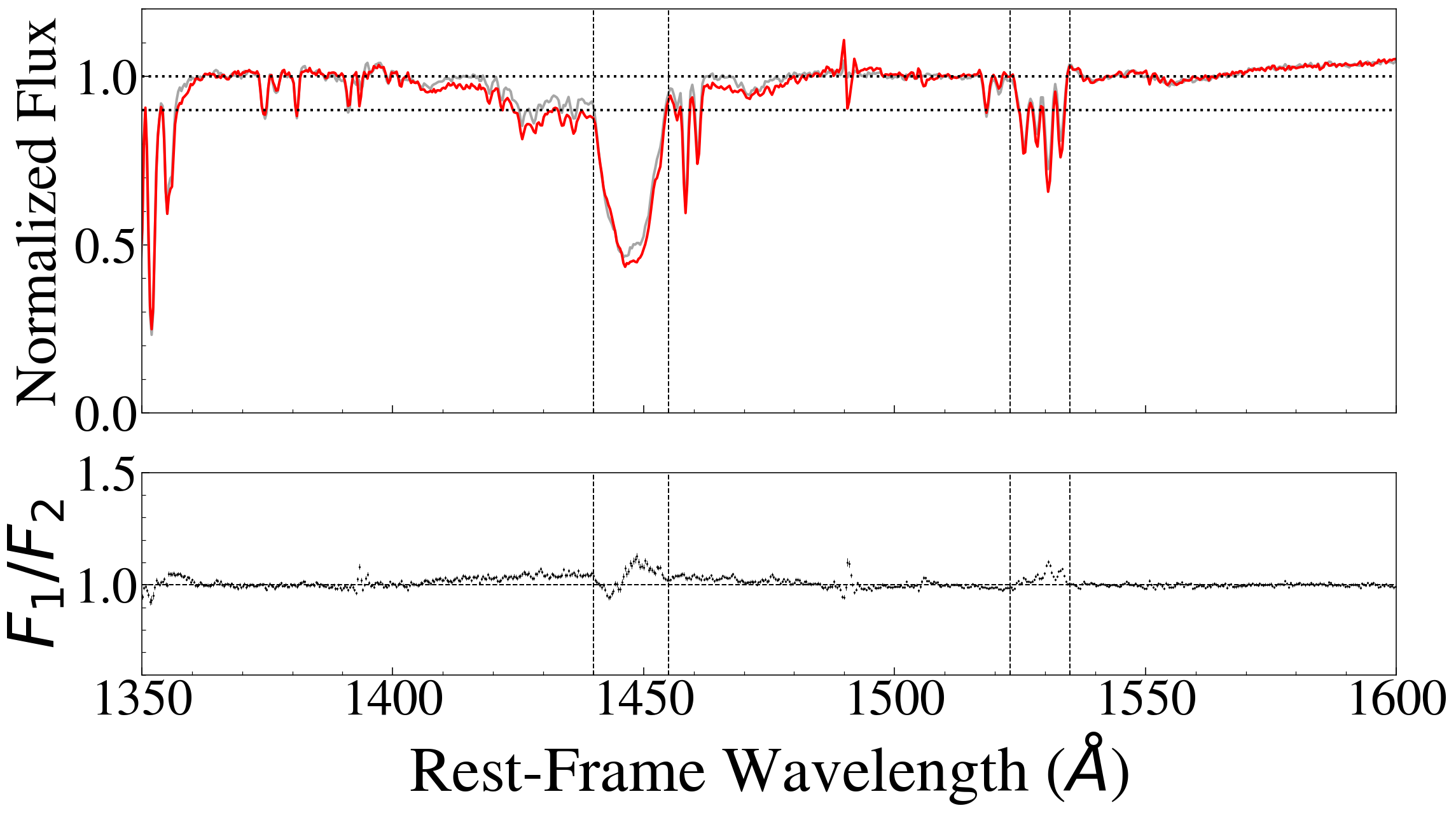}
\caption{The mean spectra from 2014 (gray line) and 2021 (red line). The top panel shows the normalized spectra, and the bottom panel shows the ratio between the two mean spectra. }
\label{fig:2014vs2021profile}
\end{figure}

While $v_{\rm cent}$ slowly decreases in magnitude over time, we have already determined from Figure~\ref{fig:BAL Upper/lower} that the edges of both BALs are very stable. This suggests that the change in $v_{\rm cent}$ is due to a change in shape of the BAL rather than global acceleration/deceleration. To explore this further, we calculated the mean normalized spectrum for each individual observing year (Figure~\ref{fig:app_mean_spec_all}) and measured the properties of the BAL in each individual season (Table~\ref{tab:mean_measurements_by_year}). These measurements average over the short-timescale variability present in each year and increase the SNR of our measurements, allowing us to see longer-term changes in line profile that are hard to discern in the individual spectra. Figure~\ref{fig:2014vs2021profile} shows the mean spectrum from 2014 plotted against the mean spectrum from 2021. During these two years, the BAL is of relatively similar strength, but we see that the redward half of the line profile has increased in strength, causing a decrease in centroid velocity. Because we find that the blueward edge of both Troughs remains constant, with a slope consistent with zero, the change in shape provides further evidence that the change in $v_{\rm cent}$ represents a slow change in outflow structure over the course of our observations rather than actual global deceleration of the outflow.  
\begin{figure}
    \centering
    \includegraphics[width=0.98\linewidth]{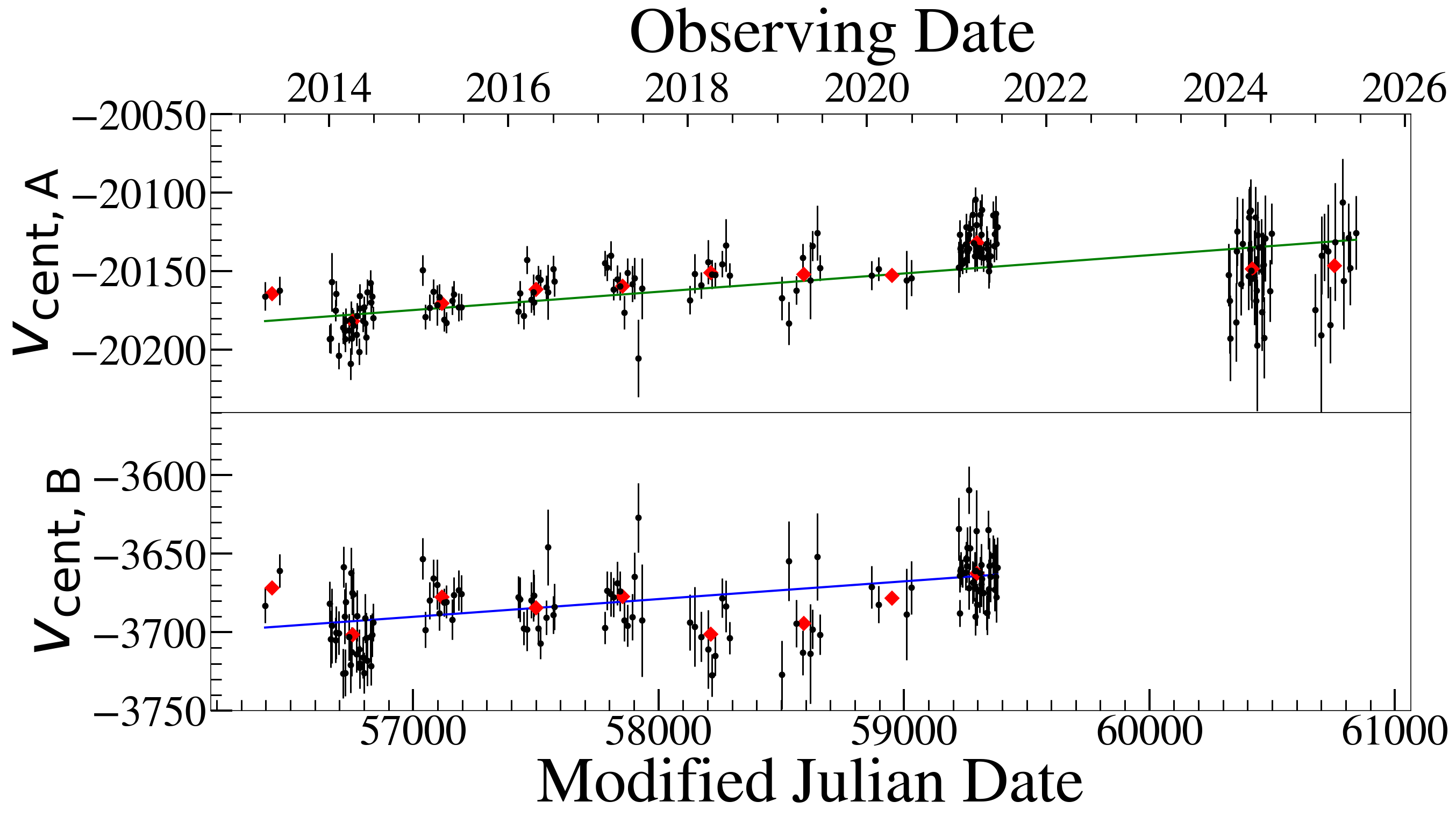}
    \caption{Measured velocity centroid of Trough A (top panel) and Trough B (bottom panel) as a function of time, from 2014--2021. Velocities are given in units of km~sec$^{-1}$. The red points indicate the mean value for each observing season/year. Lines indicate the best fit to the data. There is a noticeable slope toward smaller velocities for both troughs. }
    \label{fig:app_fig_vcent}
\end{figure}

 The bottom panel of Figure~\ref{fig:B Measurments} also shows the flux measured in the continuum as a function of time. To measure the continuum flux over time, we use photometric light curves in the $g$-band that were produced by \cite{Pena26}. These include synthetic photometric fluxes from SDSS as well as photometric flux measurements from a variety of telescopes such as the Canada-France-Hawaii Telescope, the Steward Observatory Bok telescope, and the Zwicky Transient facility (ZTF). See \cite{Pena26} for details on the photometry.


\subsection{Significant Variability on Short Timescales} 
\label{sec:sig_var}

\begin{figure}
    \centering
   \includegraphics[width=1\linewidth]{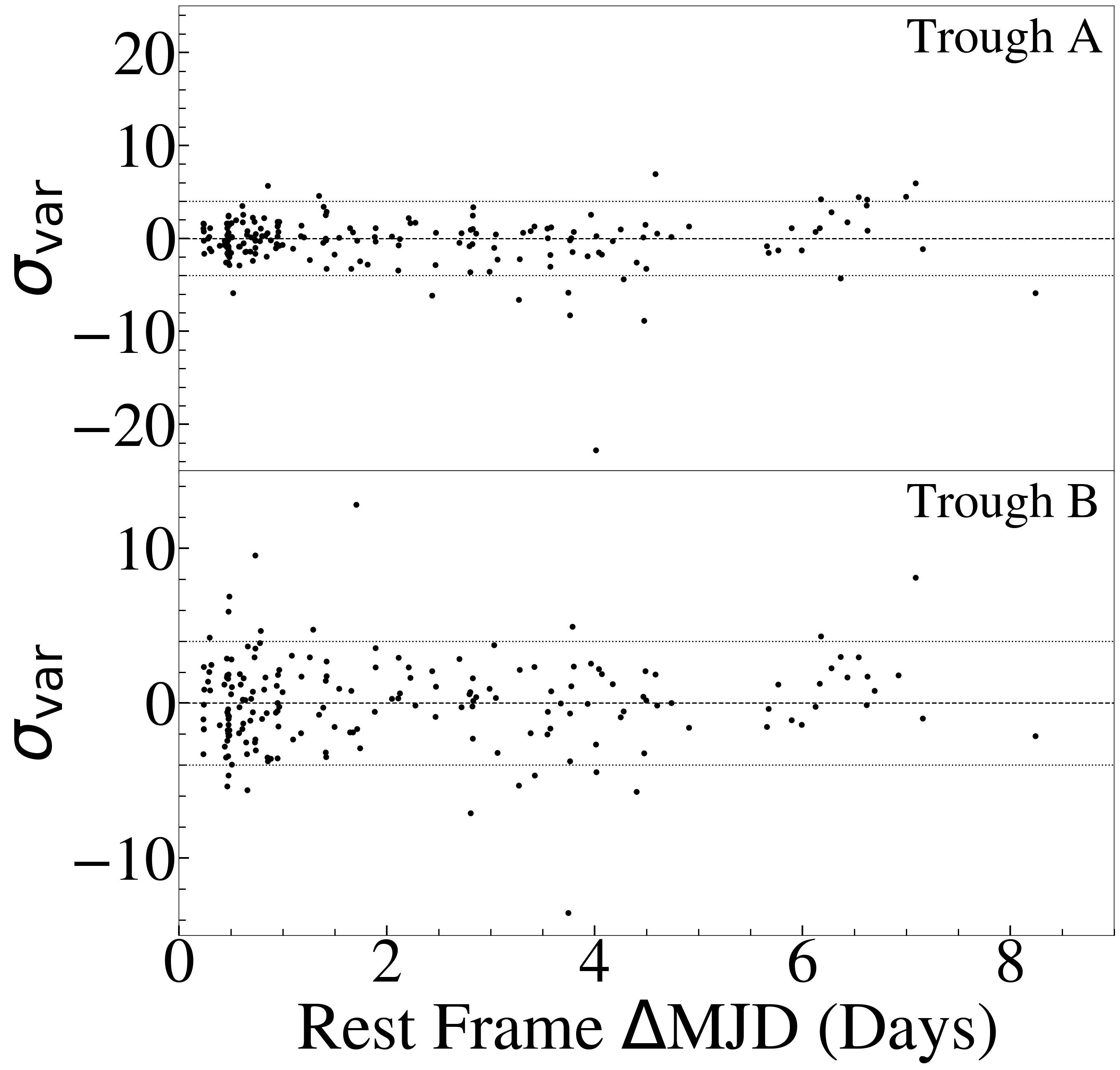}
    \caption{Observed $\Delta$EW vs $\Delta$MJD for sequential pairs of spectral epochs of trough A. The horizontal dashed line shows $\sigma_{var}$ = 0, and the horizontal dotted lines represent the 4$\sigma$ threshold; points outside that envelope are considered to show significant variability and their measurements provided in Table~\ref{tab:significant_EW_variability}.}
    \label{fig:deltaEW vs deltaMJD}
\end{figure}

\begin{figure*}
\centering
\includegraphics[width=0.49\linewidth]{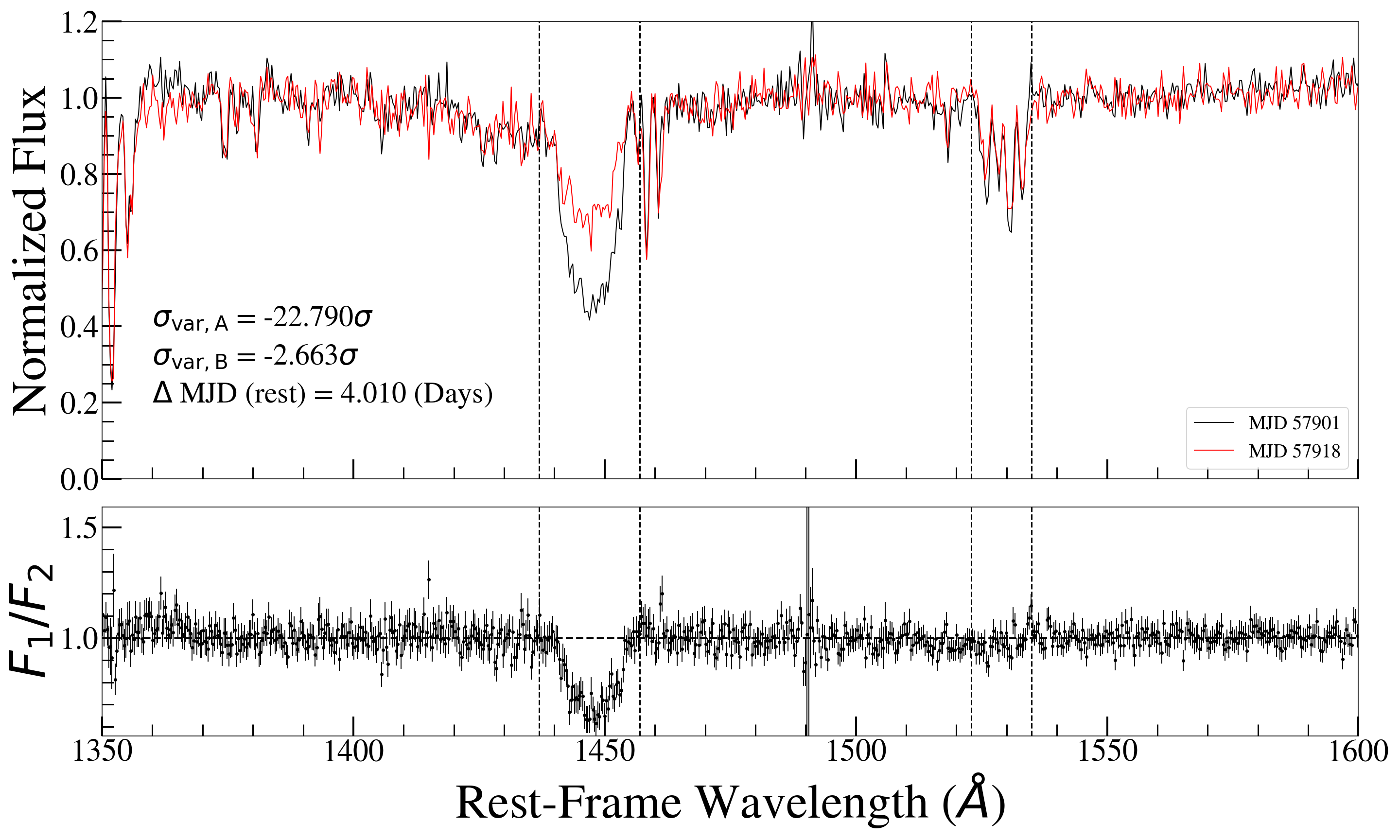}
\includegraphics[width=0.49\linewidth]{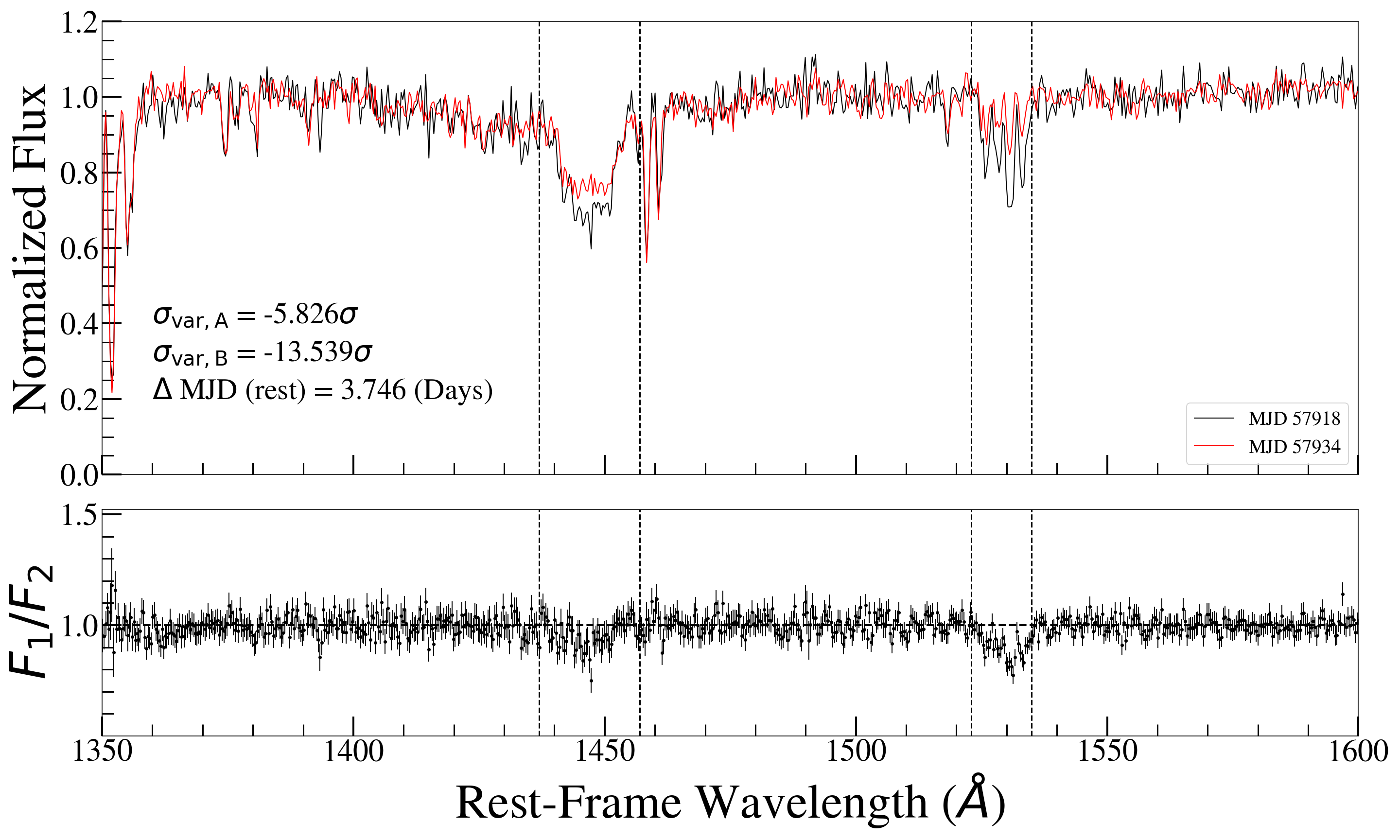}
\caption{Examples of the pairs of epochs with the most significant variability observed between them. The top subpanels show the normalized flux, and the bottom subpanels show the flux ratios between the two epochs as a function of wavelength.}
\label{fig:EW significant pairs}
\end{figure*}

The observations of both Trough A and Trough B (Figures~\ref{fig:Measurements} and \ref{fig:B Measurments}) suggest that both features undergo occasional rapid variations on short timescales ($<$10 days in the quasar rest frame). To quantify this, we follow H19 and define ``significant" variability between two epochs as a change in EW that is greater than 4$\sigma$ in significance (i.e., $\Delta EW/\sigma_{\Delta EW} > 4$, where $\sigma_{\Delta EW }$ is the uncertainty in $\Delta$EW). We thus define the quantity $\sigma_{\rm var} = \Delta EW/\sigma_{\Delta EW }$  between pairs of spectra. Figure~\ref{fig:deltaEW vs deltaMJD} shows $\sigma_{\rm var}$ for all sequential pairs of spectral epochs as a function of $\Delta$MJD in the quasar rest frame, and  Table~\ref{tab:significant_EW_variability} provides $\Delta$MJD, $\Delta$EW and $\sigma_{\rm var}$ for all pairs of epochs for which either Trough A or Trough B showed significant variability on timescales of $<$ 10 days in the quasar rest frame. We find 33 total pairs of epochs that show $>4\sigma$ variability in either Trough A or B (or both) on timescales of less than 10 days in the quasar rest-frame: Trough A is significantly variable in 18 of those epochs, and Trough B is significantly variable in 20 epochs (with 5 epochs where both were significantly variable). These results support the conclusion of H19 that short-timescale variability is common in some objects. Figure~\ref{fig:EW significant pairs} shows our two most examples of significant variability between sequential pairs for Trough A (left) and Trough B (right). 

H19 examined only Trough A in their work, and their light curves covered only the first four years of monitoring. All five of our significant-variability pairs for Trough A in these first four years were also identified by H19. In the subsequent years, we find an additional 13 pairs of significantly-variable epochs for Trough A. 

While Trough A is detached from the emission line, we note that significant variability in Trough B is likely more tentative, as Trough B lies directly on top of the \civ emission line, and its measurement therefore relies on the accuracy of the emission-line fits. We have inspected our fits thoroughly, but it is still possible that small inconsistencies in emission-line fits between sequential epochs may cause spurious identification of significant variability in Trough B. We thus consider the Trough B variable pairs to be tentative detections rather than solid detections, whereas our Trough A significant-variability pairs rely only on the stability of our continuum fits, which are more robust. 

\begin{deluxetable*}{cccrrrrrr}
\caption{Epoch pairs showing significant \civ BAL variability, with $|\sigma_{\rm var}| > 4$ in either Trough A or Trough B\tablenotemark{*}.}
\setlength{\tabcolsep}{2pt} 
\label{tab:significant_EW_variability}
\tablehead{
MJD$_1$ & MJD$_2$ & $\Delta$MJD & $\Delta_{EW, A}$ & $\sigma_{var,A}$ & $\Delta_{EW, B}$ & $\sigma_{var,B}$ \\
& & (Days) & (\AA) & & (\AA)
}
\startdata
56669.500 & 56683.480 & 3.307 & $-0.043$ & $-0.61$ & $-0.412$ & $\mathbf{-4.66}$ \\
56686.473 & 56696.778 & 2.437 & $-0.468$ & $\mathbf{-6.16}$ & $0.107$ & $2.07$ \\
56696.778 & 56715.388 & 4.402 & $-0.202$ & $-2.60$ & $-0.292$ & $\mathbf{-5.72}$ \\
57789.442 & 57805.348 & 3.762 & $-0.741$ & $\mathbf{-8.29}$ & $-0.239$ &$\mathbf{ -3.73}$ \\
57874.208 & 57892.294 & 4.278 & $-0.432$ & $\mathbf{-4.41}$ & $-0.036$ & $-0.53$ \\
57901.209 & 57918.164 & 4.010 & $-2.508$ & $\mathbf{-22.79}$ & $-0.208$ & $-2.66$ \\
57918.164 & 57933.421 & 3.609 & $-0.591$ & $\mathbf{-5.83}$ & $-0.914$ & $\mathbf{-13.54}$ \\
58127.033 & 58146.422 & 4.586 & $0.672$ & $\mathbf{6.94}$ & $0.117$ & $1.85$ \\
58146.422 & 58174.079 & 6.541 & $0.443$ & $\mathbf{4.45}$ & $0.191$ & $2.96$ \\
58502.473 & 58531.501 & 6.866 & $0.419$ & $\mathbf{4.16}$ & $0.057$ & $0.84$ \\
58531.501 & 58560.572 & 6.876 & $0.455$ & $\mathbf{4.50}$ & $0.108$ & $1.79$ \\
59230.262 & 59232.492 & 0.528 & $-0.094$ & $-1.10$ & $0.209$ & $\mathbf{4.21}$ \\
59238.492 & 59252.320 & 3.271 & $-0.460$ & $\mathbf{-6.59}$ & $-0.243$ & $\mathbf{-5.30}$ \\
59296.457 & 59308.324 & 2.807 & $0.083$ & $0.95$ & $-0.358$ & $\mathbf{-7.09}$ \\
59308.324 & 59311.434 & 0.735 & $-0.017$ & $-0.22$ & $0.438$ & $\mathbf{9.54}$ \\
59311.434 & 59314.207 & 0.656 & $0.065$ & $0.79$ & $-0.291$ & $\mathbf{-5.60}$ \\
59721.491 & 59724.198 & 0.640 & $-0.522$ & $\mathbf{-5.89}$ & $0.087$ & $1.21$ \\
59724.198 & 59729.886 & 1.345 & $0.384$ & $\mathbf{4.57}$ & $-0.035$ & $-0.74$ \\
59938.883 & 59973.731 & 8.242 & $-0.759$ & $\mathbf{-5.88}$ & $-0.181$ & $-2.13$ \\
59973.731 & 59999.841 & 6.175 & $0.579$ & $\mathbf{4.22}$ & $0.359$ & $\mathbf{4.32}$ \\
59999.841 & 60003.448 & 0.853 & $0.721$ & $\mathbf{5.68}$ & $-0.303$ & $-3.76$ \\
60003.448 & 60030.359 & 6.365 & $-0.663$ & $\mathbf{-4.32}$ & $0.276$ & $2.98$ \\
60072.766 & 60076.085 & 0.785 & $0.078$ & $1.06$ & $0.217$ & $\mathbf{4.67}$ \\
60088.186 & 60104.189 & 3.785 & $-0.166$ & $-1.46$ &$ 0.341$ & $\mathbf{4.93}$ \\
60106.199 & 60108.210 & 0.476 & $-0.146$ & $-1.14$ & $-0.366$ & $\mathbf{-4.65}$ \\
60110.227 & 60118.205 & 1.887 & $-0.253$ & $-2.82$ & $0.863$ & $\mathbf{12.82}$ \\
60373.381 & 60379.463 & 1.439 & $0.412$ & $3.43$ & $0.394$ & $\mathbf{4.76}$ \\
60432.158 & 60434.166 & 0.475 & $-0.107$ & $-0.86$ & $0.423$ & $\mathbf{5.90}$ \\
60461.168 & 60463.221 & 0.486 & $-0.192$ & $-1.99$ & $0.437$ & $\mathbf{6.87}$ \\
60463.221 & 60465.178 & 0.463 & $0.110$ & $1.11$ & $-0.370$ & $\mathbf{-5.39}$ \\
60711.430 & 60728.409 & 4.016 & $0.029$ & $0.25$ & $-0.342$ & $\mathbf{-4.44}$ \\
60737.342 & 60756.265 & 4.476 & $-0.922$ & $\mathbf{-8.83}$ & $-0.214$ & $-3.23$ \\
60756.265 & 60786.233 & 7.088 & $0.595$ & $\mathbf{5.95}$ & $0.519$ & $\mathbf{8.11}$
\enddata
\tablenotetext{*}{Bolded numbers indicate measurements that exceed the 4$\sigma$ threshold to be considered ``significant" to aid the eye in identifying which Trough varied significantly in that pair of epochs. }
\end{deluxetable*}



\section{Discussion} 
Our primary measurements/observations from the work above can be summarized as follows: 
\begin{itemize} 
\item This quasar has one \civ BAL feature (Trough A) at a velocity centered at roughly $-20,150$~km~s$^{-1}$ and a weaker, lower-velocity mini-BAL (Trough B) at a velocity centered at roughly $-3700$~km~s$^{-1}$. 
We see corresponding absorption in  O\,{\sc vi}, Ly$\alpha$, and N\,{\sc v} as well, but these systems are in a crowded and blended region of the spectrum that makes the analysis of their variability difficult, so they are not examined in this work. 
\item Both Trough A and Trough B show significant variability in EW and depth throughout the 12-year duration of monitoring; see Figures~\ref{fig:Measurements} and ~\ref{fig:B Measurments}. Due to a slow change in shape of the troughs over time, their centroid velocities decrease slowly over the course of the observations (See Figure~\ref{fig:app_fig_vcent}), but the upper/lower edges of the absorption are very stable, suggesting that the change in $v_{\rm cent}$ is due to a change in outflow structure/shape rather than evidence for actual global acceleration/deceleration of the gas. 
\item Both Trough A and Trough B vary significantly in EW on very short timescales (Figure~\ref{fig:deltaEW vs deltaMJD}, Table~\ref{tab:significant_EW_variability}). We find 33 cases of significant variability in either Trough A or B on timescales of less than 10 days in the quasar rest frame and we confirm a particularly dramatic case when Trough A weakens at a significance of 22$\sigma$ over a timescale of 4.01 days (originally identified by H19; see Figure~\ref{fig:EW significant pairs}). Trough B then undergoes a similar dramatic (13$\sigma$) weakening in the next pair of epochs over a timescale of 3.7 days.
\item  Both Troughs A and B weaken considerably in the later years of monitoring. Trough B disappears entirely during a period between 2022 and 2023 and then reappears by 2024 at roughly the same velocity as before disappearance (Figure~\ref{fig:app_mean_spec_all}). 
\end{itemize} 
\label{sec:discussion} 
 We here further explore the behavior of Troughs A and B, examine possible scenarios for what is causing the variability, and discuss the implications and possible interpretations of these various pieces of evidence. 
 
\subsection{The Relationship with the Observed Continuum}
\label{sec:con_lack}
\begin{figure}
    \centering
    \includegraphics[width=0.98\linewidth]{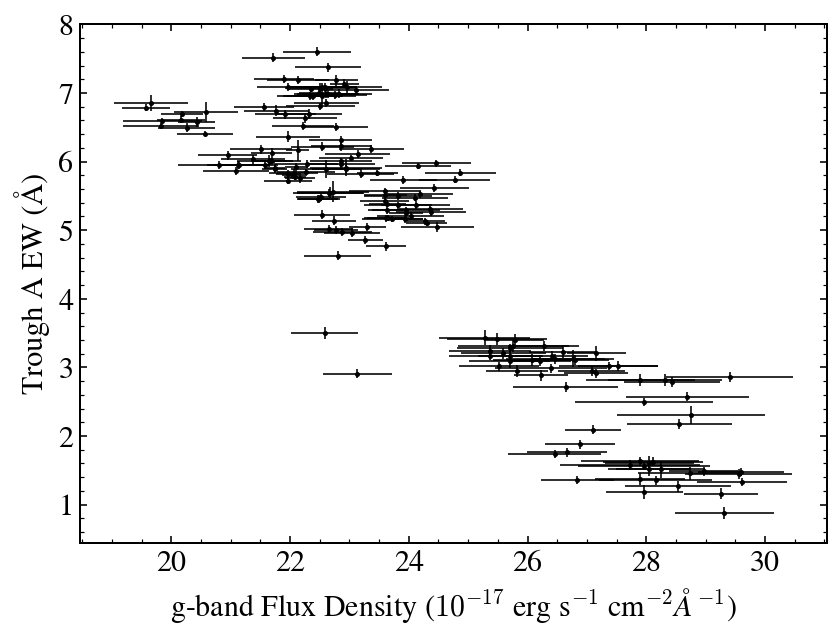}
    \caption{The equivalent width of Trough A (y-axis) as a function of the $g$-band Flux Density (x-axis). $g$-band flux densities in this plot were generated by linearly interpolating between points to match the MJDs of the Trough A EW values. }
    \label{fig:C1700 vs EW}
\end{figure}

To measure the correlation between the troughs and the continuum, we use the $g$-band flux density (Figure~\ref{fig:Measurements}) measured by \cite{Pena26} and cross-correlate with our measurements of EW. Because we have many more photometric data points that do not match the EW measurements exactly in MJD, we interpolate the $g$-band light curve to estimate the flux at the exact MJDs of our spectroscopy. We then compare the interpolated $g$-band flux with the Trough EW. Figure~\ref{fig:C1700 vs EW} shows the Trough A EW as a function of the $g$-band continuum flux density. We find a strong anticorrelation between the strength of the continuum and the strength of the BAL:  We calculate a Pearson coefficient of $r = -0.91$ with a p-value of $\approx 0$ and a Spearman rank coefficient of $\rho = -0.84$ with a p-value of $\approx 0$. We find a similarly strong correlation between the $g$-band flux density and the EW of Trough B, with a Pearson coefficient of $r = -0.90$ with a p-value of $\approx 0$  and a Spearman rank coefficient of $\rho = -0.86$ with a p-value of $\approx 0$. 

Because our EW values are measured from continuum-normalized spectra, there is a possibility that our observed correlation could be in part due to mathematical coupling between the two quantities; i.e. we are correlating $f_{\rm con}$ with something that depends on $1/f_{\rm con}$. We thus computed the integrated absorbed flux ($f_{\rm BAL}$) rather than the EW, which removes the explicit dependence on the continuum normalization, and repeated our correlation analysis. We find that the correlation between $f_{\rm BAL}$ and $g$-band flux density is similarly strong as that between EW and $g$-band flux density, suggesting that the observed correlations are not an artifact of the normalization procedure.  

Interestingly, if we use the 1700\,\AA\ continuum instead of the $g$-band photometric flux, we find no significant correlation between the continuum and the trough EWs. The 1700\,\AA\ flux density shows very little variability throughout the duration of our campaign. This is consistent with past work that suggests that the observed continuum at 1700\,\AA\ is not necessarily a good proxy for the C\,{\sc iv}-ionizing continuum, which is emitted in the extreme ultraviolet (EUV) and that other species that respond more directly to the EUV can be better proxies for the ionizing continuum (e.g., \citealt{Timlin21}). The EUV has been found to be significantly more variable than longer wavelengths (e.g., \citealt{Marshall97}); our observed continuum may thus be a poor representation of the actual variability that the BAL ``sees". In this case, we find that the $g$-band photometric light curve, which is tracing a bluer region in the quasar rest frame (see Section~\ref{sec:basic_measurements}), appears to better represent the behavior of the C\,{\sc iv}-ionizing continuum. The SDSS $g$ band covers roughly 4000-5400\,\AA\ in the observed frame, which corresponds to roughly 1000-1275\,\AA\ at $z=3.228$ (see Figure~\ref{fig:app_filter}). This band thus traces primarily the Ly$\alpha$ emission line and the region blueward. 

\subsection{Coordination between Troughs A and B} 
\label{sec:AvB}
\begin{figure*}
    \centering
    \includegraphics[width=0.9\linewidth, trim = 0 0 0 0, clip]{
    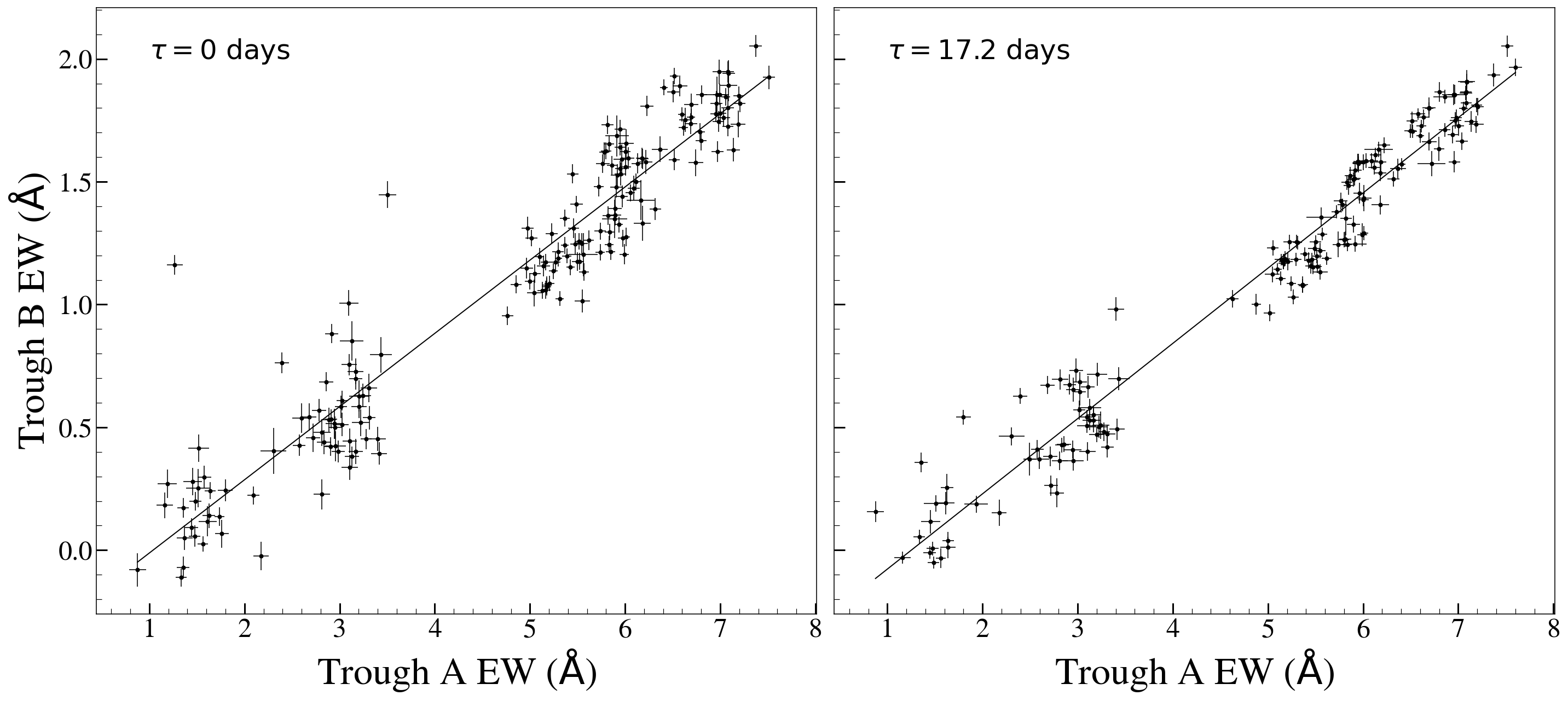}
    \caption{Equivalent width of Trough B (y-axis) vs Equivalent width of Trough A (x-axis) with and without an applied time delay. The solid line indicates the best-fit line to the data. The left plot shows the correlation between the original measurement with no time delay applied ($\tau = 0$). The major outliers in this subplot represent one point at MJD 57918, when the EW of A has dropped significantly but the EW of Trough B has not yet dropped. The second outlier is at MJD 60118, when we see an inconsistency in the emission-line fit, causing the EW of Trough B to to be overestimated (see Section~\ref{sec:sig_var}). The right panel shows the correlation after a 17.2-day (observed-frame) lag has been applied (see Section~\ref{sec:AvB}), demonstrating the reduced scatter and increased strength of the correlation when a time delay is considered. }
    \label{fig:ew_correlation_AB}
\end{figure*}

We next searched for coordination between the behavior of Troughs A and B. The two absorption troughs are separated by $\sim$16,400 km~s$^{-1}$ in velocity space; this separation is large enough that it is extremely unlikely that they are part of the same structure. However, we find a strong correlation between the two: Figure~\ref{fig:ew_correlation_AB} shows the EW of Trough A compared to Trough B, indicating a strong, significant correlation between the two quantities. Pearson and Spearman Rank correlation tests confirm this: The two quantities have a Pearson coefficient value of $r = 0.9547$ ($p \approx 0$), and a Spearman Rank coefficient of $\rho =  0.9502$ ($p \approx 0$).

The strong correlation between the EW in Troughs A and B and the fact that they disappear and reappear almost in concert is very strong evidence that something global is modulating both absorption features, such as changes in the ionizing continuum that the outflow is seeing. If the changes in observed strength of these two troughs were due to transverse motions or changes in density of the outflows as they move across the line of sight, we would not expect to see such strongly correlated behavior between two widely seprated features. The correlation between the behavior of these troughs (Figure~\ref{fig:ew_correlation_AB}), combined with the strong correlation with the $g$-band flux density, is strong evidence for ionization state changes as the driver of the observed variability. We hereby adopt this as our favored model for the discussion moving forward. Many previous studies have found variability consistent with ionization state variability (e.g., \citealt{Misawa10}; \citealt{Filizak12, Filizak13}; \citealt{Capellupo12}; \citealt{Grier15}, \citealt{Rogerson18}).

\begin{figure*}
    \centering
    \includegraphics[width=0.85\linewidth, trim = 0 0 0 65,clip]{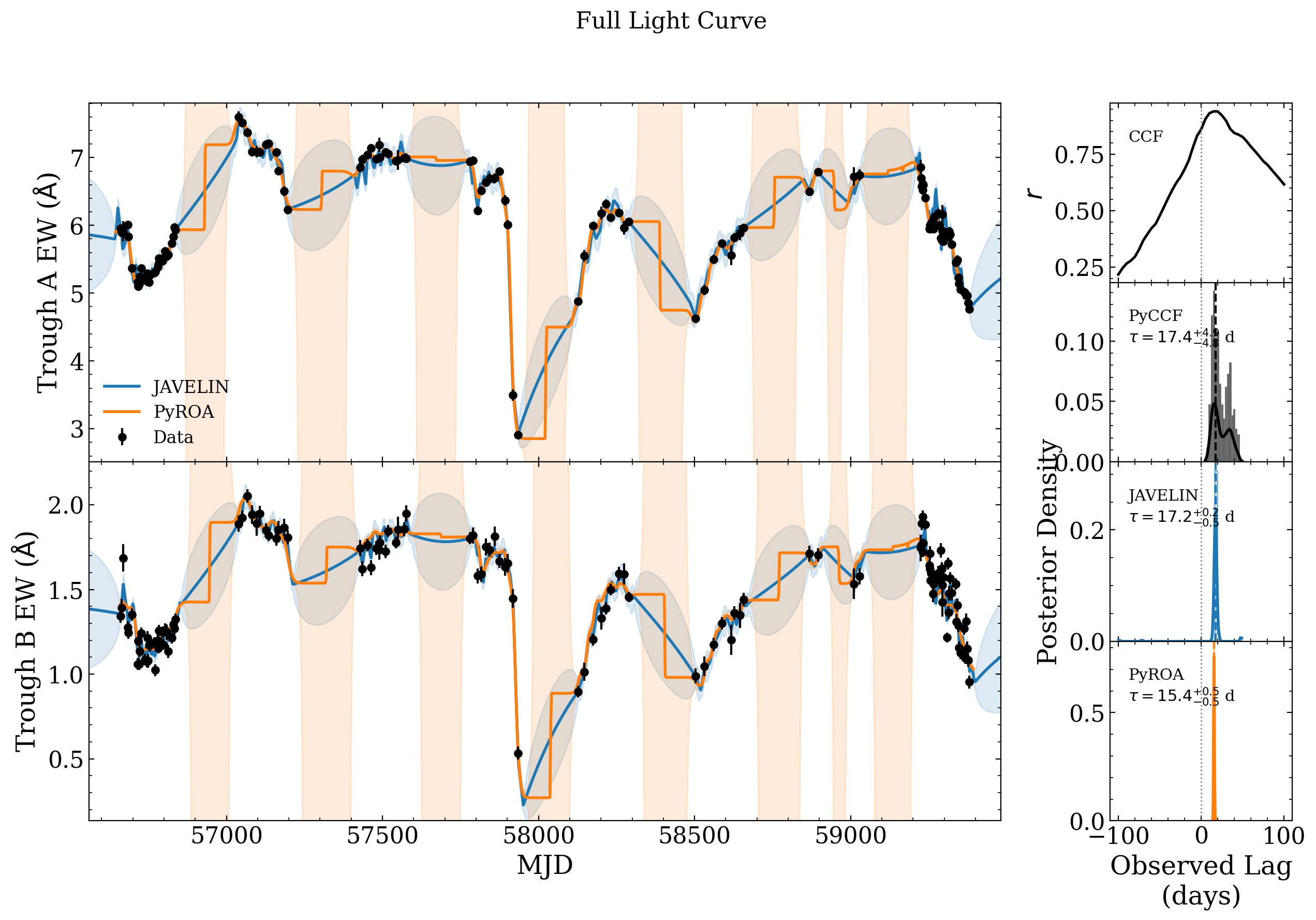}
    \caption{Lag results for Trough A and Trough B, with Trough A as the driving light curve. The two left-hand plots show the EW as a function of time for Trough A (top) and Trough B (bottom), with the best-fit \javelin and PyROA models shown in blue and orange, respectively. The right-hand plots show the CCF (top), cross correlation centroid distribution from {\tt PyCCF} (second-from-top), posterior distribution of lag samples from {\tt JAVELIN} (third-from-top), and posterior distribution of lag samples from {\tt PyROA} (bottom). The thicker lines on top of the posterior distributions show the KDE (see Section~\ref{sec:peak_selection}) generated to select samples in the primary peak.}
    \label{fig:lags_fulldata}
\end{figure*}

It is clear that the EW of the two troughs vary in a strongly correlated fashion, so we next investigate whether or not they vary simultaneously. We searched for a time delay between Troughs A and B using a variety of lag-measurement software packages that are commonly employed in reverberation mapping studies (e.g., \citealt{Shen24}). We began with a standard interpolated cross correlation function technique (ICCF; e.g., \citealt{Peterson98}) as implemented by the \pyccf software (\citealt{Sun18}). We also employed the \javelin and \pyroa software (\citealt{Zu11}; \citealt{Donnan21}) to confirm (or reject) the lag detection results using alternate methods, as each of these three lag-detection methods has different strengths and weaknesses and tends to report very different measurement uncertainties (e.g., \citealt{Edelson19}). For example, the ICCF method has been found to be less sensitive to poorly estimated uncertainties in the light curves than \javelin (\citealt{Yu20}), though \pyroa includes an algorithm to adjust the uncertainties in the process when they are found to be questionable. In addition, \javelin reports smaller uncertainties that have been found via simulations to be more representative of the actual statistical uncertainties than the ICCF method, which tends to overestimate the uncertainties (\citealt{Edelson19, Yu20}) --- \pyroa tends to report similarly small uncertainties as \javelin. See the Appendix for details on the lag-detection methodologies and procedures used in this work.

We chose the EW of Trough A as the ``driving" light curve and measured the response time of Trough B EW. We use only data spanning from 2014--2021, as both troughs effectively disappear in 2022 and remain weak (or nonexistent, in the case of Trough B) for the rest of the monitoring period. When exploring the full light curve, we allow for time lags ranging from $-500$ to 500 days. 
 
 Figure~\ref{fig:lags_fulldata} shows the results of our time series analysis. In all three cases, we detect a significant, nonzero lag (see the Appendix for our adopted lag significance criteria). \pyccf detected an observed-frame lag of 22.4$^{+12.5}_{-7.4}$ days, with \javelin reporting a lag of 17.2$^{+0.2}_{-0.5}$ days and \pyroa reporting a lag of 15.4$^{+0.5}_{-0.5}$ days (see Figure~\ref{fig:lags_fulldata}). These three measurements are formally consistent with one another to within the uncertainties. 

This lag is short enough that we should be able to detect it within a single season of observations. To further determine whether our lag measurements from the full EW ``light curve" are robust, we divided our EW light curves into separate observing seasons and re-ran our lag-detection software on each observing season individually to see if we can detect a time lag in a single season. We did not search for a lag in 2020, as there were only four data points that year, and we again note that Trough B has disappeared in 2022--2023 and is still extremely weak in 2024 and 2025, so we do not have sufficient data quality during those years to detect a lag. We thus omit those years in our analysis. For the individual years, we limit the lag range to $\pm$100 days. Table~\ref{tab:bal_lags} shows the results for each individual season for each of the three detection methods, and Figures showing the posterior distributions and models for each individual observing season are provided in the Appendix (Figures~\ref{fig:2014lag}-\ref{fig:2021lag}). We detect a significant, nonzero lag in the 2014, 2015, 2017, and 2021 observing seasons. During these years, the BAL EW light curve shows a number of inflection points that make these light curves favorable for lag measurements. In 2016, 2018, and 2019, the light curves show either very little variability (2016), or contain only monotonically rising features (2018 and 2019), which make it difficult to measure a lag. We note that we can see a hint of a visible lag in the 2018 light curve (Figure~\ref{fig:2018lag}), but its measurement was too uncertain for us to consider it a solid detection.

\begin{deluxetable}{crrr}
\tablecaption{Lag Measurements between Trough A and Trough B
\label{tab:bal_lags}}
\tablewidth{0pt}
\tablehead{
\colhead{Observing} &
\colhead{$\tau_{\rm PyCCF}$} &
\colhead{$\tau_{\rm JAVELIN}$} &
\colhead{$\tau_{\rm PyROA}$} \\
\colhead{Season} &
\colhead{(days)} &
\colhead{(days)} &
\colhead{(days)}
}
\startdata
2014 & $4.8^{+7.9}_{-7.5}$ & $\mathbf{17.1^{+3.2}_{-6.7}}$ & $\mathbf{9.7^{+2.4}_{-2.6}}$ \\
2015 & $54.8^{+30.2}_{-14.5}$ & $\mathbf{45.8^{+12.2}_{-10.9}}$ & $\mathbf{46.5^{+6.7}_{-5.8}}$ \\
2016 & $57.6^{+27.4}_{-142.4}$ & $-11.6^{+65.8}_{-60.0}$ & ${-73.0^{+39.4}_{-19.0}}$ \\
2017 & $7.2^{+13.0}_{-22.6}$ & $\mathbf{12.4^{+18.8}_{-2.6}}$ & $\mathbf{3.1^{+2.0}_{-1.8}}$ \\
2018 & ${40.0^{+12.8}_{-24.7}}$ & $-4.5^{+43.2}_{-62.9}$ & ${-8.5^{+1.1}_{-0.9}}$ \\
2019 & $-2.4^{+65.0}_{-62.6}$ & $-1.9^{+66.3}_{-67.3}$ & $3.2^{+5.9}_{-7.0}$ \\
2021 & $2.6^{+7.3}_{-9.5}$ & $\mathbf{8.1^{+2.3}_{-3.8}}$ & $\mathbf{11.2^{+0.9}_{-1.0}}$ \\
\hline
All data & $\mathbf{22.2^{+12.6}_{-9.6}}$ & $\mathbf{17.2^{+0.2}_{-0.5}}$ & $\mathbf{15.4^{+0.5}_{-0.5}}$ \\
\enddata
\tablecomments{
Lag measurements are given in the observed frame.
Bold measurements satisfy all adopted quality
criteria as discussed in the Appendix (Section~\ref{sec:lag_significance}). 
}
\end{deluxetable}

We adopt $\tau_{\rm JAVELIN}$ measurements as our fiducial lag values, as the \javelin models appear to describe the data very well for all four years (see, for example, 2017, where {\tt JAVELIN}'s DRW model is able to accommodate the dramatic decrease in EW, whereas {\tt PyROA}'s running optimal average model is not). Converting our measured lags into the quasar rest frame, we find the ``average" lag over the entire observing period to be $\tau_{\rm JAVELIN,rest} = 4.07^{+0.05}_{-0.12}$ days in the quasar rest frame, and the lag measured in individual seasons ranges is typically on the order of a few days (4.0$^{+0.8}_{-1.6}$ days in 2014, 2.9$^{+4.5}_{-0.6}$ in 2017, and 1.9$^{+0.5}_{-0.9}$ in 2021), with a longer lag of 10.8$^{+2.8}_{-2.6}$ days detected in 2015. 

In the case of ionization state variability, an observed time delay between Trough A and Trough B would mean that the Trough B absorber reaches its new ionization equilibrium roughly 4 days more slowly than Trough A. The recombination time is related to the density of the absorbing gas ($t_{\rm rec} \propto 1/\alpha_{\rm rec}n_e$), so one potential interpretation of this time lag is that the two absorbers are responding to the same ionizing continuum as one another, but the two outflow structures have different densities. In this case, Trough B would have a longer recombination time than Trough A, which would suggest that its density is lower than that of the gas producing Trough A. 
An alternative interpretation is that the two absorbers are located at different distances: When the continuum increases, the response time may instead be dominated by the photoionization rate, which scales as $L/r^2$. The measured lag could therefore also reflect different distances of the two absorbers from the central source, rather than different densities. 

\subsection{Summary and Future Work} 
We have completed an investigation of a \civ BAL in the quasar J1421, which shows dramatic short-timescale variability as well as longer-timescale changes in strength over our 12 years of monitoring. We find that the main BAL, Trough A, and its lower-velocity mini-BAL counterpart, Trough B, show significant variability on short timescales throughout the duration, consistent with previous studies that have found such short-timescale variability (\citealt{Capellupo13, Grier15, Hemler19}). The variability of the two troughs is highly correlated with the $g$-band continuum flux as well as with one another, but the signals are not simultaneous; 
we find that Trough B lags Trough A by about 4 days in the quasar rest frame on average. 
This is the first detection of time delays between two BAL troughs within the same system, and potentially indicates different physical conditions (i.e. densities or distances) of the two absorbing structures.

Past datasets have not had sufficient cadence or long enough duration to explore this topic in detail and measure time lags in multi-trough systems; future work using high-cadence monitoring data such as that from SDSS would be illuminating to determine whether or not this behavior is unique, or whether time lags in variations in BALs in multi-trough systems are commonly seen in BAL quasars. The SDSS-V Black Hole Mapper program contains many BAL quasars suitable for a similar investigation (Seaton et al, in prep; Robinson et al., in prep), and the upcoming SDSS-VI Dynamic Universe Explorer program (starting in 2027) is expected to continue to monitor many of these sources to further extend the time baseline covered. Additional multi-epoch spectroscopic programs such as 4MOST, which are also obtaining many epochs of spectroscopy of numerous quasars, may also produce datasets that are suitable for such investigations in the future. 


 

\begin{acknowledgments} 
CJG, LR, JRT and YS acknowledge support from NSF grants AST-2310211 and AST-2309930. Support for this research was provided by the Office of the Vice Chancellor for Research and Graduate Education at the University of Wisconsin–Madison with funding from the Wisconsin Alumni Research Foundation. This material is based upon work supported by the National Space Grant College and Fellowship Program and the Wisconsin Space Grant Consortium under NASA Educational Cooperative Agreement \#80NSSC25M7122. WNB acknowledges support from NSF grant AST-2407089. MEC acknowledges support from SECIHTI through a postdoctoral fellowship within the programme ``Estancias Posdoctorales por México”.

Funding for the Sloan Digital Sky Survey V has been provided by the Alfred P. Sloan Foundation, the Heising-Simons Foundation, the National Science Foundation, and the Participating Institutions. SDSS acknowledges support and resources from the Center for High-Performance Computing at the University of Utah. SDSS telescopes are located at Apache Point Observatory, funded by the Astrophysical Research Consortium and operated by New Mexico State University, and at Las Campanas Observatory, operated by the Carnegie Institution for Science. The SDSS web site is \url{www.sdss.org}.

SDSS is managed by the Astrophysical Research Consortium for the Participating Institutions of the SDSS Collaboration, including the Carnegie Institution for Science, Chilean National Time Allocation Committee (CNTAC) ratified researchers, Caltech, the Gotham Participation Group, Harvard University, Heidelberg University, The Flatiron Institute, The Johns Hopkins University, L'Ecole polytechnique f\'{e}d\'{e}rale de Lausanne (EPFL), Leibniz-Institut f\"{u}r Astrophysik Potsdam (AIP), Max-Planck-Institut f\"{u}r Astronomie (MPIA Heidelberg), Max-Planck-Institut f\"{u}r Extraterrestrische Physik (MPE), Nanjing University, National Astronomical Observatories of China (NAOC), New Mexico State University, The Ohio State University, Pennsylvania State University, Smithsonian Astrophysical Observatory, Space Telescope Science Institute (STScI), the Stellar Astrophysics Participation Group, Universidad Nacional Aut\'{o}noma de M\'{e}xico, University of Arizona, University of Colorado Boulder, University of Illinois at Urbana-Champaign, University of Toronto, University of Utah, University of Virginia, Yale University, and Yunnan University.   
\facilities{Apache Point Observatory (SDSS)}

\software{Astropy \citep{Astropy13, Astropy22}, NumPy \citep{Harris20}, SciPy \citep{Virtanen20}, Matplotlib \citep{Hunter07}}

\end{acknowledgments}

\appendix  
\label{sec:appendix} 
\renewcommand{\thetable}{A\arabic{table}}
\setcounter{table}{0}
\renewcommand{\thefigure}{A\arabic{figure}}
\setcounter{figure}{0}

\section{Additional Figures and Tables} 
\begin{figure}
    \centering
    \includegraphics[width=0.9\linewidth]{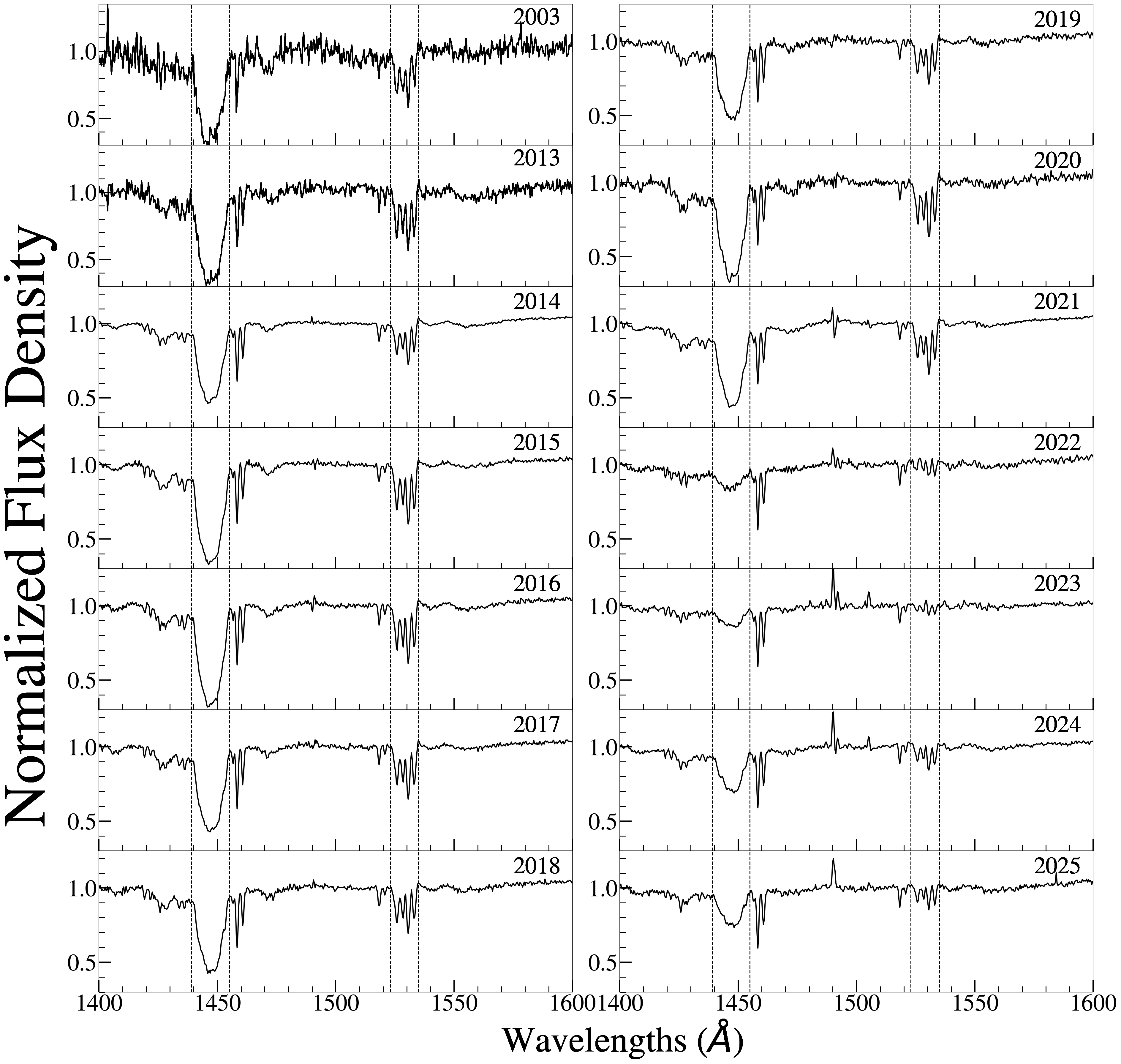}
    \caption{Mean normalized spectra from each individual year. Since we only have a single spectrum from 2003, we just show that individual epoch.}
    \label{fig:app_mean_spec_all}
\end{figure}

In this section, we provide additional figures and tables to support the work discussed in the main text. Figure~\ref{fig:app_mean_spec_all} shows the mean spectra of each individual observing season. Table~\ref{tab:spectral_data} provides BAL measurements for each individual epoch in our study and Table~\ref{tab:mean_measurements_by_year} provides measurements made in the mean spectra of each individual observing season (corresponding to the spectra shown in Figure~\ref{fig:app_mean_spec_all}). 

\begin{figure}
    \centering
    \includegraphics[width=0.9\linewidth]{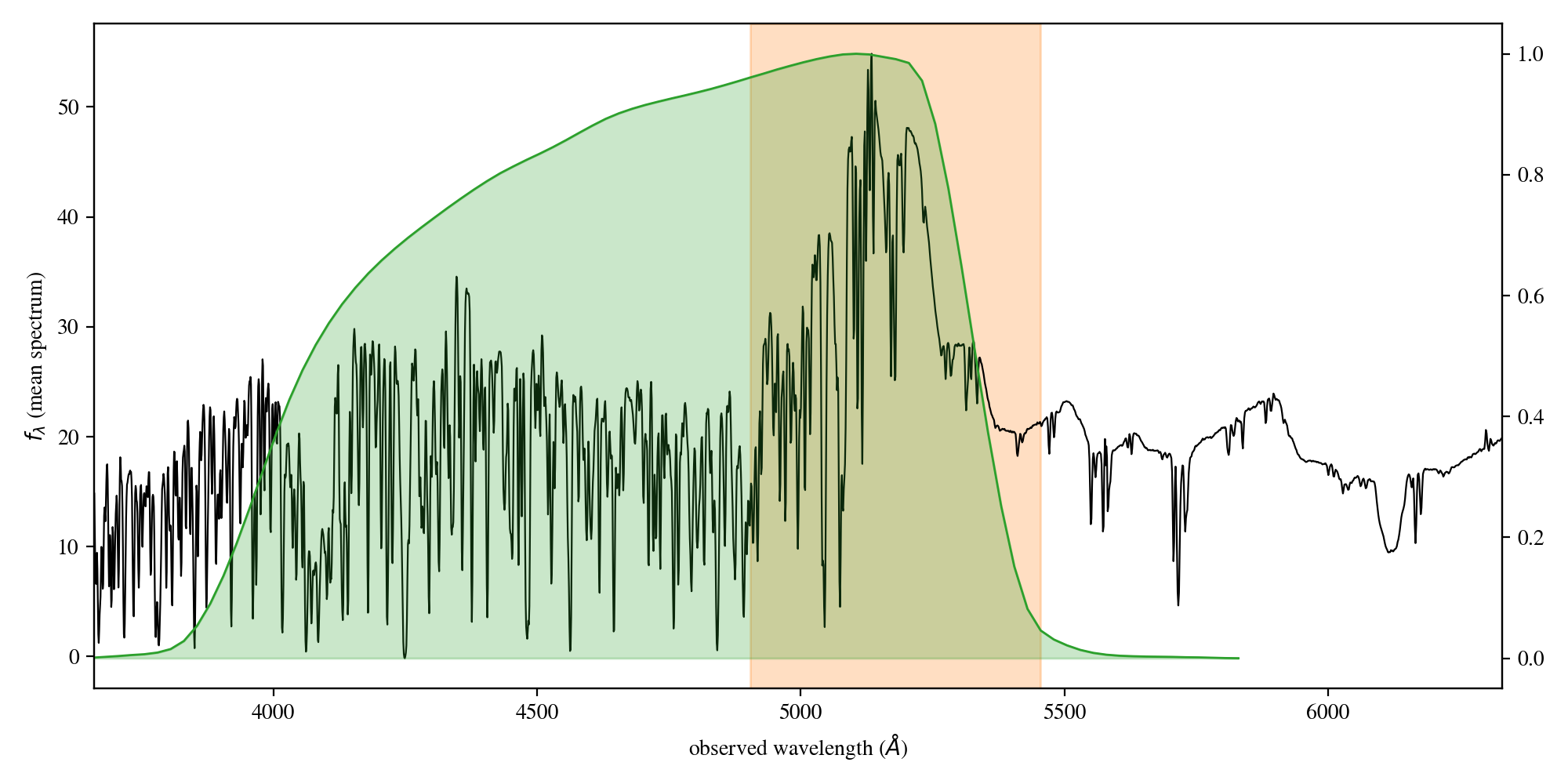}
    \caption{The mean corrected spectrum with wavelengths shifted into the observed frame, compared to the SDSS $g$-band filter bandpass. The green shaded region shows the region covered by the $g$-band filter, and the orange rectangle shows the region encompassed by Lyman-$\alpha$. We find that for our quasar, the $g$ band covers primarily the Lyman-$\alpha$ emission line and the region blueward.}
    \label{fig:app_filter}
\end{figure}

\section{Lag Measurement Methodology} 

\renewcommand{\thefigure}{B\arabic{figure}}
\setcounter{figure}{0}
In this section, we provide additional information and figures describing our lag-measurement methodology and the procedures used to measure the time delay between Trough A and Trough B in Section~\ref{sec:AvB}.

\subsection{PyCCF}
 One of the most commonly used methods to measure time delays in reverberation mapping is the Interpolated Cross Correlation Function (ICCF). With the ICCF method, the user specifies a grid over which the software linearly interpolates between data points, and the CCF is constructed by calculating the Pearson coefficient $r$ between the two light curves at each possible lag on that grid. We measure the centroid of the CCF ($\tau_{\rm cent}$) using points surrounding the maximum correlation coefficient $r_{\rm max}$ of the CCF. We used the {\tt PyCCF} code\footnote{The {\tt PyCCF} code is available for download at https://bitbucket.org/cgrier/python\_ccf\_code.} (\citealt{Peterson98}; \citealt{Sun18}) to carry out the ICCF analysis. We chose a grid spacing of 2 days. We calculate the best lag measurement and its uncertainties via the flux randomization/random subset sampling method, using Monte Carlo simulations, as discussed by \cite{Peterson04}. We perform 500 realizations to obtain the cross correlation centroid distribution (CCCD). We then isolate the primary peak of the CCCD via the method described below in Section~\ref{sec:peak_selection} and measure the median and uncertainties from the isolated primary peak in the CCCD. 

\subsection{JAVELIN}
Our secondary method for time lag detection is the \javelin code \citep{Zu11, Zu13}, which models the light curves as an autoregressive process using a damped random walk (DRW) model. DRW models have been demonstrated to be a good description of quasar continuum behavior on the relevant timescales to our study (e.g., \citealt{Kelly09, Kozlowski10, Kozlowski16, Macleod10, Macleod12}); however, we note that the DRW model has not been widely used in the past to model {\it BAL} behavior and to our knowledge, this work is the first application of \javelin to the variability of BALs. Because both Troughs A and B are highly correlated with the $g$-band photometric light curve, which can be well modeled by a DRW, we thus argue that the DRW model is applicable to the BALs as well. 

\javelin first fits a DRW model to the continuum light curve to set priors on the DRW parameters, and then \javelin models both the continuum and emission-line light curves, simultaneously fitting a transfer function, using Markov Chain Monte Carlo techniques. \javelin applies expanded uncertainties to model points that fall during gaps in the light curves to accounts for all likely behavior of the light curves. We adopt a top-hat transfer function that is parameterized by a scaling factor, width, and time delay. In all cases, we allow \javelin to explore a range of observed lags from $\pm$100 days; results from \pyccf suggest that all of our lags should be smaller than 100 days, and if we allow \javelin to explore larger ranges, it has a tendency to have trouble with the regular $\sim$180-day seasonal gaps and often returns lag measurements of $\sim$180 days. After we run \javelin, we then determine $\tau_{\rm JAVELIN}$, the best-fit time delay, from the posterior distribution of lags that is produced by the MCMC chain, after applying the posterior selection procedure described below. 

\subsection{PyROA} 
PyROA (\citealt{Donnan21}) measures time delays by simultaneously modeling the light curves using a running optimal average (ROA) rather than assuming a DRW model (or any other particular variability model). The ROA constructs an empirical representation of the light curve variability as an inverse-variance-weighted average of the observed data, with a Gaussian window that gives greater weight to observations close in time. 
PyROA simultaneously constrains the lag and the other model parameters using Markov Chain Monte Carlo (MCMC) sampling, producing a posterior probability distribution for the time delay from which the lag and its uncertainty can be determined. PyROA can also include an additional variance term for each light curve, allowing it to adjust the uncertainties to better represent errors or intrinsic scatter than the reported measurement errors.

\subsection{Posterior Sample Selection}  
\label{sec:peak_selection}
The lag posterior distributions produced by the Monte Carlo realizations of the ICCF or the MCMC iterations of \javelin and \pyroa can contain multiple peaks. To isolate the primary lag solution, we first smooth over the probability density of the posterior lag samples using a Gaussian kernel density estimate (KDE). We identify local maxima in the KDE candidate lag solutions, and select the maximum that lies closest to the centroid measured from the original, non-resampled posterior distribution as the preferred solution. The boundaries of the primary peak are taken to be the nearest local minima in the KDE on either side of the selected maximum. We then retain only the Monte Carlo lag samples lying within these boundaries and use this subset of the posterior lag samples to determine the final lag and its uncertainty. We adopt the median of the selected posterior samples, with the uncertainties  corresponding to the central 68\% interval. This procedure minimizes interference by secondary or alias peaks in the actual reported lag measurements.

\subsection{Lag Significance Criteria}
\label{sec:lag_significance}
While many of the lags identified by our procedures are robust, all three of our lag detection methods will report the ``best" lag they have identified and/or the ``best" model fit to the data regardless of whether or not the lag is clearly detected. To reject false lag detections, we impose additional requirements on our lag measurements for the lags to be considered robust or significant. We consider a lag to be ``significant" if the following criteria are met: 
\begin{enumerate} 
\item{The maximum $r$ value of the CCF between the two light curves is $>$ 0.5, indicating an actual correlation between the two light curves.}
\item{Less than 50\% of the posterior lag samples (or Monte Carlo samples, in the case of the ICCF method) are rejected in the Posterior Sample Selection stage as listed above, indicating that most of the samples lie in a single, well-defined peak in the posteriors.} 
\item{The lag is inconsistent with zero to at least 1$\sigma$. }
\item{The lags from all three methods are generally consistent; i.e. none of the methods identified significant negative lags when the others measured positive, etc. (unless one method is clearly unable to find a reasonable solution; for example, if the model fit for \pyroa is clearly not a good fit to the actual data but the other two methods look reasonable).}
\end{enumerate}

\begin{figure*}
    \centering
    \includegraphics[width=0.7\linewidth, trim = 0 0 0 50, clip] {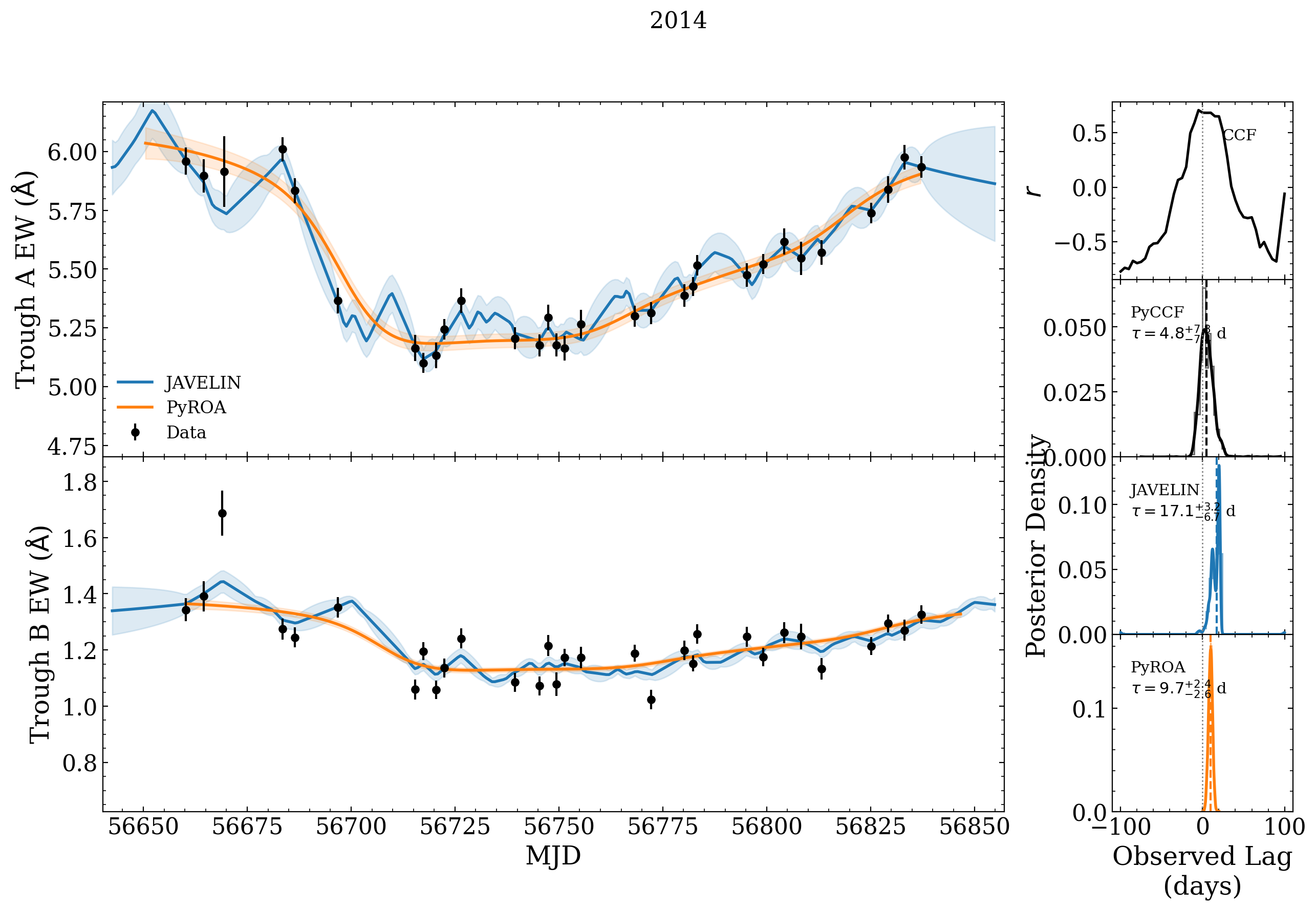}
    \caption{Lag results for Trough A and Trough B with Trough A as the driver for 2014. The two left-hand plots show the EW as a function of time for Trough A (top) and Trough B (bottom), with the best-fit \javelin and \pyroa models shown in blue and orange, respectively. The right-hand plots show the CCF (top), cross correlation centroid distribution from {\tt PyCCF}, the posterior distribution of lag samples from {\tt JAVELIN}, and the posterior distribution of lag samples from {\tt PyROA}. The thicker lines on top of the posterior distributions show the KDE (see Section~\ref{sec:peak_selection}) generated to select samples in the primary peak. }
    \label{fig:2014lag}
\end{figure*}

\begin{figure*}
    \centering
    \includegraphics[width=0.7\linewidth, trim = 0 0 0 50, clip]{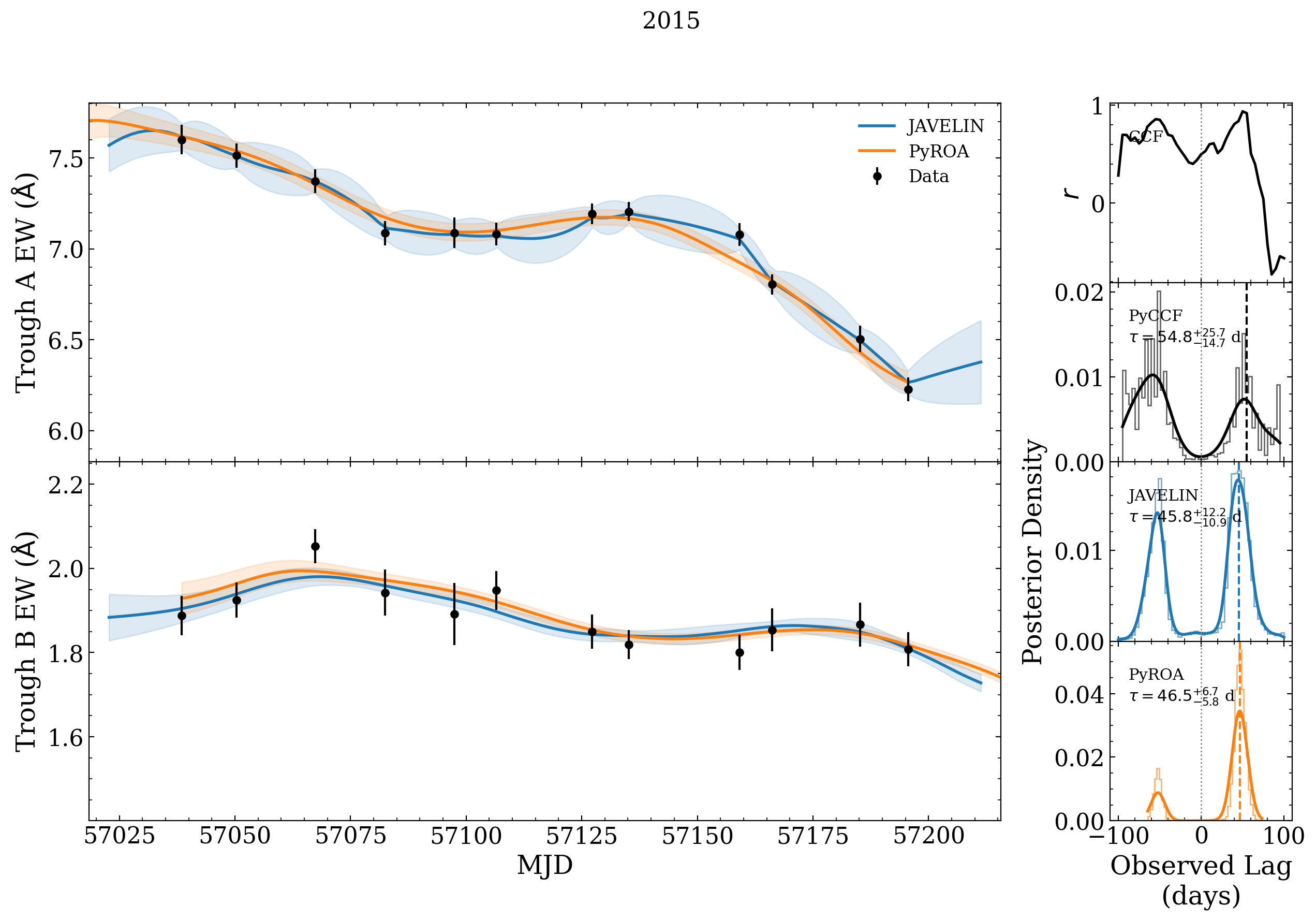}
    \caption{Lag results for Trough A and Trough B with Trough A as the driver for 2015. Panels and markers are the same as Figure~\ref{fig:2014lag}.}
    \label{fig:2015lag}
\end{figure*}

\begin{figure*}
    \centering
    \includegraphics[width=0.7\linewidth, trim = 0 0 0 50, clip]{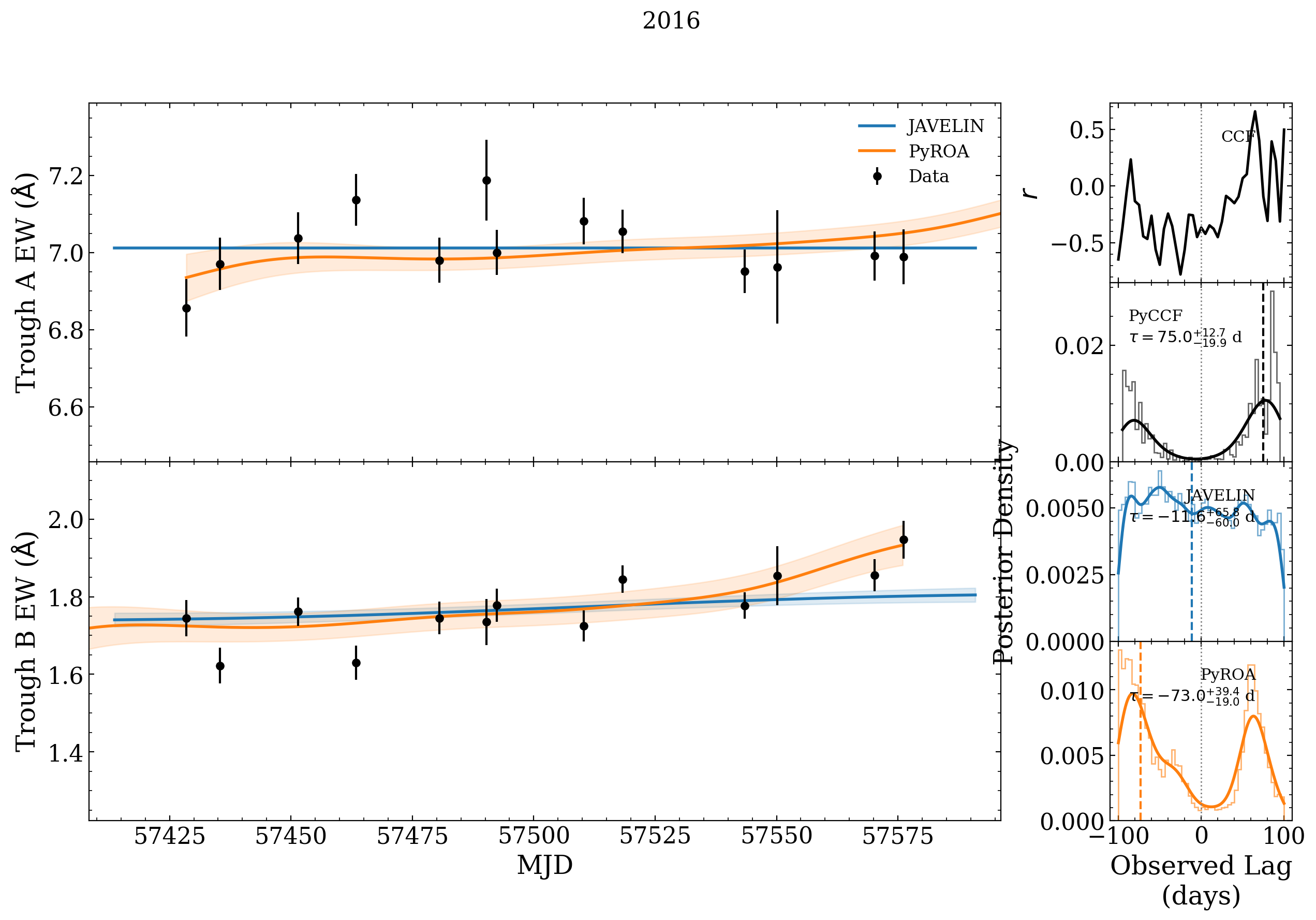}
    \caption{Lag results for Trough A and Trough B with Trough A as the driver for 2016. Panels and markers are the same as Figure~\ref{fig:2014lag}.}
    \label{fig:2016lag}
\end{figure*}

\begin{figure*}
    \centering
    \includegraphics[width=0.7\linewidth, trim = 0 0 0 50, clip]{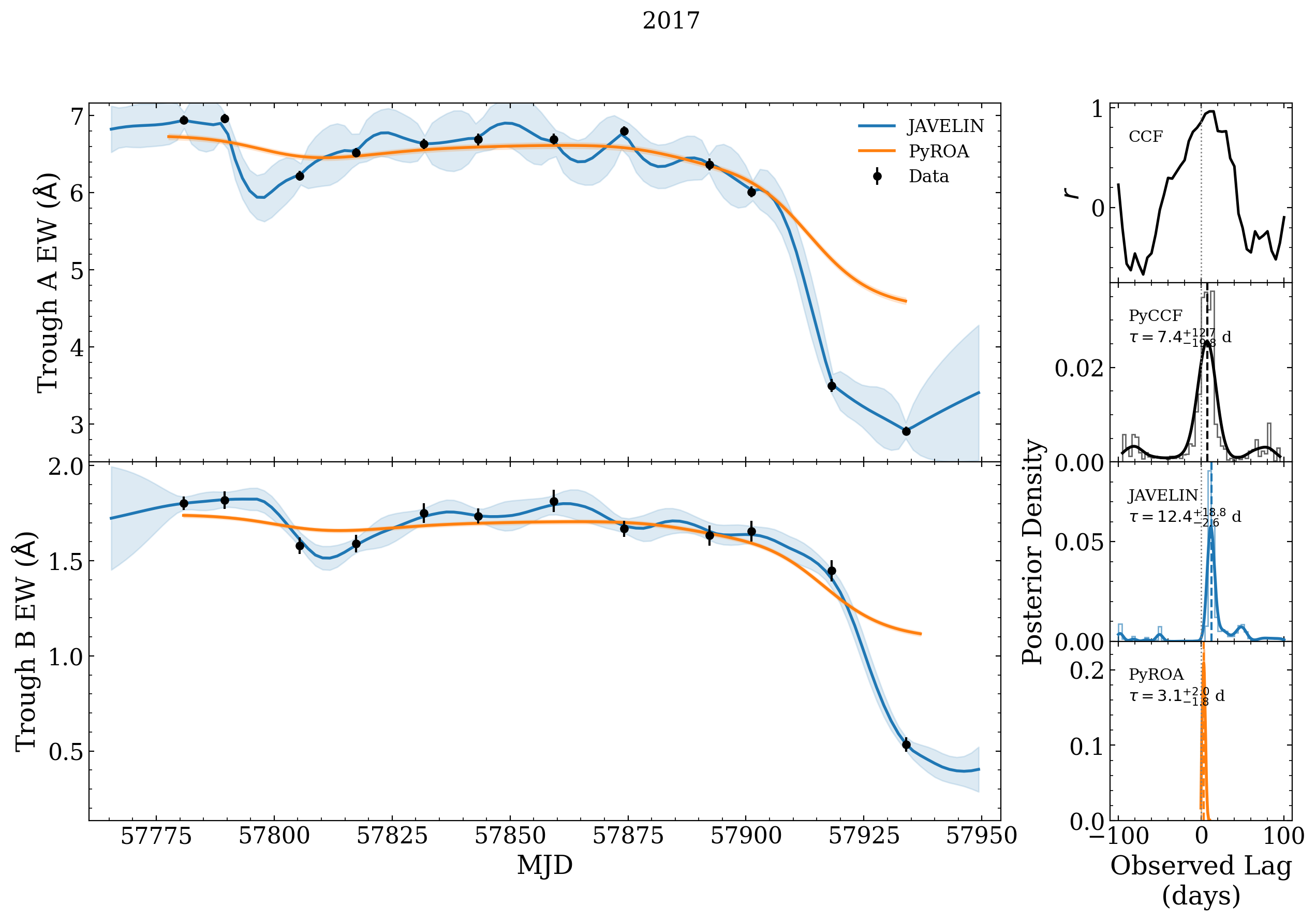}
    \caption{Lag results for Trough A and Trough B with Trough A as the driver for 2017. Panels and markers are the same as Figure~\ref{fig:2014lag}.}
    \label{fig:2017lag}
\end{figure*}

\begin{figure*}
    \centering
    \includegraphics[width=0.7\linewidth, trim = 0 0 0 50, clip]{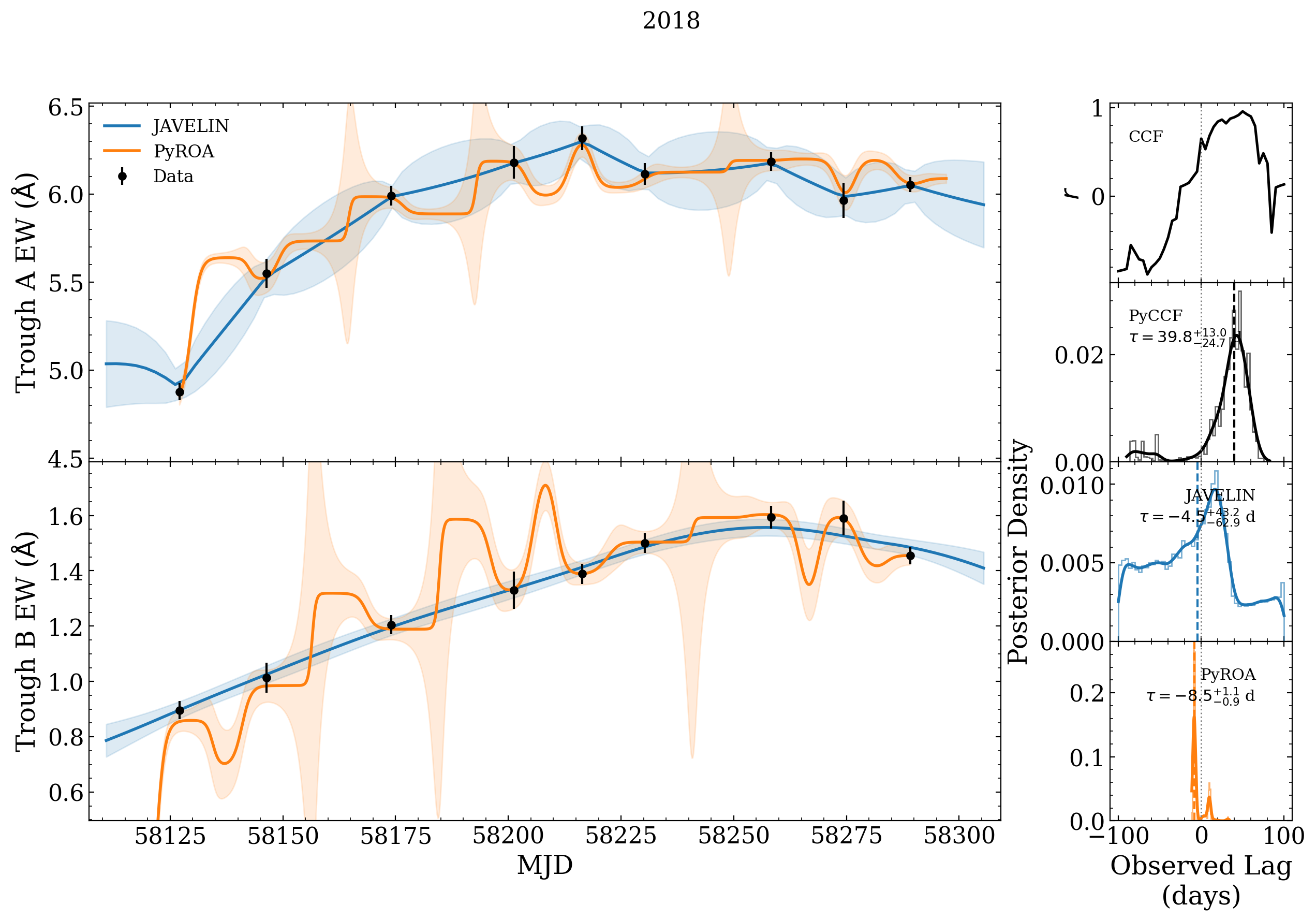}
    \caption{Lag results for Trough A and Trough B with Trough A as the driver for 2018. Panels and markers are the same as Figure~\ref{fig:2014lag}.}
    \label{fig:2018lag}
\end{figure*}

\begin{figure*}
    \centering
    \includegraphics[width=0.7\linewidth, trim = 0 0 0 50, clip]{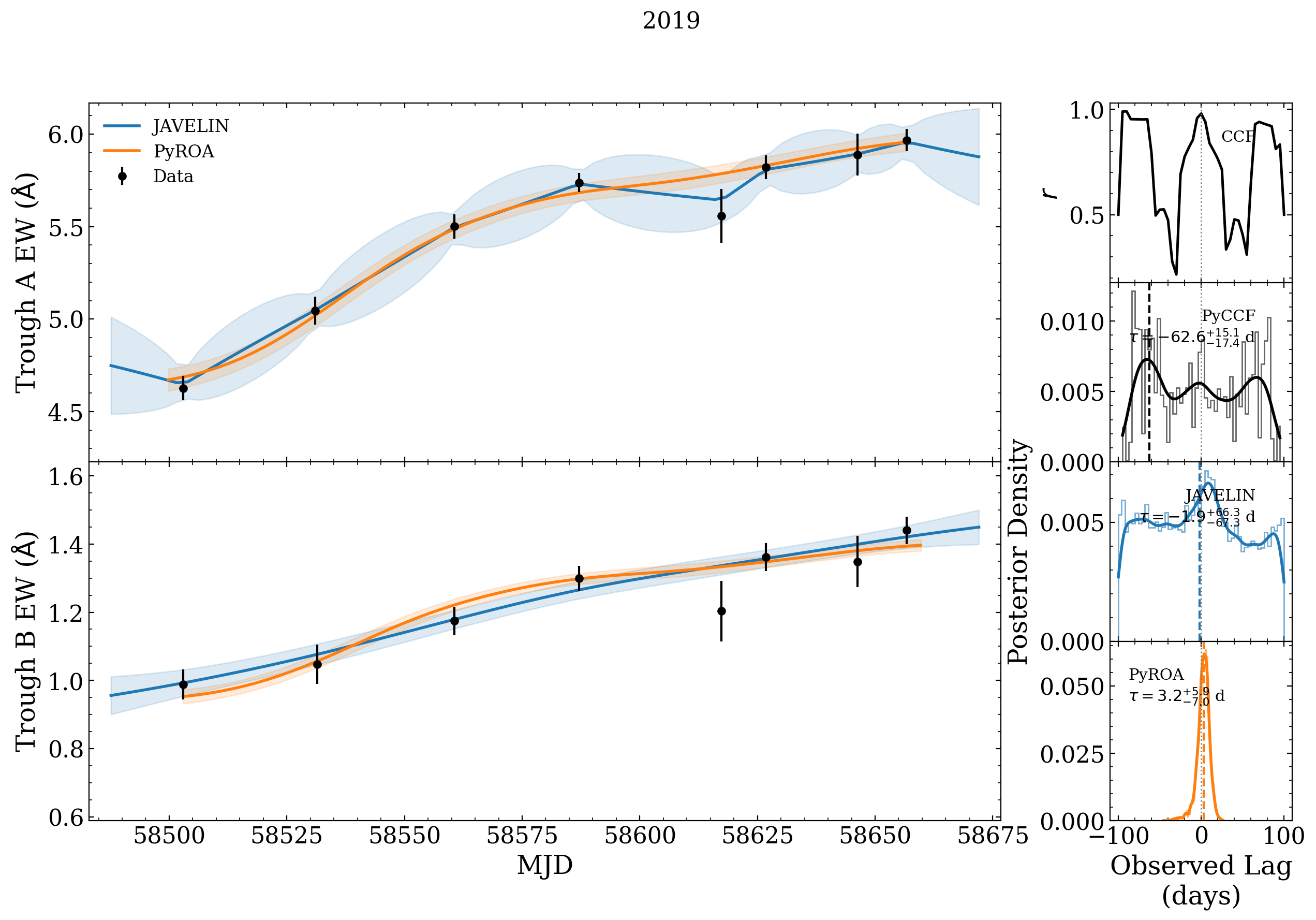}
    \caption{Lag results for Trough A and Trough B with Trough A as the driver for 2019. Panels and markers are the same as Figure~\ref{fig:2014lag}.}
    \label{fig:2019lag}
\end{figure*}

\begin{figure*}
    \centering
    \includegraphics[width=0.7\linewidth, trim = 0 0 0 50, clip]{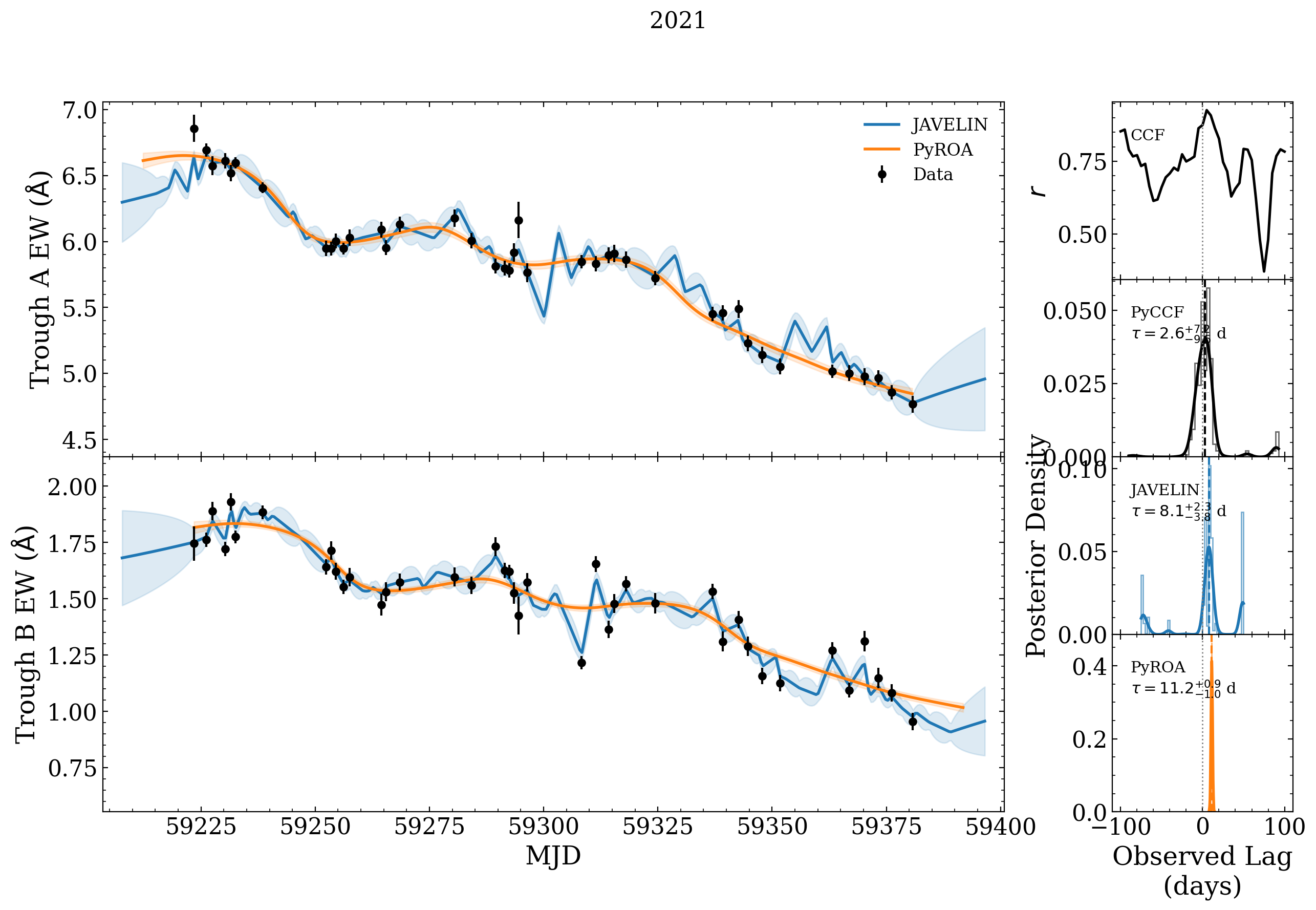}
    \caption{Lag results for Trough A and Trough B with Trough A as the driver for 2021. Panels and markers are the same as Figure~\ref{fig:2014lag}.}
    \label{fig:2021lag}
\end{figure*}

\clearpage
\startlongtable
\begin{deluxetable}{cccccccc}
\tablecaption{BAL Measurements}
\label{tab:spectral_data}
\tablehead{
 & $EW_{\rm A}$ & Mean & $V_{\rm cent,A}$ & $EW_{\rm B}$ & Mean  & $V_{\rm cent,B}$ & $C1700$\tablenotemark{a} \\
 MJD & (\AA) & Depth$_A$  & (km~s$^{-1}$) & (\AA)& Depth$_B$  & (km~s$^{-1}$) &  
}
\startdata
52764.000 & 7.159 $\pm$ 0.117 & 0.500 $\pm$ 0.008 & -20178.3 $\pm$ 15.7 & 1.916 $\pm$ 0.066 & 0.218 $\pm$ 0.008 & -3757.6 $\pm$ 19.8 & 7.34 \\
56397.000 & 7.005 $\pm$ 0.064 & 0.489 $\pm$ 0.004 & -20166.0 $\pm$ 9.1 & 1.957 $\pm$ 0.048 & 0.222 $\pm$ 0.005 & -3682.9 $\pm$ 11.1 & 10.41\\
56455.000 & 6.931 $\pm$ 0.064 & 0.484 $\pm$ 0.004 & -20162.4 $\pm$ 9.0 & 1.807 $\pm$ 0.048 & 0.205 $\pm$ 0.006 & -3660.9 $\pm$ 10.6 & 9.83\\
56660.209 & 5.959 $\pm$ 0.058 & 0.416 $\pm$ 0.004 & -20193.1 $\pm$ 8.5 & 1.343 $\pm$ 0.041 & 0.153 $\pm$ 0.005 & -3681.8 $\pm$ 14.6 & 13.35 \\
56664.513 & 5.896 $\pm$ 0.070 & 0.411 $\pm$ 0.005 & -20192.8 $\pm$ 11.7 & 1.391 $\pm$ 0.053 & 0.158 $\pm$ 0.006 & -3704.3 $\pm$ 15.3 & 13.46 \\
56669.500 & 5.914 $\pm$ 0.151 & 0.413 $\pm$ 0.011 & -20157.0 $\pm$ 18.2 & 1.687 $\pm$ 0.080 & 0.192 $\pm$ 0.009 & -3695.9 $\pm$ 23.7 & 13.37 \\
56683.480 & 6.010 $\pm$ 0.051 & 0.419 $\pm$ 0.004 & -20174.9 $\pm$ 8.0 & 1.275 $\pm$ 0.037 & 0.145 $\pm$ 0.004 & -3705.1 $\pm$ 13.5 & 14.33 \\
56686.473 & 5.833 $\pm$ 0.053 & 0.407 $\pm$ 0.004 & -20164.2 $\pm$ 7.5 & 1.245 $\pm$ 0.036 & 0.141 $\pm$ 0.004 & -3700.3 $\pm$ 14.7 & 13.71 \\
56696.778 & 5.365 $\pm$ 0.054 & 0.374 $\pm$ 0.004 & -20203.9 $\pm$ 9.7 & 1.352 $\pm$ 0.037 & 0.154 $\pm$ 0.004 & -3700.5 $\pm$ 11.1 & 13.40 \\
56715.388 & 5.164 $\pm$ 0.055 & 0.360 $\pm$ 0.004 & -20186.1 $\pm$ 10.3 & 1.059 $\pm$ 0.035 & 0.120 $\pm$ 0.004 & -3726.4 $\pm$ 18.5 & 14.32 \\
56717.334 & 5.101 $\pm$ 0.042 & 0.356 $\pm$ 0.003 & -20186.0 $\pm$ 8.3 & 1.196 $\pm$ 0.032 & 0.136 $\pm$ 0.004 & -3658.4 $\pm$ 13.9 & 13.00 \\
56720.446 & 5.132 $\pm$ 0.054 & 0.358 $\pm$ 0.004 & -20187.8 $\pm$ 8.6 & 1.058 $\pm$ 0.033 & 0.120 $\pm$ 0.004 & -3690.0 $\pm$ 16.0 & 13.80 \\
56722.387 & 5.244 $\pm$ 0.043 & 0.366 $\pm$ 0.003 & -20193.2 $\pm$ 7.6 & 1.136 $\pm$ 0.034 & 0.129 $\pm$ 0.004 & -3725.8 $\pm$ 12.7 & 13.24 \\
56726.457 & 5.365 $\pm$ 0.052 & 0.374 $\pm$ 0.004 & -20181.3 $\pm$ 8.1 & 1.241 $\pm$ 0.035 & 0.141 $\pm$ 0.004 & -3680.7 $\pm$ 13.4 & 14.18 \\
56739.410 & 5.206 $\pm$ 0.047 & 0.363 $\pm$ 0.003 & -20187.5 $\pm$ 8.6 & 1.085 $\pm$ 0.033 & 0.123 $\pm$ 0.004 & -3703.1 $\pm$ 14.0 & 13.13 \\
56745.281 & 5.175 $\pm$ 0.047 & 0.361 $\pm$ 0.003 & -20209.0 $\pm$ 9.4 & 1.072 $\pm$ 0.034 & 0.122 $\pm$ 0.004 & -3720.7 $\pm$ 16.4 & 13.14 \\
56747.416 & 5.294 $\pm$ 0.055 & 0.369 $\pm$ 0.004 & -20180.3 $\pm$ 8.3 & 1.216 $\pm$ 0.038 & 0.138 $\pm$ 0.004 & -3662.3 $\pm$ 17.0 & 12.42 \\
56749.373 & 5.177 $\pm$ 0.050 & 0.361 $\pm$ 0.003 & -20192.7 $\pm$ 9.8 & 1.079 $\pm$ 0.042 & 0.123 $\pm$ 0.005 & -3712.5 $\pm$ 18.3 & 13.31 \\
56751.340 & 5.163 $\pm$ 0.052 & 0.360 $\pm$ 0.004 & -20181.3 $\pm$ 8.4 & 1.173 $\pm$ 0.031 & 0.133 $\pm$ 0.004 & -3675.0 $\pm$ 14.6 & 13.62 \\
56755.341 & 5.266 $\pm$ 0.060 & 0.367 $\pm$ 0.004 & -20184.5 $\pm$ 9.1 & 1.172 $\pm$ 0.038 & 0.133 $\pm$ 0.004 & -3675.9 $\pm$ 18.0 & 13.18 \\
56768.227 & 5.300 $\pm$ 0.045 & 0.370 $\pm$ 0.003 & -20190.3 $\pm$ 8.4 & 1.188 $\pm$ 0.030 & 0.135 $\pm$ 0.003 & -3713.8 $\pm$ 13.4 & 14.40 \\
56772.234 & 5.313 $\pm$ 0.047 & 0.371 $\pm$ 0.003 & -20178.5 $\pm$ 7.5 & 1.024 $\pm$ 0.035 & 0.116 $\pm$ 0.004 & -3689.5 $\pm$ 15.0 & 14.10 \\
56780.234 & 5.388 $\pm$ 0.047 & 0.376 $\pm$ 0.003 & -20201.2 $\pm$ 7.2 & 1.198 $\pm$ 0.034 & 0.136 $\pm$ 0.004 & -3710.8 $\pm$ 13.8 & 14.52 \\
56782.247 & 5.427 $\pm$ 0.040 & 0.379 $\pm$ 0.003 & -20165.6 $\pm$ 8.5 & 1.152 $\pm$ 0.029 & 0.131 $\pm$ 0.003 & -3722.6 $\pm$ 13.4 & 14.27 \\
56783.248 & 5.516 $\pm$ 0.043 & 0.385 $\pm$ 0.003 & -20173.4 $\pm$ 7.5 & 1.257 $\pm$ 0.035 & 0.143 $\pm$ 0.004 & -3719.9 $\pm$ 11.7 & 14.35 \\
56795.175 & 5.475 $\pm$ 0.050 & 0.382 $\pm$ 0.003 & -20181.3 $\pm$ 9.4 & 1.247 $\pm$ 0.035 & 0.142 $\pm$ 0.004 & -3716.2 $\pm$ 15.1 & 14.23 \\
56799.212 & 5.521 $\pm$ 0.043 & 0.385 $\pm$ 0.003 & -20172.9 $\pm$ 7.2 & 1.175 $\pm$ 0.032 & 0.134 $\pm$ 0.004 & -3725.9 $\pm$ 13.7 & 13.97 \\
56804.187 & 5.617 $\pm$ 0.055 & 0.392 $\pm$ 0.004 & -20183.3 $\pm$ 9.8 & 1.262 $\pm$ 0.038 & 0.143 $\pm$ 0.004 & -3691.1 $\pm$ 14.7 & 12.06 \\
56808.259 & 5.546 $\pm$ 0.070 & 0.387 $\pm$ 0.005 & -20192.2 $\pm$ 10.5 & 1.248 $\pm$ 0.045 & 0.142 $\pm$ 0.005 & -3703.6 $\pm$ 15.7 & 12.85 \\
56813.227 & 5.570 $\pm$ 0.052 & 0.389 $\pm$ 0.004 & -20163.4 $\pm$ 8.6 & 1.133 $\pm$ 0.038 & 0.129 $\pm$ 0.004 & -3718.0 $\pm$ 15.8 & 14.84 \\
56825.186 & 5.738 $\pm$ 0.043 & 0.400 $\pm$ 0.003 & -20157.5 $\pm$ 7.5 & 1.213 $\pm$ 0.032 & 0.138 $\pm$ 0.004 & -3703.1 $\pm$ 14.0 & 14.53 \\
56829.211 & 5.838 $\pm$ 0.056 & 0.407 $\pm$ 0.004 & -20169.9 $\pm$ 8.5 & 1.295 $\pm$ 0.032 & 0.147 $\pm$ 0.004 & -3721.6 $\pm$ 14.1 & 14.55 \\
56833.210 & 5.975 $\pm$ 0.053 & 0.417 $\pm$ 0.004 & -20165.9 $\pm$ 7.2 & 1.270 $\pm$ 0.037 & 0.144 $\pm$ 0.004 & -3701.7 $\pm$ 13.3 & 14.41 \\
56837.189 & 5.934 $\pm$ 0.045 & 0.414 $\pm$ 0.003 & -20179.8 $\pm$ 7.1 & 1.326 $\pm$ 0.033 & 0.151 $\pm$ 0.004 & -3693.7 $\pm$ 12.2 & 13.36 \\
57038.487 & 7.601 $\pm$ 0.081 & 0.530 $\pm$ 0.006 & -20149.3 $\pm$ 8.8 & 1.888 $\pm$ 0.047 & 0.214 $\pm$ 0.005 & -3653.2 $\pm$ 14.0 & 13.28 \\
57050.294 & 7.514 $\pm$ 0.067 & 0.524 $\pm$ 0.005 & -20179.2 $\pm$ 7.1 & 1.925 $\pm$ 0.042 & 0.219 $\pm$ 0.005 & -3698.5 $\pm$ 10.6 & 14.19 \\
57067.372 & 7.372 $\pm$ 0.065 & 0.514 $\pm$ 0.005 & -20173.2 $\pm$ 8.5 & 2.053 $\pm$ 0.040 & 0.233 $\pm$ 0.005 & -3679.8 $\pm$ 10.8 & 13.97 \\
57082.473 & 7.087 $\pm$ 0.068 & 0.494 $\pm$ 0.005 & -20162.9 $\pm$ 8.8 & 1.942 $\pm$ 0.055 & 0.221 $\pm$ 0.006 & -3665.7 $\pm$ 13.0 & 14.36 \\
57097.480 & 7.089 $\pm$ 0.084 & 0.495 $\pm$ 0.006 & -20171.7 $\pm$ 12.2 & 1.892 $\pm$ 0.073 & 0.215 $\pm$ 0.008 & -3669.6 $\pm$ 16.3 & 13.81 \\
57106.472 & 7.082 $\pm$ 0.064 & 0.494 $\pm$ 0.004 & -20166.5 $\pm$ 8.1 & 1.948 $\pm$ 0.046 & 0.221 $\pm$ 0.005 & -3687.9 $\pm$ 11.7 & 13.58 \\
57127.217 & 7.193 $\pm$ 0.056 & 0.502 $\pm$ 0.004 & -20180.7 $\pm$ 7.4 & 1.850 $\pm$ 0.041 & 0.210 $\pm$ 0.005 & -3681.1 $\pm$ 11.0 & 13.69 \\
57135.173 & 7.206 $\pm$ 0.052 & 0.503 $\pm$ 0.004 & -20182.7 $\pm$ 7.8 & 1.819 $\pm$ 0.034 & 0.207 $\pm$ 0.004 & -3680.6 $\pm$ 9.0 & 13.60 \\
57159.165 & 7.080 $\pm$ 0.063 & 0.494 $\pm$ 0.004 & -20168.7 $\pm$ 9.4 & 1.800 $\pm$ 0.042 & 0.204 $\pm$ 0.005 & -3691.9 $\pm$ 12.6 & 13.43 \\
57166.165 & 6.804 $\pm$ 0.057 & 0.475 $\pm$ 0.004 & -20164.6 $\pm$ 8.0 & 1.854 $\pm$ 0.051 & 0.211 $\pm$ 0.006 & -3676.3 $\pm$ 11.2 & 14.37 \\
57185.171 & 6.505 $\pm$ 0.072 & 0.454 $\pm$ 0.005 & -20173.0 $\pm$ 9.5 & 1.866 $\pm$ 0.052 & 0.212 $\pm$ 0.006 & -3673.2 $\pm$ 12.6 & 14.57 \\
57195.598 & 6.228 $\pm$ 0.064 & 0.435 $\pm$ 0.004 & -20172.7 $\pm$ 9.4 & 1.808 $\pm$ 0.041 & 0.205 $\pm$ 0.005 & -3675.5 $\pm$ 10.1 & 14.22 \\
57428.450 & 6.857 $\pm$ 0.075 & 0.478 $\pm$ 0.005 & -20175.5 $\pm$ 9.9 & 1.745 $\pm$ 0.046 & 0.198 $\pm$ 0.005 & -3677.6 $\pm$ 12.5 & 14.45 \\
57435.404 & 6.971 $\pm$ 0.067 & 0.486 $\pm$ 0.005 & -20164.2 $\pm$ 8.5 & 1.622 $\pm$ 0.046 & 0.184 $\pm$ 0.005 & -3679.1 $\pm$ 13.8 & 14.84 \\
57451.456 & 7.037 $\pm$ 0.067 & 0.491 $\pm$ 0.005 & -20178.4 $\pm$ 8.9 & 1.762 $\pm$ 0.037 & 0.200 $\pm$ 0.004 & -3697.4 $\pm$ 10.8 & 15.04 \\
57463.401 & 7.137 $\pm$ 0.067 & 0.498 $\pm$ 0.005 & -20142.7 $\pm$ 8.6 & 1.630 $\pm$ 0.044 & 0.185 $\pm$ 0.005 & -3698.3 $\pm$ 13.8 & 14.19 \\
57480.602 & 6.980 $\pm$ 0.059 & 0.487 $\pm$ 0.004 & -20168.2 $\pm$ 8.8 & 1.745 $\pm$ 0.042 & 0.198 $\pm$ 0.005 & -3679.6 $\pm$ 10.6 & 14.02 \\
57490.225 & 7.188 $\pm$ 0.105 & 0.502 $\pm$ 0.007 & -20163.3 $\pm$ 12.5 & 1.735 $\pm$ 0.059 & 0.197 $\pm$ 0.007 & -3676.5 $\pm$ 16.6 & 14.43 \\
57492.342 & 7.001 $\pm$ 0.058 & 0.488 $\pm$ 0.004 & -20169.8 $\pm$ 9.7 & 1.778 $\pm$ 0.042 & 0.202 $\pm$ 0.005 & -3676.5 $\pm$ 12.9 & 14.00 \\
57510.319 & 7.082 $\pm$ 0.061 & 0.494 $\pm$ 0.004 & -20154.1 $\pm$ 6.7 & 1.725 $\pm$ 0.039 & 0.196 $\pm$ 0.004 & -3697.4 $\pm$ 11.9 & 14.67 \\
57518.313 & 7.055 $\pm$ 0.057 & 0.492 $\pm$ 0.004 & -20155.5 $\pm$ 7.3 & 1.845 $\pm$ 0.035 & 0.210 $\pm$ 0.004 & -3707.1 $\pm$ 10.8 & 14.30 \\
57543.453 & 6.952 $\pm$ 0.056 & 0.485 $\pm$ 0.004 & -20160.4 $\pm$ 6.2 & 1.777 $\pm$ 0.034 & 0.202 $\pm$ 0.004 & -3690.6 $\pm$ 10.4 & 14.44 \\
57550.163 & 6.963 $\pm$ 0.147 & 0.486 $\pm$ 0.010 & -20163.3 $\pm$ 19.0 & 1.854 $\pm$ 0.076 & 0.211 $\pm$ 0.009 & -3645.9 $\pm$ 20.3 & 14.30 \\
57570.201 & 6.991 $\pm$ 0.063 & 0.488 $\pm$ 0.004 & -20148.7 $\pm$ 9.2 & 1.856 $\pm$ 0.041 & 0.211 $\pm$ 0.005 & -3688.9 $\pm$ 10.6 & 14.29 \\
57576.163 & 6.990 $\pm$ 0.071 & 0.488 $\pm$ 0.005 & -20156.4 $\pm$ 9.3 & 1.947 $\pm$ 0.049 & 0.221 $\pm$ 0.006 & -3683.8 $\pm$ 13.7 & 13.57 \\
57780.773 & 6.941 $\pm$ 0.057 & 0.484 $\pm$ 0.004 & -20145.0 $\pm$ 6.8 & 1.803 $\pm$ 0.035 & 0.205 $\pm$ 0.004 & -3697.3 $\pm$ 9.9 & 13.85 \\
57789.442 & 6.959 $\pm$ 0.061 & 0.486 $\pm$ 0.004 & -20147.7 $\pm$ 9.1 & 1.819 $\pm$ 0.047 & 0.207 $\pm$ 0.005 & -3673.6 $\pm$ 11.8 & 13.94 \\
57805.348 & 6.218 $\pm$ 0.065 & 0.434 $\pm$ 0.005 & -20140.1 $\pm$ 9.6 & 1.580 $\pm$ 0.043 & 0.179 $\pm$ 0.005 & -3675.7 $\pm$ 14.9 & 13.81 \\
57817.317 & 6.516 $\pm$ 0.059 & 0.455 $\pm$ 0.004 & -20161.5 $\pm$ 8.7 & 1.590 $\pm$ 0.046 & 0.181 $\pm$ 0.005 & -3677.8 $\pm$ 14.6 & 13.91 \\
57831.769 & 6.632 $\pm$ 0.065 & 0.463 $\pm$ 0.005 & -20155.0 $\pm$ 9.4 & 1.751 $\pm$ 0.051 & 0.199 $\pm$ 0.006 & -3668.8 $\pm$ 14.5 & 13.60 \\
57843.264 & 6.691 $\pm$ 0.077 & 0.467 $\pm$ 0.005 & -20159.5 $\pm$ 7.7 & 1.736 $\pm$ 0.040 & 0.197 $\pm$ 0.005 & -3673.8 $\pm$ 11.5 & 13.80 \\
57859.220 & 6.695 $\pm$ 0.073 & 0.467 $\pm$ 0.005 & -20176.4 $\pm$ 10.9 & 1.814 $\pm$ 0.059 & 0.206 $\pm$ 0.007 & -3692.3 $\pm$ 14.6 & 13.59 \\
57874.208 & 6.798 $\pm$ 0.062 & 0.474 $\pm$ 0.004 & -20151.0 $\pm$ 8.6 & 1.668 $\pm$ 0.042 & 0.189 $\pm$ 0.005 & -3695.7 $\pm$ 14.0 & 13.74 \\
57892.294 & 6.366 $\pm$ 0.076 & 0.444 $\pm$ 0.005 & -20158.1 $\pm$ 11.5 & 1.633 $\pm$ 0.052 & 0.185 $\pm$ 0.006 & -3690.3 $\pm$ 15.9 & 12.99 \\
57901.209 & 6.009 $\pm$ 0.071 & 0.419 $\pm$ 0.005 & -20154.6 $\pm$ 11.4 & 1.656 $\pm$ 0.055 & 0.188 $\pm$ 0.006 & -3664.6 $\pm$ 16.9 & 13.69 \\
57918.164 & 3.501 $\pm$ 0.084 & 0.244 $\pm$ 0.006 & -20205.5 $\pm$ 27.4 & 1.447 $\pm$ 0.055 & 0.164 $\pm$ 0.006 & -3627.0 $\pm$ 20.2 & 13.55 \\
57934.000 & 2.910 $\pm$ 0.057 & 0.203 $\pm$ 0.004 & -20161.0 $\pm$ 21.4 & 0.534 $\pm$ 0.039 & 0.061 $\pm$ 0.004 & -3692.4 $\pm$ 37.7 & 14.74 \\
58127.033 & 4.877 $\pm$ 0.050 & 0.340 $\pm$ 0.003 & -20168.3 $\pm$ 9.0 & 0.896 $\pm$ 0.033 & 0.102 $\pm$ 0.004 & -3693.8 $\pm$ 18.8 & 13.61 \\
58146.422 & 5.549 $\pm$ 0.083 & 0.387 $\pm$ 0.006 & -20151.5 $\pm$ 12.3 & 1.014 $\pm$ 0.054 & 0.115 $\pm$ 0.006 & -3696.4 $\pm$ 21.4 & 14.40 \\
58174.079 & 5.992 $\pm$ 0.055 & 0.418 $\pm$ 0.004 & -20159.0 $\pm$ 7.2 & 1.205 $\pm$ 0.035 & 0.137 $\pm$ 0.004 & -3702.8 $\pm$ 14.7 & 14.11 \\
58201.269 & 6.181 $\pm$ 0.093 & 0.431 $\pm$ 0.007 & -20144.2 $\pm$ 13.4 & 1.330 $\pm$ 0.067 & 0.151 $\pm$ 0.008 & -3710.9 $\pm$ 25.1 & 14.80 \\
58216.413 & 6.317 $\pm$ 0.068 & 0.441 $\pm$ 0.005 & -20152.0 $\pm$ 8.3 & 1.389 $\pm$ 0.037 & 0.158 $\pm$ 0.004 & -3727.5 $\pm$ 13.1 & 14.81 \\
58230.276 & 6.115 $\pm$ 0.062 & 0.427 $\pm$ 0.004 & -20152.4 $\pm$ 8.4 & 1.500 $\pm$ 0.036 & 0.170 $\pm$ 0.004 & -3714.8 $\pm$ 13.8 & 15.04 \\
58258.285 & 6.185 $\pm$ 0.053 & 0.432 $\pm$ 0.004 & -20145.3 $\pm$ 9.4 & 1.593 $\pm$ 0.041 & 0.181 $\pm$ 0.005 & -3678.2 $\pm$ 13.9 & 14.47 \\
58274.302 & 5.965 $\pm$ 0.101 & 0.416 $\pm$ 0.007 & -20133.5 $\pm$ 13.1 & 1.591 $\pm$ 0.063 & 0.181 $\pm$ 0.007 & -3683.5 $\pm$ 17.2 & 14.87 \\
58289.206 & 6.054 $\pm$ 0.043 & 0.422 $\pm$ 0.003 & -20152.6 $\pm$ 8.0 & 1.455 $\pm$ 0.031 & 0.165 $\pm$ 0.004 & -3703.6 $\pm$ 12.0 & 12.99 \\
58503.000 & 4.627 $\pm$ 0.065 & 0.323 $\pm$ 0.005 & -20167.1 $\pm$ 13.3 & 0.989 $\pm$ 0.044 & 0.112 $\pm$ 0.005 & -3727.1 $\pm$ 23.7 & 14.68 \\
58531.000 & 5.046 $\pm$ 0.077 & 0.352 $\pm$ 0.005 & -20183.2 $\pm$ 15.1 & 1.047 $\pm$ 0.058 & 0.119 $\pm$ 0.007 & -3654.8 $\pm$ 23.5 & 14.69 \\
58560.572 & 5.501 $\pm$ 0.066 & 0.384 $\pm$ 0.005 & -20162.2 $\pm$ 9.2 & 1.175 $\pm$ 0.041 & 0.133 $\pm$ 0.005 & -3694.4 $\pm$ 18.2 & 15.10 \\
58587.123 & 5.738 $\pm$ 0.052 & 0.400 $\pm$ 0.004 & -20141.3 $\pm$ 9.6 & 1.299 $\pm$ 0.037 & 0.148 $\pm$ 0.004 & -3713.1 $\pm$ 16.6 & 13.77 \\
58617.376 & 5.559 $\pm$ 0.145 & 0.388 $\pm$ 0.010 & -20155.8 $\pm$ 26.9 & 1.203 $\pm$ 0.089 & 0.137 $\pm$ 0.010 & -3713.8 $\pm$ 34.7 & 14.19 \\
58626.784 & 5.821 $\pm$ 0.065 & 0.406 $\pm$ 0.005 & -20133.8 $\pm$ 9.6 & 1.362 $\pm$ 0.041 & 0.155 $\pm$ 0.005 & -3698.0 $\pm$ 12.8 & 13.76 \\
58646.231 & 5.890 $\pm$ 0.112 & 0.411 $\pm$ 0.008 & -20125.5 $\pm$ 18.7 & 1.348 $\pm$ 0.075 & 0.153 $\pm$ 0.009 & -3652.1 $\pm$ 32.0 & 13.82 \\
58656.686 & 5.967 $\pm$ 0.060 & 0.416 $\pm$ 0.004 & -20148.0 $\pm$ 8.7 & 1.440 $\pm$ 0.040 & 0.164 $\pm$ 0.005 & -3701.5 $\pm$ 12.5 & 12.17 \\
58867.438 & 6.497 $\pm$ 0.061 & 0.453 $\pm$ 0.004 & -20152.6 $\pm$ 9.1 & 1.711 $\pm$ 0.048 & 0.194 $\pm$ 0.005 & -3671.2 $\pm$ 11.2 & 14.57 \\
58895.415 & 6.789 $\pm$ 0.055 & 0.474 $\pm$ 0.004 & -20148.5 $\pm$ 7.3 & 1.703 $\pm$ 0.040 & 0.193 $\pm$ 0.004 & -3682.3 $\pm$ 9.6 & 12.53 \\
59010.299 & 6.720 $\pm$ 0.140 & 0.469 $\pm$ 0.010 & -20155.6 $\pm$ 16.9 & 1.533 $\pm$ 0.090 & 0.174 $\pm$ 0.010 & -3688.7 $\pm$ 25.7 & 14.10 \\
59029.196 & 6.743 $\pm$ 0.083 & 0.470 $\pm$ 0.006 & -20154.2 $\pm$ 11.4 & 1.577 $\pm$ 0.048 & 0.179 $\pm$ 0.005 & -3671.6 $\pm$ 15.7 & 13.18 \\
59223.480 & 6.858 $\pm$ 0.103 & 0.478 $\pm$ 0.007 & -20147.4 $\pm$ 16.2 & 1.745 $\pm$ 0.077 & 0.198 $\pm$ 0.009 & -3634.2 $\pm$ 20.9 & 14.08 \\
59226.180 & 6.694 $\pm$ 0.050 & 0.467 $\pm$ 0.003 & -20147.8 $\pm$ 6.4 & 1.762 $\pm$ 0.032 & 0.200 $\pm$ 0.004 & -3687.9 $\pm$ 9.2 & 13.70 \\
59227.496 & 6.575 $\pm$ 0.071 & 0.459 $\pm$ 0.005 & -20126.7 $\pm$ 10.2 & 1.890 $\pm$ 0.041 & 0.215 $\pm$ 0.005 & -3664.0 $\pm$ 11.4 & 13.68 \\
59230.262 & 6.612 $\pm$ 0.058 & 0.461 $\pm$ 0.004 & -20135.5 $\pm$ 7.6 & 1.721 $\pm$ 0.031 & 0.195 $\pm$ 0.004 & -3661.0 $\pm$ 8.2 & 14.05 \\
59231.504 & 6.518 $\pm$ 0.062 & 0.455 $\pm$ 0.004 & -20137.7 $\pm$ 8.0 & 1.930 $\pm$ 0.038 & 0.219 $\pm$ 0.004 & -3659.4 $\pm$ 9.1 & 13.78 \\
59232.492 & 6.598 $\pm$ 0.043 & 0.460 $\pm$ 0.003 & -20134.5 $\pm$ 5.2 & 1.775 $\pm$ 0.028 & 0.202 $\pm$ 0.003 & -3662.0 $\pm$ 8.6 & 14.41 \\
59238.492 & 6.407 $\pm$ 0.040 & 0.447 $\pm$ 0.003 & -20145.2 $\pm$ 6.2 & 1.884 $\pm$ 0.030 & 0.214 $\pm$ 0.003 & -3662.6 $\pm$ 8.5 & 12.35 \\
59252.320 & 5.948 $\pm$ 0.057 & 0.415 $\pm$ 0.004 & -20132.9 $\pm$ 8.4 & 1.642 $\pm$ 0.034 & 0.186 $\pm$ 0.004 & -3661.8 $\pm$ 9.9 & 13.42 \\
59253.488 & 5.947 $\pm$ 0.052 & 0.415 $\pm$ 0.004 & -20142.4 $\pm$ 7.5 & 1.714 $\pm$ 0.040 & 0.195 $\pm$ 0.005 & -3653.7 $\pm$ 10.7 & 13.50 \\
59254.492 & 6.003 $\pm$ 0.056 & 0.419 $\pm$ 0.004 & -20121.9 $\pm$ 8.7 & 1.622 $\pm$ 0.038 & 0.184 $\pm$ 0.004 & -3653.0 $\pm$ 10.5 & 13.53 \\
59256.141 & 5.947 $\pm$ 0.044 & 0.415 $\pm$ 0.003 & -20136.3 $\pm$ 6.4 & 1.553 $\pm$ 0.031 & 0.176 $\pm$ 0.004 & -3657.9 $\pm$ 10.8 & 13.16 \\
59257.500 & 6.031 $\pm$ 0.062 & 0.421 $\pm$ 0.004 & -20137.6 $\pm$ 11.1 & 1.596 $\pm$ 0.041 & 0.181 $\pm$ 0.005 & -3646.0 $\pm$ 12.6 & 12.69 \\
59264.480 & 6.090 $\pm$ 0.061 & 0.425 $\pm$ 0.004 & -20126.8 $\pm$ 9.1 & 1.474 $\pm$ 0.050 & 0.167 $\pm$ 0.006 & -3609.4 $\pm$ 15.8 & 13.17 \\
59265.496 & 5.952 $\pm$ 0.056 & 0.415 $\pm$ 0.004 & -20135.5 $\pm$ 7.8 & 1.531 $\pm$ 0.043 & 0.174 $\pm$ 0.005 & -3671.9 $\pm$ 13.7 & 13.19 \\
59268.492 & 6.131 $\pm$ 0.058 & 0.428 $\pm$ 0.004 & -20122.7 $\pm$ 8.6 & 1.574 $\pm$ 0.037 & 0.179 $\pm$ 0.004 & -3646.3 $\pm$ 11.9 & 12.65 \\
59280.480 & 6.178 $\pm$ 0.067 & 0.431 $\pm$ 0.005 & -20114.0 $\pm$ 9.3 & 1.597 $\pm$ 0.042 & 0.181 $\pm$ 0.005 & -3668.2 $\pm$ 13.2 & 12.70 \\
59284.148 & 6.004 $\pm$ 0.058 & 0.419 $\pm$ 0.004 & -20132.2 $\pm$ 8.2 & 1.560 $\pm$ 0.039 & 0.177 $\pm$ 0.004 & -3667.9 $\pm$ 12.3 & 12.79 \\
59289.469 & 5.813 $\pm$ 0.057 & 0.406 $\pm$ 0.004 & -20140.5 $\pm$ 8.8 & 1.732 $\pm$ 0.043 & 0.197 $\pm$ 0.005 & -3661.0 $\pm$ 11.7 & 12.16 \\
59291.465 & 5.798 $\pm$ 0.056 & 0.404 $\pm$ 0.004 & -20104.4 $\pm$ 8.6 & 1.625 $\pm$ 0.035 & 0.185 $\pm$ 0.004 & -3690.0 $\pm$ 12.6 & 12.21 \\
59292.469 & 5.781 $\pm$ 0.053 & 0.403 $\pm$ 0.004 & -20134.4 $\pm$ 8.6 & 1.621 $\pm$ 0.031 & 0.184 $\pm$ 0.003 & -3660.7 $\pm$ 11.3 & 12.05 \\
59293.473 & 5.917 $\pm$ 0.069 & 0.413 $\pm$ 0.005 & -20137.3 $\pm$ 10.9 & 1.526 $\pm$ 0.048 & 0.173 $\pm$ 0.005 & -3672.1 $\pm$ 14.0 & 12.63 \\
59294.457 & 6.163 $\pm$ 0.137 & 0.430 $\pm$ 0.010 & -20120.5 $\pm$ 17.4 & 1.425 $\pm$ 0.085 & 0.162 $\pm$ 0.010 & -3635.6 $\pm$ 27.9 & 12.24 \\
59296.457 & 5.764 $\pm$ 0.071 & 0.402 $\pm$ 0.005 & -20140.5 $\pm$ 10.5 & 1.574 $\pm$ 0.041 & 0.179 $\pm$ 0.005 & -3682.5 $\pm$ 13.9 & 15.22 \\
59308.324 & 5.847 $\pm$ 0.052 & 0.408 $\pm$ 0.004 & -20135.2 $\pm$ 8.8 & 1.216 $\pm$ 0.030 & 0.138 $\pm$ 0.003 & -3673.7 $\pm$ 14.1 & 14.70 \\
59311.434 & 5.830 $\pm$ 0.057 & 0.407 $\pm$ 0.004 & -20113.8 $\pm$ 9.1 & 1.654 $\pm$ 0.035 & 0.188 $\pm$ 0.004 & -3662.3 $\pm$ 12.0 & 14.85 \\
59314.207 & 5.896 $\pm$ 0.060 & 0.411 $\pm$ 0.004 & -20138.0 $\pm$ 8.8 & 1.364 $\pm$ 0.038 & 0.155 $\pm$ 0.004 & -3657.2 $\pm$ 13.1 & 14.63 \\
59315.438 & 5.909 $\pm$ 0.065 & 0.412 $\pm$ 0.005 & -20126.7 $\pm$ 7.6 & 1.478 $\pm$ 0.042 & 0.168 $\pm$ 0.005 & -3669.3 $\pm$ 15.1 & 14.59 \\
59318.062 & 5.863 $\pm$ 0.061 & 0.409 $\pm$ 0.004 & -20111.0 $\pm$ 7.8 & 1.567 $\pm$ 0.035 & 0.178 $\pm$ 0.004 & -3666.1 $\pm$ 11.2 & 14.82 \\
59324.387 & 5.723 $\pm$ 0.053 & 0.399 $\pm$ 0.004 & -20141.3 $\pm$ 10.6 & 1.480 $\pm$ 0.045 & 0.168 $\pm$ 0.005 & -3674.7 $\pm$ 13.7 & 13.79 \\
59337.012 & 5.449 $\pm$ 0.055 & 0.380 $\pm$ 0.004 & -20132.5 $\pm$ 10.0 & 1.532 $\pm$ 0.034 & 0.174 $\pm$ 0.004 & -3687.5 $\pm$ 10.3 & 13.69 \\
59339.152 & 5.459 $\pm$ 0.058 & 0.381 $\pm$ 0.004 & -20135.5 $\pm$ 12.2 & 1.310 $\pm$ 0.044 & 0.149 $\pm$ 0.005 & -3687.7 $\pm$ 15.5 & 13.65 \\
59342.668 & 5.487 $\pm$ 0.069 & 0.383 $\pm$ 0.005 & -20139.7 $\pm$ 10.4 & 1.408 $\pm$ 0.038 & 0.160 $\pm$ 0.004 & -3634.9 $\pm$ 14.0 & 13.75 \\
59344.711 & 5.228 $\pm$ 0.059 & 0.365 $\pm$ 0.004 & -20149.9 $\pm$ 11.7 & 1.289 $\pm$ 0.043 & 0.146 $\pm$ 0.005 & -3672.4 $\pm$ 14.4 & 13.93 \\
59347.816 & 5.140 $\pm$ 0.063 & 0.359 $\pm$ 0.004 & -20146.2 $\pm$ 10.7 & 1.157 $\pm$ 0.037 & 0.131 $\pm$ 0.004 & -3657.8 $\pm$ 17.4 & 13.92 \\
59351.762 & 5.052 $\pm$ 0.059 & 0.352 $\pm$ 0.004 & -20140.3 $\pm$ 11.0 & 1.125 $\pm$ 0.036 & 0.128 $\pm$ 0.004 & -3664.6 $\pm$ 14.5 & 13.17 \\
59363.160 & 5.016 $\pm$ 0.051 & 0.350 $\pm$ 0.004 & -20114.2 $\pm$ 9.7 & 1.271 $\pm$ 0.037 & 0.144 $\pm$ 0.004 & -3651.4 $\pm$ 12.9 & 13.71 \\
59366.902 & 5.000 $\pm$ 0.061 & 0.349 $\pm$ 0.004 & -20133.9 $\pm$ 10.4 & 1.095 $\pm$ 0.033 & 0.124 $\pm$ 0.004 & -3656.8 $\pm$ 18.9 & 13.29 \\
59370.199 & 4.975 $\pm$ 0.064 & 0.347 $\pm$ 0.004 & -20126.3 $\pm$ 10.6 & 1.312 $\pm$ 0.045 & 0.149 $\pm$ 0.005 & -3672.7 $\pm$ 14.3 & 14.03 \\
59373.281 & 4.966 $\pm$ 0.056 & 0.346 $\pm$ 0.004 & -20113.1 $\pm$ 11.5 & 1.149 $\pm$ 0.046 & 0.130 $\pm$ 0.005 & -3664.6 $\pm$ 18.8 & 13.44 \\
59376.184 & 4.856 $\pm$ 0.054 & 0.339 $\pm$ 0.004 & -20132.5 $\pm$ 9.6 & 1.082 $\pm$ 0.039 & 0.123 $\pm$ 0.004 & -3677.7 $\pm$ 17.6 & 13.83 \\
59380.816 & 4.765 $\pm$ 0.063 & 0.332 $\pm$ 0.004 & -20121.8 $\pm$ 10.9 & 0.954 $\pm$ 0.039 & 0.108 $\pm$ 0.004 & -3658.7 $\pm$ 16.7 & 5.52 \\
59648.453 & 1.934 $\pm$ 0.133 & 0.135 $\pm$ 0.009 & -20219.1 $\pm$ 54.6 & 0.118 $\pm$ 0.073 & 0.013 $\pm$ 0.008 & -1677.9 $\pm$ 14701.1 & 14.07 \\
59674.322 & 2.093 $\pm$ 0.057 & 0.146 $\pm$ 0.004 & -20183.9 $\pm$ 26.2 & 0.222 $\pm$ 0.037 & 0.025 $\pm$ 0.004 & -3666.3 $\pm$ 80.8 & 10.31 \\
59722.000 & 1.876 $\pm$ 0.062 & 0.131 $\pm$ 0.004 & -20115.5 $\pm$ 29.2 & 0.107 $\pm$ 0.043 & 0.012 $\pm$ 0.005 & -2820.1 $\pm$ 980.3 & 16.99 \\
59724.198 & 1.354 $\pm$ 0.063 & 0.095 $\pm$ 0.004 & -20239.9 $\pm$ 37.9 & 0.172 $\pm$ 0.033 & 0.019 $\pm$ 0.004 & -3484.7 $\pm$ 121.3 & 14.91 \\
59729.886 & 1.738 $\pm$ 0.055 & 0.121 $\pm$ 0.004 & -20168.7 $\pm$ 27.7 & 0.137 $\pm$ 0.034 & 0.016 $\pm$ 0.004 & -3709.4 $\pm$ 155.3 & 12.56 \\
59733.264 & 1.762 $\pm$ 0.075 & 0.123 $\pm$ 0.005 & -20149.7 $\pm$ 43.3 & 0.067 $\pm$ 0.060 & 0.008 $\pm$ 0.007 & -3958.1 $\pm$ 56004.5 & 13.90 \\
59938.883 & 1.634 $\pm$ 0.087 & 0.114 $\pm$ 0.006 & -20120.1 $\pm$ 48.3 & 0.100 $\pm$ 0.058 & 0.011 $\pm$ 0.007 & -3846.5 $\pm$ 1558.6 & 8.37 \\
59973.731 & 0.875 $\pm$ 0.096 & 0.061 $\pm$ 0.007 & -20163.4 $\pm$ 83.5 & -0.080 $\pm$ 0.061 & -0.009 $\pm$ 0.007 & -3967.9 $\pm$ 2413.6 & 9.33 \\
59999.841 & 1.454 $\pm$ 0.098 & 0.101 $\pm$ 0.007 & -20226.6 $\pm$ 53.9 & 0.279 $\pm$ 0.056 & 0.032 $\pm$ 0.006 & -3614.8 $\pm$ 128.9 & 10.33 \\
60003.448 & 2.175 $\pm$ 0.081 & 0.152 $\pm$ 0.006 & -20119.0 $\pm$ 33.5 & -0.024 $\pm$ 0.058 & -0.003 $\pm$ 0.007 & -7681.7 $\pm$ 12598.7 & 5.23 \\
60030.359 & 1.511 $\pm$ 0.131 & 0.105 $\pm$ 0.009 & -19862.5 $\pm$ 78.6 & 0.252 $\pm$ 0.072 & 0.029 $\pm$ 0.008 & -3623.6 $\pm$ 154.5 & 15.40 \\
60046.260 & 1.484 $\pm$ 0.063 & 0.104 $\pm$ 0.004 & -20208.0 $\pm$ 36.3 & 0.199 $\pm$ 0.037 & 0.023 $\pm$ 0.004 & -3590.1 $\pm$ 114.7 & 7.65 \\
60048.721 & 1.160 $\pm$ 0.093 & 0.081 $\pm$ 0.006 & -20165.0 $\pm$ 68.4 & 0.182 $\pm$ 0.058 & 0.021 $\pm$ 0.007 & -3343.5 $\pm$ 395.6 & 12.12 \\
60051.329 & 1.444 $\pm$ 0.060 & 0.101 $\pm$ 0.004 & -20131.6 $\pm$ 35.9 & 0.093 $\pm$ 0.035 & 0.011 $\pm$ 0.004 & -3269.8 $\pm$ 540.8 & 11.04 \\
60053.293 & 1.476 $\pm$ 0.073 & 0.103 $\pm$ 0.005 & -20169.9 $\pm$ 41.6 & 0.057 $\pm$ 0.051 & 0.007 $\pm$ 0.006 & -3468.3 $\pm$ 163661.2 & 11.29 \\
60056.015 & 1.335 $\pm$ 0.069 & 0.093 $\pm$ 0.005 & -20225.6 $\pm$ 45.7 & -0.112 $\pm$ 0.043 & -0.013 $\pm$ 0.005 & -3204.7 $\pm$ 420.9 & 16.88 \\
60072.766 & 1.561 $\pm$ 0.056 & 0.109 $\pm$ 0.004 & -20222.1 $\pm$ 31.0 & 0.025 $\pm$ 0.031 & 0.003 $\pm$ 0.004 & -2937.8 $\pm$ 5146.3 & 13.17 \\
60076.085 & 1.639 $\pm$ 0.047 & 0.114 $\pm$ 0.003 & -20171.4 $\pm$ 34.7 & 0.242 $\pm$ 0.034 & 0.027 $\pm$ 0.004 & -3723.4 $\pm$ 93.3 & 9.69 \\
60083.333 & 1.612 $\pm$ 0.094 & 0.113 $\pm$ 0.007 & -20180.7 $\pm$ 49.3 & 0.118 $\pm$ 0.067 & 0.013 $\pm$ 0.008 & -3240.5 $\pm$ 9665.3 & 10.77 \\
60086.269 & 1.626 $\pm$ 0.078 & 0.113 $\pm$ 0.005 & -20170.4 $\pm$ 37.1 & 0.140 $\pm$ 0.039 & 0.016 $\pm$ 0.004 & -3496.9 $\pm$ 174.4 & 12.87 \\
60088.186 & 1.354 $\pm$ 0.070 & 0.095 $\pm$ 0.005 & -20233.0 $\pm$ 38.8 & -0.071 $\pm$ 0.046 & -0.008 $\pm$ 0.005 & -3877.2 $\pm$ 2257.3 & 7.02 \\
60104.189 & 1.189 $\pm$ 0.090 & 0.083 $\pm$ 0.006 & -20257.0 $\pm$ 64.4 & 0.270 $\pm$ 0.052 & 0.031 $\pm$ 0.006 & -3678.8 $\pm$ 101.5 & 5.88 \\
60106.199 & 1.517 $\pm$ 0.104 & 0.106 $\pm$ 0.007 & -20155.4 $\pm$ 66.6 & 0.415 $\pm$ 0.059 & 0.047 $\pm$ 0.007 & -3678.4 $\pm$ 66.0 & 8.82 \\
60108.210 & 1.371 $\pm$ 0.075 & 0.096 $\pm$ 0.005 & -20224.3 $\pm$ 50.4 & 0.050 $\pm$ 0.052 & 0.006 $\pm$ 0.006 & -3180.0 $\pm$ 13557.2 & 10.63 \\
60110.227 & 1.578 $\pm$ 0.075 & 0.110 $\pm$ 0.005 & -20227.5 $\pm$ 44.0 & 0.298 $\pm$ 0.043 & 0.034 $\pm$ 0.005 & -3739.1 $\pm$ 75.5 & 8.67 \\
60118.205 & 1.268 $\pm$ 0.080 & 0.088 $\pm$ 0.006 & -20098.5 $\pm$ 59.9 & 1.161 $\pm$ 0.052 & 0.132 $\pm$ 0.006 & -3690.8 $\pm$ 18.9 & 14.58 \\
60324.000 & 2.712 $\pm$ 0.075 & 0.189 $\pm$ 0.005 & -20152.5 $\pm$ 24.5 & 0.393 $\pm$ 0.048 & 0.045 $\pm$ 0.005 & -3567.5 $\pm$ 60.4 & 10.38 \\
60326.464 & 2.599 $\pm$ 0.106 & 0.181 $\pm$ 0.007 & -20168.8 $\pm$ 34.8 & 0.537 $\pm$ 0.060 & 0.061 $\pm$ 0.007 & -3617.6 $\pm$ 69.3 & 12.98 \\
60328.475 & 2.955 $\pm$ 0.097 & 0.206 $\pm$ 0.007 & -20192.7 $\pm$ 29.6 & 0.423 $\pm$ 0.057 & 0.048 $\pm$ 0.006 & -3578.0 $\pm$ 75.4 & 13.14 \\
60353.405 & 3.103 $\pm$ 0.089 & 0.217 $\pm$ 0.006 & -20182.5 $\pm$ 23.8 & 0.337 $\pm$ 0.055 & 0.038 $\pm$ 0.006 & -3421.6 $\pm$ 108.7 & 15.39 \\
60356.467 & 3.312 $\pm$ 0.072 & 0.231 $\pm$ 0.005 & -20137.2 $\pm$ 18.0 & 0.539 $\pm$ 0.040 & 0.061 $\pm$ 0.005 & -3645.5 $\pm$ 44.7 & 13.66 \\
60358.477 & 3.169 $\pm$ 0.075 & 0.221 $\pm$ 0.005 & -20124.6 $\pm$ 21.7 & 0.402 $\pm$ 0.051 & 0.046 $\pm$ 0.006 & -3560.5 $\pm$ 60.8 & 16.10 \\
60373.381 & 2.985 $\pm$ 0.073 & 0.208 $\pm$ 0.005 & -20158.2 $\pm$ 20.2 & 0.401 $\pm$ 0.048 & 0.046 $\pm$ 0.005 & -3587.5 $\pm$ 58.1 & 13.50 \\
60379.463 & 3.433 $\pm$ 0.108 & 0.240 $\pm$ 0.008 & -20132.5 $\pm$ 29.6 & 0.795 $\pm$ 0.067 & 0.090 $\pm$ 0.008 & -3685.0 $\pm$ 36.3 & 11.33 \\
60403.376 & 3.309 $\pm$ 0.100 & 0.231 $\pm$ 0.007 & -20153.0 $\pm$ 26.6 & 0.660 $\pm$ 0.058 & 0.075 $\pm$ 0.007 & -3544.2 $\pm$ 44.6 & 17.72 \\
60405.240 & 3.277 $\pm$ 0.068 & 0.229 $\pm$ 0.005 & -20115.8 $\pm$ 16.4 & 0.453 $\pm$ 0.045 & 0.051 $\pm$ 0.005 & -3581.0 $\pm$ 47.2 & 13.40 \\
60407.274 & 3.414 $\pm$ 0.068 & 0.238 $\pm$ 0.005 & -20112.3 $\pm$ 19.2 & 0.394 $\pm$ 0.051 & 0.045 $\pm$ 0.006 & -3660.1 $\pm$ 55.5 & 20.38 \\
60410.383 & 3.242 $\pm$ 0.080 & 0.226 $\pm$ 0.006 & -20136.1 $\pm$ 22.6 & 0.629 $\pm$ 0.043 & 0.071 $\pm$ 0.005 & -3556.8 $\pm$ 44.4 & 17.55 \\
60412.216 & 3.169 $\pm$ 0.073 & 0.221 $\pm$ 0.005 & -20111.1 $\pm$ 20.5 & 0.699 $\pm$ 0.039 & 0.079 $\pm$ 0.004 & -3631.2 $\pm$ 32.7 & 14.36 \\
60414.268 & 3.015 $\pm$ 0.070 & 0.210 $\pm$ 0.005 & -20149.5 $\pm$ 20.7 & 0.584 $\pm$ 0.053 & 0.066 $\pm$ 0.006 & -3724.6 $\pm$ 39.1 & 14.58 \\
60416.850 & 3.400 $\pm$ 0.084 & 0.237 $\pm$ 0.006 & -20154.6 $\pm$ 23.2 & 0.453 $\pm$ 0.059 & 0.051 $\pm$ 0.007 & -3556.9 $\pm$ 74.1 & 9.56 \\
60428.694 & 2.945 $\pm$ 0.092 & 0.205 $\pm$ 0.006 & -20154.0 $\pm$ 29.4 & 0.515 $\pm$ 0.063 & 0.058 $\pm$ 0.007 & -3637.0 $\pm$ 52.3 & 13.52 \\
60432.158 & 3.203 $\pm$ 0.071 & 0.223 $\pm$ 0.005 & -20115.8 $\pm$ 22.0 & 0.584 $\pm$ 0.048 & 0.066 $\pm$ 0.005 & -3645.9 $\pm$ 34.2 & 10.96 \\
60434.166 & 3.095 $\pm$ 0.104 & 0.216 $\pm$ 0.007 & -20168.9 $\pm$ 28.7 & 1.006 $\pm$ 0.053 & 0.114 $\pm$ 0.006 & -3713.1 $\pm$ 32.0 & 15.16 \\
60436.168 & 3.100 $\pm$ 0.070 & 0.216 $\pm$ 0.005 & -20162.2 $\pm$ 18.0 & 0.755 $\pm$ 0.051 & 0.086 $\pm$ 0.006 & -3731.1 $\pm$ 30.9 & 7.19 \\
60438.320 & 3.126 $\pm$ 0.142 & 0.218 $\pm$ 0.010 & -20197.2 $\pm$ 42.2 & 0.851 $\pm$ 0.077 & 0.097 $\pm$ 0.009 & -3536.8 $\pm$ 53.6 & 11.66 \\
60441.924 & 3.222 $\pm$ 0.078 & 0.225 $\pm$ 0.005 & -20127.0 $\pm$ 24.7 & 0.519 $\pm$ 0.055 & 0.059 $\pm$ 0.006 & -3655.7 $\pm$ 51.2 & 17.09 \\
60444.366 & 3.127 $\pm$ 0.077 & 0.218 $\pm$ 0.005 & -20134.9 $\pm$ 20.0 & 0.381 $\pm$ 0.046 & 0.043 $\pm$ 0.005 & -3544.5 $\pm$ 60.9 & 16.22 \\
60457.190 & 3.025 $\pm$ 0.064 & 0.211 $\pm$ 0.004 & -20134.5 $\pm$ 18.9 & 0.609 $\pm$ 0.040 & 0.069 $\pm$ 0.005 & -3590.0 $\pm$ 36.8 & 15.17 \\
60459.170 & 2.953 $\pm$ 0.071 & 0.206 $\pm$ 0.005 & -20176.1 $\pm$ 23.9 & 0.499 $\pm$ 0.049 & 0.057 $\pm$ 0.006 & -3592.3 $\pm$ 45.6 & 16.95 \\
60461.168 & 3.107 $\pm$ 0.066 & 0.217 $\pm$ 0.005 & -20149.4 $\pm$ 19.4 & 0.445 $\pm$ 0.050 & 0.051 $\pm$ 0.006 & -3652.7 $\pm$ 50.4 & 14.25 \\
60463.221 & 2.915 $\pm$ 0.071 & 0.203 $\pm$ 0.005 & -20146.4 $\pm$ 23.4 & 0.882 $\pm$ 0.039 & 0.100 $\pm$ 0.004 & -3657.0 $\pm$ 25.4 & 14.76 \\
60465.178 & 3.025 $\pm$ 0.069 & 0.211 $\pm$ 0.005 & -20145.8 $\pm$ 22.9 & 0.511 $\pm$ 0.056 & 0.058 $\pm$ 0.006 & -3644.8 $\pm$ 44.8 & 12.75 \\
60467.180 & 2.814 $\pm$ 0.096 & 0.196 $\pm$ 0.007 & -20192.3 $\pm$ 24.1 & 0.479 $\pm$ 0.058 & 0.054 $\pm$ 0.007 & -3579.9 $\pm$ 62.2 & 13.58 \\
60471.187 & 3.206 $\pm$ 0.095 & 0.224 $\pm$ 0.007 & -20129.0 $\pm$ 24.6 & 0.627 $\pm$ 0.061 & 0.071 $\pm$ 0.007 & -3444.7 $\pm$ 54.8 & 15.97 \\
60490.831 & 3.166 $\pm$ 0.086 & 0.221 $\pm$ 0.006 & -20162.5 $\pm$ 20.9 & 0.728 $\pm$ 0.053 & 0.083 $\pm$ 0.006 & -3651.4 $\pm$ 30.3 & 13.45 \\
60498.202 & 2.884 $\pm$ 0.076 & 0.201 $\pm$ 0.005 & -20126.0 $\pm$ 24.4 & 0.532 $\pm$ 0.042 & 0.060 $\pm$ 0.005 & -3557.8 $\pm$ 44.3 & 15.45 \\
60675.106 & 2.495 $\pm$ 0.057 & 0.174 $\pm$ 0.004 & -20174.7 $\pm$ 25.0 & 0.288 $\pm$ 0.045 & 0.033 $\pm$ 0.005 & -3769.3 $\pm$ 68.9 & 15.20 \\
60699.488 & 2.305 $\pm$ 0.139 & 0.161 $\pm$ 0.010 & -20190.7 $\pm$ 51.8 & 0.404 $\pm$ 0.086 & 0.046 $\pm$ 0.010 & -3569.0 $\pm$ 124.7 & 14.71 \\
60702.084 & 2.575 $\pm$ 0.066 & 0.180 $\pm$ 0.005 & -20140.0 $\pm$ 22.5 & 0.427 $\pm$ 0.041 & 0.048 $\pm$ 0.005 & -3576.2 $\pm$ 51.8 & 15.29 \\
60711.430 & 2.783 $\pm$ 0.067 & 0.194 $\pm$ 0.005 & -20134.7 $\pm$ 21.8 & 0.569 $\pm$ 0.046 & 0.065 $\pm$ 0.005 & -3587.8 $\pm$ 43.7 & 15.15 \\
60728.409 & 2.812 $\pm$ 0.094 & 0.196 $\pm$ 0.007 & -20137.2 $\pm$ 26.8 & 0.227 $\pm$ 0.062 & 0.026 $\pm$ 0.007 & -3536.3 $\pm$ 173.3 & 12.01 \\
60737.342 & 2.720 $\pm$ 0.079 & 0.190 $\pm$ 0.006 & -20184.2 $\pm$ 26.9 & 0.458 $\pm$ 0.048 & 0.052 $\pm$ 0.005 & -3817.7 $\pm$ 58.9 & 11.76 \\
60756.265 & 1.798 $\pm$ 0.068 & 0.125 $\pm$ 0.005 & -20131.4 $\pm$ 37.0 & 0.244 $\pm$ 0.045 & 0.028 $\pm$ 0.005 & -3506.6 $\pm$ 106.0 & 12.98 \\
60786.233 & 2.393 $\pm$ 0.073 & 0.167 $\pm$ 0.005 & -20105.9 $\pm$ 25.4 & 0.763 $\pm$ 0.045 & 0.087 $\pm$ 0.005 & -3657.4 $\pm$ 32.2 & 12.83 \\
60792.192 & 2.682 $\pm$ 0.089 & 0.187 $\pm$ 0.006 & -20156.0 $\pm$ 27.8 & 0.542 $\pm$ 0.052 & 0.062 $\pm$ 0.006 & -3710.3 $\pm$ 55.7 & 12.11 \\
60811.169 & 2.857 $\pm$ 0.077 & 0.199 $\pm$ 0.005 & -20128.7 $\pm$ 23.1 & 0.686 $\pm$ 0.045 & 0.078 $\pm$ 0.005 & -3650.2 $\pm$ 32.9 & 14.07 \\
60817.160 & 2.835 $\pm$ 0.074 & 0.198 $\pm$ 0.005 & -20147.9 $\pm$ 24.1 & 0.439 $\pm$ 0.055 & 0.050 $\pm$ 0.006 & -3712.9 $\pm$ 49.8 & 13.27 \\
60842.053 & 2.904 $\pm$ 0.065 & 0.203 $\pm$ 0.005 & -20125.5 $\pm$ 19.9 & 0.422 $\pm$ 0.048 & 0.048 $\pm$ 0.005 & -3529.3 $\pm$ 50.5 & 15.14 \\
\enddata
\tablenotetext{a}{Continuum fluxes are in units of 10$^{-17}$ erg~s$^{-1}$~cm$^{-2}$~\AA$^{-1}$.}
\end{deluxetable}

\begin{deluxetable*}{ccccccccccccccc}
\tablecaption{BAL Measurements in Mean Spectra by Observing Season}
\label{tab:mean_measurements_by_year}
\tablehead{
Observing & Median & $\Delta t$ & Upper$\_A$ & Lower$\_A$ & $EW_A$ & Depth$\_A$ & $v_{\rm cent,A}$ & Upper$\_B$ & Lower$\_B$ & $EW_B$ & Depth$\_B$ & $v_{\rm cent,B}$ \\
Season & MJD & (days) & (\AA) & (\AA) & (\AA) & & (km s$^{-1}$) & (\AA) & (\AA) & (\AA) & & (km s$^{-1}$) \\
}
\startdata
2003 & 52764.258 & -- & $1454.01 \pm 0.08$ & $1439.68 \pm 0.04$ & $7.15 \pm 0.11$ & $0.499 \pm 0.008$ & $-20178 \pm 9$ & $1534.14 \pm 0.04$ & $1524.98 \pm 0.03$ & $1.91 \pm 0.06$ & $0.210 \pm 0.007$ & $-3757 \pm 19$ \\
2013 & 56426.346 & 3662.088 & $1454.67 \pm 0.03$ & $1440.01 \pm 0.04$ & $6.99 \pm 0.06$ & $0.485 \pm 0.004$ & $-20164 \pm 9$ & $1532.65 \pm 0.06$ & $1524.89 \pm 0.04$ & $1.88 \pm 0.04$ & $0.210 \pm 0.005$ & $-3671 \pm 11$ \\
2014 & 56751.340 & 324.994 & $1454.67 \pm 0.04$ & $1440.01 \pm 0.02$ & $5.61 \pm 0.05$ & $0.370 \pm 0.004$ & $-20218 \pm 9$ & $1532.73 \pm 0.05$ & $1524.98 \pm 0.02$ & $1.09 \pm 0.04$ & $0.140 \pm 0.004$ & $-3791 \pm 15$ \\
2015 & 57116.845 & 365.50 & $1455.01 \pm 0.02$ & $1433.72 \pm 0.03$ & $8.00 \pm 0.07$ & $0.370 \pm 0.005$ & $-20416 \pm 9$ & $1534.14 \pm 0.05$ & $1524.63 \pm 0.03$ & $2.03 \pm 0.05$ & $0.210 \pm 0.005$ & $-3652 \pm 12$ \\
2016 & 57492.342 & 375.50 & $1454.67 \pm 0.03$ & $1439.68 \pm 0.01$ & $7.19 \pm 0.07$ & $0.470 \pm 0.005$ & $-20201 \pm 10$ & $1534.14 \pm 0.02$ & $1524.63 \pm 0.04$ & $1.90 \pm 0.05$ & $0.190 \pm 0.005$ & $-3662 \pm 13$ \\
2017 & 57851.242 & 358.90 & $1454.67 \pm 0.02$ & $1439.35 \pm 0.01$ & $6.21 \pm 0.07$ & $0.400 \pm 0.008$ & $-20213 \pm 12$ & $1534.14 \pm 0.01$ & $1524.98 \pm 0.04$ & $1.67 \pm 0.05$ & $0.180 \pm 0.004$ & $-3634 \pm 16$ \\
2018 & 58216.413 & 365.17 & $1454.67 \pm 0.02$ & $1439.68 \pm 0.05$ & $6.05 \pm 0.07$ & $0.390 \pm 0.005$ & $-20195 \pm 10$ & $1533.08 \pm 0.02$ & $1524.98 \pm 0.04$ & $1.26 \pm 0.04$ & $0.150 \pm 0.005$ & $-3745 \pm 17$ \\
2019 & 58602.250 & 385.84 & $1454.67 \pm 0.01$ & $1439.68 \pm 0.03$ & $5.66 \pm 0.08$ & $0.370 \pm 0.006$ & $-20195 \pm 14$ & $1533.08 \pm 0.04$ & $1528.85 \pm 0.04$ & $0.64 \pm 0.05$ & $0.140 \pm 0.006$ & $-3344 \pm 22$ \\
2020 & 58952.857 & 350.61 & $1454.67 \pm 0.05$ & $1432.40 \pm 0.04$ & $7.80 \pm 0.09$ & $0.340 \pm 0.006$ & $-20474 \pm 11$ & $1533.79 \pm 0.04$ & $1524.98 \pm 0.04$ & $1.68 \pm 0.06$ & $0.180 \pm 0.006$ & $-3652 \pm 16$ \\
2021 & 59293.473 & 340.62 & $1455.01 \pm 0.02$ & $1430.09 \pm 0.03$ & $7.16 \pm 0.06$ & $0.280 \pm 0.004$ & $-20599 \pm 10$ & $1533.79 \pm 0.04$ & $1524.98 \pm 0.04$ & $1.55 \pm 0.04$ & $0.170 \pm 0.007$ & $-3636 \pm 13$ \\
2022 & 59724.198 & 430.73 & $1462.06 \pm 0.02$ & $1457.02 \pm 0.04$ & $0.90 \pm 0.08$ & $0.170 \pm 0.005$ & $-17706 \pm 38$ & $1533.61 \pm 0.01$ & $1524.81 \pm 0.04$ & $0.14 \pm 0.05$ & $0.020 \pm 0.005$ & $-3496 \pm 10515$ \\
2023 & 60072.766 & 348.57 & $1461.72 \pm 0.01$ & $1457.02 \pm 0.02$ & $0.82 \pm 0.08$ & $0.160 \pm 0.006$ & $-17718 \pm 50$ & $1533.61 \pm 0.04$ & $1524.81 \pm 0.03$ & $0.18 \pm 0.05$ & $0.020 \pm 0.006$ & $-3603 \pm 11136$ \\
2024 & 60430.426 & 357.66 & $1453.67 \pm 0.02$ & $1440.67 \pm 0.05$ & $3.06 \pm 0.08$ & $0.230 \pm 0.006$ & $-20197 \pm 24$ & $1533.61 \pm 0.04$ & $1524.81 \pm 0.03$ & $0.57 \pm 0.05$ & $0.060 \pm 0.006$ & $-3614 \pm 51$ \\
2025 & 60756.265 & 325.84 & $1453.00 \pm 0.01$ & $1441.34 \pm 0.02$ & $2.43 \pm 0.08$ & $0.200 \pm 0.006$ & $-20196 \pm 28$ & $1533.61 \pm 0.05$ & $1524.81 \pm 0.02$ & $0.46 \pm 0.05$ & $0.050 \pm 0.006$ & $-3636 \pm 71$ \\
\enddata
\end{deluxetable*}


\begin{thebibliography}{}
\expandafter\ifx\csname natexlab\endcsname\relax\def\natexlab#1{#1}\fi
\providecommand{\url}[1]{\href{#1}{#1}}
\providecommand{\dodoi}[1]{doi:~\href{http://doi.org/#1}{\nolinkurl{#1}}}
\providecommand{\doeprint}[1]{\href{http://ascl.net/#1}{\nolinkurl{http://ascl.net/#1}}}
\providecommand{\doarXiv}[1]{\href{https://arxiv.org/abs/#1}{\nolinkurl{https://arxiv.org/abs/#1}}}

\bibitem[{C.~P. {Ahn} {et~al.}(2014){Ahn}, {Alexandroff}, {Allende Prieto},
  {Anders}, {Anderson}, {Anderton}, {Andrews}, {Aubourg}, {Bailey}, {Bastien},
  \& et~al.}]{Ahn14}
{Ahn}, C.~P., {Alexandroff}, R., {Allende Prieto}, C., {et~al.} 2014,
  \bibinfo{title}{{The Tenth Data Release of the Sloan Digital Sky Survey:
  First Spectroscopic Data from the SDSS-III Apache Point Observatory Galactic
  Evolution Experiment},} \apjs, 211, 17, \dodoi{10.1088/0067-0049/211/2/17}

\bibitem[{ {Astropy Collaboration} {et~al.}(2013){Astropy Collaboration},
  {Robitaille}, {Tollerud}, {Greenfield}, {Droettboom}, {Bray}, {Aldcroft},
  {Davis}, {Ginsburg}, {Price-Whelan}, {Kerzendorf}, {Conley}, {Crighton},
  {Barbary}, {Muna}, {Ferguson}, {Grollier}, {Parikh}, {Nair}, {Unther},
  {Deil}, {Woillez}, {Conseil}, {Kramer}, {Turner}, {Singer}, {Fox}, {Weaver},
  {Zabalza}, {Edwards}, {Azalee Bostroem}, {Burke}, {Casey}, {Crawford},
  {Dencheva}, {Ely}, {Jenness}, {Labrie}, {Lim}, {Pierfederici}, {Pontzen},
  {Ptak}, {Refsdal}, {Servillat}, \& {Streicher}}]{Astropy13}
{Astropy Collaboration}, {Robitaille}, T.~P., {Tollerud}, E.~J., {et~al.} 2013,
  {Astropy: A community Python package for astronomy},
  \dodoi{10.1051/0004-6361/201322068}

\bibitem[{ {Astropy Collaboration} {et~al.}(2022){Astropy Collaboration},
  {Price-Whelan}, {Lim}, {Earl}, {Starkman}, {Bradley}, {Shupe}, {Patil},
  {Corrales}, {Brasseur}, {N{\"o}the}, {Donath}, {Tollerud}, {Morris},
  {Ginsburg}, {Vaher}, {Weaver}, {Tocknell}, {Jamieson}, {van Kerkwijk},
  {Robitaille}, {Merry}, {Bachetti}, {G{\"u}nther}, {Aldcroft},
  {Alvarado-Montes}, {Archibald}, {B{\'o}di}, {Bapat}, {Barentsen},
  {Baz{\'a}n}, {Biswas}, {Boquien}, {Burke}, {Cara}, {Cara}, {Conroy},
  {Conseil}, {Craig}, {Cross}, {Cruz}, {D'Eugenio}, {Dencheva}, {Devillepoix},
  {Dietrich}, {Eigenbrot}, {Erben}, {Ferreira}, {Foreman-Mackey}, {Fox},
  {Freij}, {Garg}, {Geda}, {Glattly}, {Gondhalekar}, {Gordon}, {Grant},
  {Greenfield}, {Groener}, {Guest}, {Gurovich}, {Handberg}, {Hart},
  {Hatfield-Dodds}, {Homeier}, {Hosseinzadeh}, {Jenness}, {Jones}, {Joseph},
  {Kalmbach}, {Karamehmetoglu}, {Ka{\l}uszy{\'n}ski}, {Kelley}, {Kern},
  {Kerzendorf}, {Koch}, {Kulumani}, {Lee}, {Ly}, {Ma}, {MacBride}, {Maljaars},
  {Muna}, {Murphy}, {Norman}, {O'Steen}, {Oman}, {Pacifici}, {Pascual},
  {Pascual-Granado}, {Patil}, {Perren}, {Pickering}, {Rastogi}, {Roulston},
  {Ryan}, {Rykoff}, {Sabater}, {Sakurikar}, {Salgado}, {Sanghi}, {Saunders},
  {Savchenko}, {Schwardt}, {Seifert-Eckert}, {Shih}, {Jain}, {Shukla}, {Sick},
  {Simpson}, {Singanamalla}, {Singer}, {Singhal}, {Sinha}, {Sip{\H{o}}cz},
  {Spitler}, {Stansby}, {Streicher}, {{\v{S}}umak}, {Swinbank}, {Taranu},
  {Tewary}, {Tremblay}, {de Val-Borro}, {Van Kooten}, {Vasovi{\'c}}, {Verma},
  {de Miranda Cardoso}, {Williams}, {Wilson}, {Winkel}, {Wood-Vasey}, {Xue},
  {Yoachim}, {Zhang}, {Zonca}, \& {Astropy Project Contributors}}]{Astropy22}
{Astropy Collaboration}, {Price-Whelan}, A.~M., {Lim}, P.~L., {et~al.} 2022,
  {The Astropy Project: Sustaining and Growing a Community-oriented Open-source
  Project and the Latest Major Release (v5.0) of the Core Package},
  \dodoi{10.3847/1538-4357/ac7c74}

\bibitem[{T.~A. {Barlow}(1993){Barlow}}]{Barlow93}
{Barlow}, T.~A. 1993, \bibinfo{title}{{Time variability of broad
  absorption-line QSO's},} PhD thesis, California University

\bibitem[{M.~R. {Blanton} {et~al.}(2017){Blanton}, {Bershady}, {Abolfathi},
  {Albareti}, {Allende Prieto}, {Almeida}, {Alonso-Garc{\'{\i}}a}, {Anders},
  {Anderson}, {Andrews}, \& et~al.}]{Blanton17}
{Blanton}, M.~R., {Bershady}, M.~A., {Abolfathi}, B., {et~al.} 2017,
  \bibinfo{title}{{Sloan Digital Sky Survey IV: Mapping the Milky Way, Nearby
  Galaxies, and the Distant Universe},} \aj, 154, 28,
  \dodoi{10.3847/1538-3881/aa7567}

\bibitem[{A.~S. {Bolton} {et~al.}(2012){Bolton}, {Schlegel}, {Aubourg},
  {Bailey}, {Bhardwaj}, {Brownstein}, {Burles}, {Chen}, {Dawson}, {Eisenstein},
  {Gunn}, {Knapp}, {Loomis}, {Lupton}, {Maraston}, {Muna}, {Myers}, {Olmstead},
  {Padmanabhan}, {P{\^a}ris}, {Percival}, {Petitjean}, {Rockosi}, {Ross},
  {Schneider}, {Shu}, {Strauss}, {Thomas}, {Tremonti}, {Wake}, {Weaver}, \&
  {Wood-Vasey}}]{Bolton12}
{Bolton}, A.~S., {Schlegel}, D.~J., {Aubourg}, {\'E}., {et~al.} 2012,
  \bibinfo{title}{{Spectral Classification and Redshift Measurement for the
  SDSS-III Baryon Oscillation Spectroscopic Survey},} \aj, 144, 144,
  \dodoi{10.1088/0004-6256/144/5/144}

\bibitem[{D.~M. {Capellupo} {et~al.}(2013){Capellupo}, {Hamann}, {Shields},
  {Halpern}, \& {Barlow}}]{Capellupo13}
{Capellupo}, D.~M., {Hamann}, F., {Shields}, J.~C., {Halpern}, J.~P., \&
  {Barlow}, T.~A. 2013, \bibinfo{title}{{Variability in quasar broad absorption
  line outflows - III. What happens on the shortest time-scales?},} \mnras,
  429, 1872, \dodoi{10.1093/mnras/sts427}

\bibitem[{D.~M. {Capellupo} {et~al.}(2011){Capellupo}, {Hamann}, {Shields},
  {Rodr{\'{\i}}guez Hidalgo}, \& {Barlow}}]{Capellupo11}
{Capellupo}, D.~M., {Hamann}, F., {Shields}, J.~C., {Rodr{\'{\i}}guez Hidalgo},
  P., \& {Barlow}, T.~A. 2011, \bibinfo{title}{{Variability in quasar broad
  absorption line outflows - I. Trends in the short-term versus long-term
  data},} \mnras, 413, 908, \dodoi{10.1111/j.1365-2966.2010.18185.x}

\bibitem[{D.~M. {Capellupo} {et~al.}(2012){Capellupo}, {Hamann}, {Shields},
  {Rodr{\'{\i}}guez Hidalgo}, \& {Barlow}}]{Capellupo12}
{Capellupo}, D.~M., {Hamann}, F., {Shields}, J.~C., {Rodr{\'{\i}}guez Hidalgo},
  P., \& {Barlow}, T.~A. 2012, \bibinfo{title}{{Variability in quasar broad
  absorption line outflows - II. Multi-epoch monitoring of Si IV and C IV broad
  absorption line variability},} \mnras, 422, 3249,
  \dodoi{10.1111/j.1365-2966.2012.20846.x}

\bibitem[{J.~A. {Cardelli} {et~al.}(1989){Cardelli}, {Clayton}, \&
  {Mathis}}]{Cardelli89}
{Cardelli}, J.~A., {Clayton}, G.~C., \& {Mathis}, J.~S. 1989,
  \bibinfo{title}{{The relationship between infrared, optical, and ultraviolet
  extinction},} \apj, 345, 245, \dodoi{10.1086/167900}

\bibitem[{C.-H. {Chan} \& J.~H. {Krolik}(2016){Chan} \& {Krolik}}]{Chan16}
{Chan}, C.-H., \& {Krolik}, J.~H. 2016, \bibinfo{title}{{Radiation-driven
  Outflows from and Radiative Support in Dusty Tori of Active Galactic
  Nuclei},} \apj, 825, 67, \dodoi{10.3847/0004-637X/825/1/67}

\bibitem[{H. {Choi} {et~al.}(2022){Choi}, {Leighly}, {Terndrup}, {Dabbieri},
  {Gallagher}, \& {Richards}}]{Choi22}
{Choi}, H., {Leighly}, K.~M., {Terndrup}, D.~M., {et~al.} 2022,
  \bibinfo{title}{{The Physical Properties of Low-redshift FeLoBAL Quasars. I.
  Spectral-synthesis Analysis of the Broad Absorption-line (BAL) Outflows Using
  SimBAL},} \apj, 937, 74, \dodoi{10.3847/1538-4357/ac61d9}

\bibitem[{K.~S. {Dawson} {et~al.}(2013){Dawson}, {Schlegel}, {Ahn}, {Anderson},
  {Aubourg}, {Bailey}, {Barkhouser}, {Bautista}, {Beifiori}, {Berlind},
  {Bhardwaj}, {Bizyaev}, {Blake}, {Blanton}, {Blomqvist}, {Bolton}, {Borde},
  {Bovy}, {Brandt}, {Brewington}, {Brinkmann}, {Brown}, {Brownstein}, {Bundy},
  {Busca}, {Carithers}, {Carnero}, {Carr}, {Chen}, {Comparat}, {Connolly},
  {Cope}, {Croft}, {Cuesta}, {da Costa}, {Davenport}, {Delubac}, {de Putter},
  {Dhital}, {Ealet}, {Ebelke}, {Eisenstein}, {Escoffier}, {Fan}, {Filiz Ak},
  {Finley}, {Font-Ribera}, {G{\'e}nova-Santos}, {Gunn}, {Guo}, {Haggard},
  {Hall}, {Hamilton}, {Harris}, {Harris}, {Ho}, {Hogg}, {Holder}, {Honscheid},
  {Huehnerhoff}, {Jordan}, {Jordan}, {Kauffmann}, {Kazin}, {Kirkby}, {Klaene},
  {Kneib}, {Le Goff}, {Lee}, {Long}, {Loomis}, {Lundgren}, {Lupton}, {Maia},
  {Makler}, {Malanushenko}, {Malanushenko}, {Mandelbaum}, {Manera}, {Maraston},
  {Margala}, {Masters}, {McBride}, {McDonald}, {McGreer}, {McMahon}, {Mena},
  {Miralda-Escud{\'e}}, {Montero-Dorta}, {Montesano}, {Muna}, {Myers},
  {Naugle}, {Nichol}, {Noterdaeme}, {Nuza}, {Olmstead}, {Oravetz}, {Oravetz},
  {Owen}, {Padmanabhan}, {Palanque-Delabrouille}, {Pan}, {Parejko},
  {P{\^a}ris}, {Percival}, {P{\'e}rez-Fournon}, {P{\'e}rez-R{\`a}fols},
  {Petitjean}, {Pfaffenberger}, {Pforr}, {Pieri}, {Prada}, {Price-Whelan},
  {Raddick}, {Rebolo}, {Rich}, {Richards}, {Rockosi}, {Roe}, {Ross}, {Ross},
  {Rossi}, {Rubi{\~n}o-Martin}, {Samushia}, {S{\'a}nchez}, {Sayres}, {Schmidt},
  {Schneider}, {Sc{\'o}ccola}, {Seo}, {Shelden}, {Sheldon}, {Shen}, {Shu},
  {Slosar}, {Smee}, {Snedden}, {Stauffer}, {Steele}, {Strauss}, {Streblyanska},
  {Suzuki}, {Swanson}, {Tal}, {Tanaka}, {Thomas}, {Tinker}, {Tojeiro},
  {Tremonti}, {Vargas Maga{\~n}a}, {Verde}, {Viel}, {Wake}, {Watson}, {Weaver},
  {Weinberg}, {Weiner}, {West}, {White}, {Wood-Vasey}, {Yeche}, {Zehavi},
  {Zhao}, \& {Zheng}}]{Dawson13}
{Dawson}, K.~S., {Schlegel}, D.~J., {Ahn}, C.~P., {et~al.} 2013,
  \bibinfo{title}{{The Baryon Oscillation Spectroscopic Survey of SDSS-III},}
  \aj, 145, 10, \dodoi{10.1088/0004-6256/145/1/10}

\bibitem[{T. {Di~Matteo} {et~al.}(2005){Di~Matteo}, {Springel}, \&
  {Hernquist}}]{Dimatteo05}
{Di~Matteo}, T., {Springel}, V., \& {Hernquist}, L. 2005,
  \bibinfo{title}{{Energy input from quasars regulates the growth and activity
  of black holes and their host galaxies},} \nat, 433, 604,
  \dodoi{10.1038/nature03335}

\bibitem[{F.~R. {Donnan} {et~al.}(2021){Donnan}, {Horne}, \& {Hern{\'a}ndez
  Santisteban}}]{Donnan21}
{Donnan}, F.~R., {Horne}, K., \& {Hern{\'a}ndez Santisteban}, J.~V. 2021,
  \bibinfo{title}{{Bayesian analysis of quasar lightcurves with a running
  optimal average: new time delay measurements of COSMOGRAIL gravitationally
  lensed quasars},} \mnras, 508, 1458, \dodoi{10.1093/mnras/stab2832}

\bibitem[{R. {Edelson} {et~al.}(2019){Edelson}, {Gelbord}, {Cackett},
  {Peterson}, {Horne}, {Barth}, {Starkey}, {Bentz}, {Brandt}, {Goad}, \&
  et~al.}]{Edelson19}
{Edelson}, R., {Gelbord}, J., {Cackett}, E., {et~al.} 2019,
  \bibinfo{title}{{The First Swift Intensive AGN Accretion Disk Reverberation
  Mapping Survey},} \apj, 870, 123, \dodoi{10.3847/1538-4357/aaf3b4}

\bibitem[{L. {Ferrarese} \& D. {Merritt}(2000){Ferrarese} \&
  {Merritt}}]{Ferrarese00}
{Ferrarese}, L., \& {Merritt}, D. 2000, \bibinfo{title}{{A Fundamental Relation
  between Supermassive Black Holes and Their Host Galaxies},} \apjl, 539, L9

\bibitem[{N. {Filiz Ak} {et~al.}(2012){Filiz Ak}, {Brandt}, {Hall},
  {Schneider}, {Anderson}, {Gibson}, {Lundgren}, {Myers}, {Petitjean}, {Ross},
  {Shen}, {York}, {Bizyaev}, {Brinkmann}, {Malanushenko}, {Oravetz}, {Pan},
  {Simmons}, \& {Weaver}}]{Filizak12}
{Filiz Ak}, N., {Brandt}, W.~N., {Hall}, P.~B., {et~al.} 2012,
  \bibinfo{title}{{Broad Absorption Line Disappearance on Multi-year Timescales
  in a Large Quasar Sample},} \apj, 757, 114,
  \dodoi{10.1088/0004-637X/757/2/114}

\bibitem[{N. {Filiz Ak} {et~al.}(2013){Filiz Ak}, {Brandt}, {Hall},
  {Schneider}, {Anderson}, {Hamann}, {Lundgren}, {Myers}, {P{\^a}ris},
  {Petitjean}, {Ross}, {Shen}, \& {York}}]{Filizak13}
{Filiz Ak}, N., {Brandt}, W.~N., {Hall}, P.~B., {et~al.} 2013,
  \bibinfo{title}{{Broad Absorption Line Variability on Multi-year Timescales
  in a Large Quasar Sample},} \apj, 777, 168,
  \dodoi{10.1088/0004-637X/777/2/168}

\bibitem[{K. {Gebhardt} {et~al.}(2000){Gebhardt}, {Bender}, {Bower},
  {Dressler}, {Faber}, {Filippenko}, {Green}, {Grillmair}, {Ho}, {Kormendy},
  {Lauer}, {Magorrian}, {Pinkney}, {Richstone}, \& {Tremaine}}]{Gebhardt00a}
{Gebhardt}, K., {Bender}, R., {Bower}, G., {et~al.} 2000, \bibinfo{title}{{A
  Relationship between Nuclear Black Hole Mass and Galaxy Velocity
  Dispersion},} \apjl, 539, L13

\bibitem[{R.~R. {Gibson} {et~al.}(2008){Gibson}, {Brandt}, {Schneider}, \&
  {Gallagher}}]{Gibson08b}
{Gibson}, R.~R., {Brandt}, W.~N., {Schneider}, D.~P., \& {Gallagher}, S.~C.
  2008, \bibinfo{title}{{Quasar Broad Absorption Line Variability on Multiyear
  Timescales},} \apj, 675, 985, \dodoi{10.1086/527462}

\bibitem[{R.~R. {Gibson} {et~al.}(2009){Gibson}, {Jiang}, {Brandt}, {Hall},
  {Shen}, {Wu}, {Anderson}, {Schneider}, {Vanden Berk}, {Gallagher}, {Fan}, \&
  {York}}]{Gibson09b}
{Gibson}, R.~R., {Jiang}, L., {Brandt}, W.~N., {et~al.} 2009,
  \bibinfo{title}{{A Catalog of Broad Absorption Line Quasars in Sloan Digital
  Sky Survey Data Release 5},} \apj, 692, 758,
  \dodoi{10.1088/0004-637X/692/1/758}

\bibitem[{C.~J. {Grier} {et~al.}(2015){Grier}, {Hall}, {Brandt}, {Trump},
  {Shen}, {Vivek}, {Filiz Ak}, {Chen}, {Dawson}, {Denney}, {Green}, {Jiang},
  {Kochanek}, {McGreer}, {P{\^a}ris}, {Peterson}, {Schneider}, {Tao},
  {Wood-Vasey}, {Bizyaev}, {Ge}, {Kinemuchi}, {Oravetz}, {Pan}, \&
  {Simmons}}]{Grier15}
{Grier}, C.~J., {Hall}, P.~B., {Brandt}, W.~N., {et~al.} 2015,
  \bibinfo{title}{{The Sloan Digital Sky Survey Reverberation Mapping Project:
  Rapid CIV Broad Absorption Line Variability},} \apj, 806, 111,
  \dodoi{10.1088/0004-637X/806/1/111}

\bibitem[{C.~J. {Grier} {et~al.}(2016){Grier}, {Brandt}, {Hall}, {Trump},
  {Filiz Ak}, {Anderson}, {Green}, {Schneider}, {Sun}, {Vivek}, {Beatty},
  {Brownstein}, \& {Roman-Lopes}}]{Grier16}
{Grier}, C.~J., {Brandt}, W.~N., {Hall}, P.~B., {et~al.} 2016,
  \bibinfo{title}{{C IV Broad Absorption Line Acceleration in Sloan Digital Sky
  Survey Quasars},} \apj, 824, 130, \dodoi{10.3847/0004-637X/824/2/130}

\bibitem[{K. {G{\"u}ltekin} {et~al.}(2009){G{\"u}ltekin}, {Richstone},
  {Gebhardt}, {Lauer}, {Tremaine}, {Aller}, {Bender}, {Dressler}, {Faber},
  {Filippenko}, {Green}, {Ho}, {Kormend}, {Magorrian}, {Pinkney}, \&
  {Siopis}}]{Gultekin09}
{G{\"u}ltekin}, K., {Richstone}, D.~O., {Gebhardt}, K., {et~al.} 2009,
  \bibinfo{title}{{The M-{$\sigma$} and M-L Relations in Galactic Bulges, and
  Determinations of Their Intrinsic Scatter},} \apj, 698, 198,
  \dodoi{10.1088/0004-637X/698/1/198}

\bibitem[{J.~E. {Gunn} {et~al.}(2006){Gunn}, {Siegmund}, {Mannery}, {Owen},
  {Hull}, {Leger}, {Carey}, {Knapp}, {York}, {Boroski}, {Kent}, {Lupton},
  {Rockosi}, {Evans}, {Waddell}, {Anderson}, {Annis}, {Barentine}, {Bartoszek},
  {Bastian}, {Bracker}, {Brewington}, {Briegel}, {Brinkmann}, {Brown}, {Carr},
  {Czarapata}, {Drennan}, {Dombeck}, {Federwitz}, {Gillespie}, {Gonzales},
  {Hansen}, {Harvanek}, {Hayes}, {Jordan}, {Kinney}, {Klaene}, {Kleinman},
  {Kron}, {Kresinski}, {Lee}, {Limmongkol}, {Lindenmeyer}, {Long}, {Loomis},
  {McGehee}, {Mantsch}, {Neilsen}, {Neswold}, {Newman}, {Nitta}, {Peoples},
  {Pier}, {Prieto}, {Prosapio}, {Rivetta}, {Schneider}, {Snedden}, \&
  {Wang}}]{Gunn06}
{Gunn}, J.~E., {Siegmund}, W.~A., {Mannery}, E.~J., {et~al.} 2006,
  \bibinfo{title}{{The 2.5 m Telescope of the Sloan Digital Sky Survey},} \aj,
  131, 2332, \dodoi{10.1086/500975}

\bibitem[{C.~R. Harris {et~al.}(2020)Harris, Millman, van~der Walt, Gommers,
  Virtanen, Cournapeau, Wieser, Taylor, Berg, Smith, Kern, Picus, Hoyer, van
  Kerkwijk, Brett, Haldane, del Río, Wiebe, Peterson, Gérard-Marchant,
  Sheppard, Reddy, Weckesser, Abbasi, Gohlke, \& Oliphant}]{Harris20}
Harris, C.~R., Millman, K.~J., van~der Walt, S.~J., {et~al.} 2020, Array
  programming with NumPy, Springer Science and Business Media LLC,
  \dodoi{10.1038/s41586-020-2649-2}

\bibitem[{Z.~S. {Hemler} {et~al.}(2019){Hemler}, {Grier}, {Brandt}, {Hall},
  {Horne}, {Shen}, {Trump}, {Schneider}, {Vivek}, {Bizyaev}, {Oravetz},
  {Oravetz}, \& {Pan}}]{Hemler19}
{Hemler}, Z.~S., {Grier}, C.~J., {Brandt}, W.~N., {et~al.} 2019,
  \bibinfo{title}{{The Sloan Digital Sky Survey Reverberation Mapping Project:
  Systematic Investigations of Short-timescale C IV Broad Absorption Line
  Variability},} \apj, 872, 21, \dodoi{10.3847/1538-4357/aaf1bf}

\bibitem[{N. {Higginbottom} {et~al.}(2014){Higginbottom}, {Proga}, {Knigge},
  {Long}, {Matthews}, \& {Sim}}]{Higginbottom14}
{Higginbottom}, N., {Proga}, D., {Knigge}, C., {et~al.} 2014,
  \bibinfo{title}{{Line-driven Disk Winds in Active Galactic Nuclei: The
  Critical Importance of Ionization and Radiative Transfer},} \apj, 789, 19,
  \dodoi{10.1088/0004-637X/789/1/19}

\bibitem[{J.~D. Hunter(2007)Hunter}]{Hunter07}
Hunter, J.~D. 2007, Matplotlib: A 2D graphics environment, IEEE Computer
  Society, \dodoi{10.1109/MCSE.2007.55}

\bibitem[{B.~C. {Kelly} {et~al.}(2009){Kelly}, {Bechtold}, \&
  {Siemiginowska}}]{Kelly09}
{Kelly}, B.~C., {Bechtold}, J., \& {Siemiginowska}, A. 2009,
  \bibinfo{title}{{Are the Variations in Quasar Optical Flux Driven by Thermal
  Fluctuations?},} \apj, 698, 895, \dodoi{10.1088/0004-637X/698/1/895}

\bibitem[{A. {King} \& K. {Pounds}(2015){King} \& {Pounds}}]{KingPounds15}
{King}, A., \& {Pounds}, K. 2015, \bibinfo{title}{{Powerful Outflows and
  Feedback from Active Galactic Nuclei},} \araa, 53, 115,
  \dodoi{10.1146/annurev-astro-082214-122316}

\bibitem[{J.~A. {Kollmeier} {et~al.}(2026){Kollmeier}, {Rix}, {Aerts}, {Aird},
  {Vera Alfaro}, {Almeida}, {Anderson}, {Arseneau}, {Assef}, {Aviram}, {Aydar},
  {Badenes}, {Bandyopadhyay}, {Barger}, {Barkhouser}, {Bauer}, {Behmard},
  {Bender}, {Besser}, {Bhattarai}, {Bilgi}, {Bird}, {Bizyaev}, {Blanc},
  {Blanton}, {Bochanski}, {Bovy}, {Brandon}, {Brandt}, {Brownstein}, {Buchner},
  {Burchett}, {Carlberg}, {Casey}, {Castaneda-Carlos}, {Chakraborty},
  {Chanam{\'e}}, {Chandra}, {Cherinka}, {Chilingarian}, {Comparat}, {Cosens},
  {Covey}, {Crane}, {Crumpler}, {Cruz-Gonzalez}, {Cunha}, {Cunningham}, {Dai},
  {Darling}, {Davidson}, {Davis}, {De Lee}, {Deacon}, {M{\'e}ndez Delgado},
  {Demasi}, {Demianenko}, {Derwent}, {D'Onghia}, {Di Mille}, {Dias}, {Donor},
  {Dow}, {Drory}, {Dwelly}, {Egorov}, {Egorova}, {El-Badry}, {Engelman},
  {Eracleous}, {Fan}, {Farr}, {Fries}, {Frinchaboy}, {Froning}, {G{\"a}nsicke},
  {Garc{\'\i}a}, {Gelfand}, {Gentile Fusillo}, {Glover}, {Grabowski}, {Grebel},
  {Green}, {Grier}, {Gupta}, {Gray}, {H{\"a}berle}, {Hall}, {Hammond},
  {Hawkins}, {Harding}, {Heged{\H{u}}s}, {Herbst}, {Hermes}, {Rodr{\'\i}guez
  Hidalgo}, {Hilder}, {Hogg}, {Holtzman}, {Horta}, {Huang}, {Hwang},
  {Ibarra-Medel}, {Imig}, {Inight}, {Jana}, {Ji}, {Jim{\'e}nez-Arranz},
  {Jofre}, {Johns}, {Johnson}, {Johnson}, {Johnston}, {Jones}, {Katkov},
  {Knapp}, {Koekemoer}, {Kounkel}, {Kreckel}, {Krishnarao}, {Krumpe}, {Kumari},
  {Kupfer}, {Lacerna}, {Laporte}, {Lepine}, {Li}, {Liu}, {Loebman}, {Long},
  {Roman-Lopes}, {Lu}, {Majewski}, {Maoz}, {McKinnon}, {Medan}, {Merloni},
  {Minniti}, {Morrison}, {Myers}, {M{\'e}sz{\'a}ros}, {Nandra}, {Nayak},
  {Ness}, {Nidever}, {O'Brien}, {Oeur}, {Oravetz}, {Oravetz}, {Otto},
  {Pallathadka}, {Palunas}, {Pan}, {Pappalardo}, {Pandey}, {Negrete
  Pe{\~n}aloza}, {Pinsonneault}, {Pogge}, {Taghizadeh Popp}, {Price-Whelan},
  {Pulatova}, {Qiu}, {Ramirez}, {Rankine}, {Ricci}, {Runnoe}, {Sanchez},
  {Salvato}, {Sarbadhicary}, {Sattler}, {Saydjari}, {Sayres}, {Schinnerer},
  {Schlaufman}, {Schneider}, {Schreiber}, {Schwope}, {Serna}, {Shen},
  {Sif{\'o}n}, {Singh}, {Sinha}, {Smee}, {Song}, {Souto}, {Stassun},
  {Steinmetz}, {Stone-Martinez}, {Stringfellow}, {Stutz},
  {S{\'a}nchez-Gallego}, {Tan}, {Tayar}, {Thai}, {Thakar}, {Ting}, {Tkachenko},
  {Tovmassian}, {Trakhtenbrot}, {Fern{\'a}ndez-Trincado}, \&
  {Troup}}]{Kollmeier26}
{Kollmeier}, J.~A., {Rix}, H.-W., {Aerts}, C., {et~al.} 2026,
  \bibinfo{title}{{Sloan Digital Sky Survey. V. Pioneering Panoptic
  Spectroscopy},} \aj, 171, 52, \dodoi{10.3847/1538-3881/ae0576}

\bibitem[{J. {Kormendy} \& L.~C. {Ho}(2013){Kormendy} \& {Ho}}]{Kormendy13}
{Kormendy}, J., \& {Ho}, L.~C. 2013, \bibinfo{title}{{Coevolution (Or Not) of
  Supermassive Black Holes and Host Galaxies},} \araa, 51, 511,
  \dodoi{10.1146/annurev-astro-082708-101811}

\bibitem[{S. {Koz{\l}owski}(2016){Koz{\l}owski}}]{Kozlowski16}
{Koz{\l}owski}, S. 2016, \bibinfo{title}{{A degeneracy in DRW modelling of AGN
  light curves},} \mnras, 459, 2787, \dodoi{10.1093/mnras/stw819}

\bibitem[{S. {Koz{\l}owski} {et~al.}(2010){Koz{\l}owski}, {Kochanek},
  {Udalski}, {Wyrzykowski}, {Soszy{\'n}ski}, {Szyma{\'n}ski}, {Kubiak},
  {Pietrzy{\'n}ski}, {Szewczyk}, {Ulaczyk}, {Poleski}, \& {The OGLE
  Collaboration}}]{Kozlowski10}
{Koz{\l}owski}, S., {Kochanek}, C.~S., {Udalski}, A., {et~al.} 2010,
  \bibinfo{title}{{Quantifying Quasar Variability as Part of a General Approach
  to Classifying Continuously Varying Sources},} \apj, 708, 927,
  \dodoi{10.1088/0004-637X/708/2/927}

\bibitem[{B.~F. {Lundgren} {et~al.}(2007){Lundgren}, {Wilhite}, {Brunner},
  {Hall}, {Schneider}, {York}, {Vanden Berk}, \& {Brinkmann}}]{Lundgren07}
{Lundgren}, B.~F., {Wilhite}, B.~C., {Brunner}, R.~J., {et~al.} 2007,
  \bibinfo{title}{{Broad Absorption Line Variability in Repeat Quasar
  Observations from the Sloan Digital Sky Survey},} \apj, 656, 73,
  \dodoi{10.1086/510202}

\bibitem[{C.~L. {MacLeod} {et~al.}(2010){MacLeod}, {Ivezi{\'c}}, {Kochanek},
  {Koz{\l}owski}, {Kelly}, {Bullock}, {Kimball}, {Sesar}, {Westman}, {Brooks},
  {Gibson}, {Becker}, \& {de Vries}}]{Macleod10}
{MacLeod}, C.~L., {Ivezi{\'c}}, {\v Z}., {Kochanek}, C.~S., {et~al.} 2010,
  \bibinfo{title}{{Modeling the Time Variability of SDSS Stripe 82 Quasars as a
  Damped Random Walk},} \apj, 721, 1014, \dodoi{10.1088/0004-637X/721/2/1014}

\bibitem[{C.~L. {MacLeod} {et~al.}(2012){MacLeod}, {Ivezi{\'c}}, {Sesar}, {de
  Vries}, {Kochanek}, {Kelly}, {Becker}, {Lupton}, {Hall}, {Richards},
  {Anderson}, \& {Schneider}}]{Macleod12}
{MacLeod}, C.~L., {Ivezi{\'c}}, {\v Z}., {Sesar}, B., {et~al.} 2012,
  \bibinfo{title}{{A Description of Quasar Variability Measured Using Repeated
  SDSS and POSS Imaging},} \apj, 753, 106, \dodoi{10.1088/0004-637X/753/2/106}

\bibitem[{J. {Magorrian} {et~al.}(1998){Magorrian} {et~al.}}]{Magorrian98}
{Magorrian}, J., {et~al.} 1998, \bibinfo{title}{{The Demography of Massive Dark
  Objects in Galaxy Centers},} \aj, 115, 2285, \dodoi{10.1086/300353}

\bibitem[{H.~L. {Marshall} {et~al.}(1997){Marshall}, {Carone}, {Peterson},
  {Clavel}, {Crenshaw}, {Korista}, {Kriss}, {Krolik}, {Malkan}, {Morris}, \&
  {Reichert}}]{Marshall97}
{Marshall}, H.~L., {Carone}, T.~E., {Peterson}, B.~M., {et~al.} 1997,
  \bibinfo{title}{{The Variability and Spectrum of NGC 5548 in the Extreme
  Ultraviolet},} \apj, 479, 222, \dodoi{10.1086/303850}

\bibitem[{T. {Misawa} {et~al.}(2010){Misawa}, {Kawabata}, {Eracleous},
  {Charlton}, \& {Kashikawa}}]{Misawa10}
{Misawa}, T., {Kawabata}, K.~S., {Eracleous}, M., {Charlton}, J.~C., \&
  {Kashikawa}, N. 2010, \bibinfo{title}{{A Spectropolarimetric Test of the
  Structure of the Intrinsic Absorbers in the Quasar HS 1603+3820},} \apj, 719,
  1890, \dodoi{10.1088/0004-637X/719/2/1890}

\bibitem[{R. {Moll} {et~al.}(2007){Moll}, {Schindler}, {Domainko}, {Kapferer},
  {Mair}, {van Kampen}, {Kronberger}, {Kimeswenger}, \& {Ruffert}}]{Moll07}
{Moll}, R., {Schindler}, S., {Domainko}, W., {et~al.} 2007,
  \bibinfo{title}{{Simulations of metal enrichment in galaxy clusters by AGN
  outflows},} \aap, 463, 513, \dodoi{10.1051/0004-6361:20066386}

\bibitem[{N. {Murray} {et~al.}(1995){Murray}, {Chiang}, {Grossman}, \&
  {Voit}}]{Murray95}
{Murray}, N., {Chiang}, J., {Grossman}, S.~A., \& {Voit}, G.~M. 1995,
  \bibinfo{title}{{Accretion Disk Winds from Active Galactic Nuclei},} \apj,
  451, 498, \dodoi{10.1086/176238}

\bibitem[{D. {Park} {et~al.}(2013){Park}, {Woo}, {Denney}, \& {Shin}}]{Park13}
{Park}, D., {Woo}, J.-H., {Denney}, K.~D., \& {Shin}, J. 2013,
  \bibinfo{title}{{Calibrating C-IV-based Black Hole Mass Estimators},} \apj,
  770, 87, \dodoi{10.1088/0004-637X/770/2/87}

\bibitem[{Y.~C. {Pei}(1992){Pei}}]{Pei92}
{Pei}, Y.~C. 1992, \bibinfo{title}{{Interstellar dust from the Milky Way to the
  Magellanic Clouds},} \apj, 395, 130, \dodoi{10.1086/171637}

\bibitem[{T. {Pena} {et~al.}(2026){Pena}, {Grier}, {Hall}, {Shen}, {Horne},
  {Brandt}, {Trump}, {Stone}, \& coauthors}]{Pena26}
{Pena}, T., {Grier}, C.~J., {Hall}, P.~B., {et~al.} 2026, \bibinfo{title}{{The
  Black Hole Mapper Reverberation Mapping Project: 17-Year Photometric Light
  Curves of the SDSS-RM Quasar Sample},} \apjs, {submitted}

\bibitem[{B.~M. {Peterson} {et~al.}(1998){Peterson}, {Wanders}, {Bertram},
  {Hunley}, {Pogge}, \& {Wagner}}]{Peterson98}
{Peterson}, B.~M., {Wanders}, I., {Bertram}, R., {et~al.} 1998,
  \bibinfo{title}{{Optical Continuum and Emission-Line Variability of Seyfert 1
  Galaxies},} \apj, 501, 82

\bibitem[{B.~M. {Peterson} {et~al.}(2004){Peterson}, {Ferrarese}, {Gilbert},
  {Kaspi}, {Malkan}, {Maoz}, {Merritt}, {Netzer}, {Onken}, {Pogge},
  {Vestergaard}, \& {Wandel}}]{Peterson04}
{Peterson}, B.~M., {Ferrarese}, L., {Gilbert}, K.~M., {et~al.} 2004,
  \bibinfo{title}{{Central Masses and Broad-Line Region Sizes of Active
  Galactic Nuclei. II. A Homogeneous Analysis of a Large Reverberation-Mapping
  Database},} \apj, 613, 682, \dodoi{10.1086/423269}

\bibitem[{D. {Proga}(2000){Proga}}]{Proga00}
{Proga}, D. 2000, \bibinfo{title}{{Winds from Accretion Disks Driven by
  Radiation and Magnetocentrifugal Force},} \apj, 538, 684,
  \dodoi{10.1086/309154}

\bibitem[{J.~A. {Rogerson} {et~al.}(2018){Rogerson}, {Hall}, {Ahmed},
  {Rodr{\'\i}guez Hidalgo}, {Brandt}, \& {Filiz Ak}}]{Rogerson18}
{Rogerson}, J.~A., {Hall}, P.~B., {Ahmed}, N.~S., {et~al.} 2018,
  \bibinfo{title}{{Emergence and Variability of Broad Absorption Line Quasar
  Outflows},} \apj, 862, 22, \dodoi{10.3847/1538-4357/aabfe5}

\bibitem[{E.~F. Schlafly \& D.~P. Finkbeiner(2011)Schlafly \&
  Finkbeiner}]{Schlafly11}
Schlafly, E.~F., \& Finkbeiner, D.~P. 2011, \bibinfo{title}{MEASURING REDDENING
  WITH SLOAN DIGITAL SKY SURVEY STELLAR SPECTRA AND RECALIBRATING SFD,} The
  Astrophysical Journal, 737, 103, \dodoi{10.1088/0004-637X/737/2/103}

\bibitem[{Y. {Shen} {et~al.}(2015){Shen}, {Brandt}, {Dawson}, {Hall},
  {McGreer}, {Anderson}, {Chen}, {Denney}, {Eftekharzadeh}, {Fan}, {Gao},
  {Green}, {Greene}, {Ho}, {Horne}, {Jiang}, {Kelly}, {Kinemuchi}, {Kochanek},
  {P{\^a}ris}, {Peters}, {Peterson}, {Petitjean}, {Ponder}, {Richards},
  {Schneider}, {Seth}, {Smith}, {Strauss}, {Tao}, {Trump}, {Wood-Vasey}, {Zu},
  {Eisenstein}, {Pan}, {Bizyaev}, {Malanushenko}, {Malanushenko}, \&
  {Oravetz}}]{Shen15a}
{Shen}, Y., {Brandt}, W.~N., {Dawson}, K.~S., {et~al.} 2015,
  \bibinfo{title}{{The Sloan Digital Sky Survey Reverberation Mapping Project:
  Technical Overview},} \apjs, 216, 4, \dodoi{10.1088/0067-0049/216/1/4}

\bibitem[{Y. {Shen} {et~al.}(2019){Shen}, {Hall}, {Horne}, {Zhu}, {McGreer},
  {Simm}, {Trump}, {Kinemuchi}, {Brandt}, {Green}, {Grier}, {Guo}, {Ho},
  {Homayouni}, {Jiang}, {I-Hsiu Li}, {Morganson}, {Petitjean}, {Richards},
  {Schneider}, {Starkey}, {Wang}, {Chambers}, {Kaiser}, {Kudritzki}, {Magnier},
  \& {Waters}}]{Shen19}
{Shen}, Y., {Hall}, P.~B., {Horne}, K., {et~al.} 2019, \bibinfo{title}{{The
  Sloan Digital Sky Survey Reverberation Mapping Project: Sample
  Characterization},} \apjs, 241, 34, \dodoi{10.3847/1538-4365/ab074f}

\bibitem[{Y. {Shen} {et~al.}(2024){Shen}, {Grier}, {Horne}, {Stone}, {Li},
  {Yang}, {Homayouni}, {Trump}, {Anderson}, {Brandt}, {Hall}, {Ho}, {Jiang},
  {Petitjean}, {Schneider}, {Tao}, {Donnan}, {AlSayyad}, {Bershady}, {Blanton},
  {Bizyaev}, {Bundy}, {Chen}, {Davis}, {Dawson}, {Fan}, {Greene},
  {Gr{\"o}ller}, {Guo}, {Ibarra-Medel}, {Jiang}, {Keenan}, {Kollmeier},
  {Lejoly}, {Li}, {de la Macorra}, {Moe}, {Nie}, {Rossi}, {Smith}, {Tee},
  {Weijmans}, {Xu}, {Yue}, {Zhou}, {Zhou}, \& {Zou}}]{Shen24}
{Shen}, Y., {Grier}, C.~J., {Horne}, K., {et~al.} 2024, \bibinfo{title}{{The
  Sloan Digital Sky Survey Reverberation Mapping Project: Key Results},} \apjs,
  272, 26, \dodoi{10.3847/1538-4365/ad3936}

\bibitem[{S.~A. {Smee} {et~al.}(2013){Smee}, {Gunn}, {Uomoto}, {Roe},
  {Schlegel}, {Rockosi}, {Carr}, {Leger}, {Dawson}, {Olmstead}, {Brinkmann},
  {Owen}, {Barkhouser}, {Honscheid}, {Harding}, {Long}, {Lupton}, {Loomis},
  {Anderson}, {Annis}, {Bernardi}, {Bhardwaj}, {Bizyaev}, {Bolton},
  {Brewington}, {Briggs}, {Burles}, {Burns}, {Castander}, {Connolly},
  {Davenport}, {Ebelke}, {Epps}, {Feldman}, {Friedman}, {Frieman}, {Heckman},
  {Hull}, {Knapp}, {Lawrence}, {Loveday}, {Mannery}, {Malanushenko},
  {Malanushenko}, {Merrelli}, {Muna}, {Newman}, {Nichol}, {Oravetz}, {Pan},
  {Pope}, {Ricketts}, {Shelden}, {Sandford}, {Siegmund}, {Simmons}, {Smith},
  {Snedden}, {Schneider}, {SubbaRao}, {Tremonti}, {Waddell}, \&
  {York}}]{Smee13}
{Smee}, S.~A., {Gunn}, J.~E., {Uomoto}, A., {et~al.} 2013, \bibinfo{title}{{The
  Multi-object, Fiber-fed Spectrographs for the Sloan Digital Sky Survey and
  the Baryon Oscillation Spectroscopic Survey},} \aj, 146, 32,
  \dodoi{10.1088/0004-6256/146/2/32}

\bibitem[{M. {Sun} {et~al.}(2018){Sun}, {Grier}, \& {Peterson}}]{Sun18}
{Sun}, M., {Grier}, C.~J., \& {Peterson}, B.~M. 2018, {PyCCF: Python Cross
  Correlation Function for reverberation mapping studies},, Astrophysics Source
  Code Library, record ascl:1805.032

\bibitem[{J.~D. {Timlin} {et~al.}(2021){Timlin}, {Brandt}, \&
  {Laor}}]{Timlin21}
{Timlin}, III, J.~D., {Brandt}, W.~N., \& {Laor}, A. 2021,
  \bibinfo{title}{{What controls the UV-to-X-ray continuum shape in quasars?},}
  \mnras, 504, 5556, \dodoi{10.1093/mnras/stab1217}

\bibitem[{R.~P. {van der Marel} \& M. {Franx}(1993){van der Marel} \&
  {Franx}}]{vanderMarel93}
{van der Marel}, R.~P., \& {Franx}, M. 1993, \bibinfo{title}{{A new method for
  the identification of non-Gaussian line profiles in elliptical galaxies},}
  \apj, 407, 525, \dodoi{10.1086/172534}

\bibitem[{P. {Virtanen} {et~al.}(2020){Virtanen}, {Gommers}, {Oliphant},
  {Haberland}, {Reddy}, {Cournapeau}, {Burovski}, {Peterson}, {Weckesser},
  {Bright}, {van der Walt}, {Brett}, {Wilson}, {Millman}, {Mayorov}, {Nelson},
  {Jones}, {Kern}, {Larson}, {Carey}, {Polat}, {Feng}, {Moore}, {VanderPlas},
  {Laxalde}, {Perktold}, {Cimrman}, {Henriksen}, {Quintero}, {Harris},
  {Archibald}, {Ribeiro}, {Pedregosa}, {van Mulbregt}, \& {SciPy 1. 0
  Contributors}}]{Virtanen20}
{Virtanen}, P., {Gommers}, R., {Oliphant}, T.~E., {et~al.} 2020, {SciPy 1.0:
  fundamental algorithms for scientific computing in Python},
  \dodoi{10.1038/s41592-019-0686-2}

\bibitem[{R.~J. {Weymann} {et~al.}(1991){Weymann}, {Morris}, {Foltz}, \&
  {Hewett}}]{Weymann91}
{Weymann}, R.~J., {Morris}, S.~L., {Foltz}, C.~B., \& {Hewett}, P.~C. 1991,
  \bibinfo{title}{{Comparisons of the emission-line and continuum properties of
  broad absorption line and normal quasi-stellar objects},} \apj, 373, 23,
  \dodoi{10.1086/170020}

\bibitem[{R. {Wheatley} {et~al.}(2024){Wheatley}, {Grier}, {Hall}, {Brandt},
  {Lotz}, {Schneider}, {Trump}, {Shen}, {Seaton}, {Anderson}, {Temple},
  {Assef}, {Fries}, {Homayouni}, {Kakkad}, {Koekemoer}, {Mart{\'\i}nez-Aldama},
  {Negrete}, {Ricci}, {Bizyaev}, {Brownstein}, {Morrison}, \&
  {Pan}}]{Wheatley24}
{Wheatley}, R., {Grier}, C.~J., {Hall}, P.~B., {et~al.} 2024,
  \bibinfo{title}{{The SDSS-V Black Hole Mapper Reverberation Mapping Project:
  C IV Broad Absorption Line Acceleration in the Quasar SBS 1408+544},} \apj,
  968, 49, \dodoi{10.3847/1538-4357/ad429e}

\bibitem[{Z. {Yu} {et~al.}(2020){Yu}, {Kochanek}, {Peterson}, {Zu}, {Brandt},
  {Cackett}, {Fausnaugh}, \& {McHardy}}]{Yu20}
{Yu}, Z., {Kochanek}, C.~S., {Peterson}, B.~M., {et~al.} 2020,
  \bibinfo{title}{{On reverberation mapping lag uncertainties},} \mnras, 491,
  6045, \dodoi{10.1093/mnras/stz3464}

\bibitem[{Y. {Zu} {et~al.}(2013){Zu}, {Kochanek}, {Koz{\l}owski}, \&
  {Udalski}}]{Zu13}
{Zu}, Y., {Kochanek}, C.~S., {Koz{\l}owski}, S., \& {Udalski}, A. 2013,
  \bibinfo{title}{{Is Quasar Optical Variability a Damped Random Walk?},} \apj,
  765, 106, \dodoi{10.1088/0004-637X/765/2/106}

\bibitem[{Y. {Zu} {et~al.}(2011){Zu}, {Kochanek}, \& {Peterson}}]{Zu11}
{Zu}, Y., {Kochanek}, C.~S., \& {Peterson}, B.~M. 2011, \bibinfo{title}{{An
  Alternative Approach to Measuring Reverberation Lags in Active Galactic
  Nuclei},} \apj, 735, 80, \dodoi{10.1088/0004-637X/735/2/80}

\end{thebibliography}

\end{document}